\documentclass[12pt]{extarticle}
\usepackage{amsmath}
\usepackage{amssymb}
\usepackage{stmaryrd}
\usepackage{color}
\usepackage{ wasysym }
\usepackage{color}

\newcommand{\tn}{\textnormal}

\newcommand{\wt}{\widetilde}

   \newtheorem{theorem}{Theorem}[section]
   \newtheorem{lemma}[theorem]{Lemma}
   \newtheorem{proposition}[theorem]{Proposition}
   \newtheorem{corollary}[theorem]{Corollary}

\newcommand{\tql}{\textquoteleft} 
\newcommand{\tqr}{\textquoteright}

\newcommand{\ben}{\begin{enumerate}}
\newcommand{\een}{\end{enumerate}}

\newcommand{\bt}{\begin{theorem}}
\newcommand{\et}{\end{theorem}}
\newcommand{\bl}{\begin{lemma}}
\newcommand{\el}{\end{lemma}}
\newcommand{\bc}{\begin{corollary}}
\newcommand{\ec}{\end{corollary}}
\newcommand{\bp}{\begin{proposition}}
\newcommand{\ep}{\end{proposition}}
\newcommand{\br}{\begin{remark}}
\newcommand{\er}{\end{remark}}

\newcommand{\bpf}{\begin{proof}}
\newcommand{\epf}{\end{proof}}

\newcommand{\be}{\begin{equation}} 
\newcommand{\ee}{\end{equation}}
\newcommand{\beq}{\begin{eqnarray}}
\newcommand{\eeq}{\end{eqnarray}}
\newcommand{\ba}{\begin{array}}
\newcommand{\ea}{\end{array}}
\newcommand{\bi}{\begin{itemize}}
\newcommand{\ei}{\end{itemize}}

\newcommand{\comm}[1]{}

\newcommand \qed {\hskip 6pt\vrule height6pt width5pt depth1pt \bigskip}

\newcommand{\cE}{{\cal E}}

\newcommand{\cR}{{\cal R}}

\newcommand{\cN}{{\cal N}}

\newcommand{\cW}{{\cal W}}

\newcommand{\cZ}{{\cal Z}}

\newcommand{\bbC}{{\mathbb C}}
\newcommand{\bbD}{{\mathbb D}}

\newcommand{\bbI}{{\mathbb I}}

\newcommand{\bbR}{{\mathbb R}}

\newcommand{\bbZ}{{\mathbb Z}}

\newcommand{\fh}{{\mathfrak h}}

\newcommand{\fz}{{\mathfrak z}}

\newfont{\msbm}{msbm10 scaled\magstep1}
\newfont{\msbms}{msbm7 scaled\magstep1} 

   \newenvironment{proof}[1][Proof]{\begin{trivlist}
   \item[\hskip \labelsep {\bfseries #1}]}{\end{trivlist}}

   \newenvironment{remark}[1][Remark]{\begin{trivlist}
   \item[\hskip \labelsep {\bfseries #1}]}{\end{trivlist}}

   \comm{\newcommand{\qed}{\nobreak \ifvmode \relax \else
      \ifdim\lastskip<1.5em \hskip-\lastskip
      \hskip1.5em plus0em minus0.5em \fi \nobreak
      \vrule height0.75em width0.5em depth0.25em\fi}}

 \numberwithin{equation}{section}

\begin{document}
\title{Can we trust the relationship between resonance poles and lifetimes?}
\author{Ira Herbst and Rajinder Mavi}

\maketitle
%

\begin{abstract}
We show that the shape resonances induced by a one dimensional well of delta functions 
disappear as soon as  
a small constant electric field is applied. 
In particular, in any compact subset of $\{ z : \text{Re} z >0, \text{Im z} <0\}$ there are no resonances if the non-zero field is small enough.  In contrast to the lack of convergence of the lifetimes computed from the widths of the resonances we show that the ``experimental lifetimes" are continuous at zero field. The shape resonances are replaced by an infinite set of other resonances whose location and number we analyze.
\end{abstract}

\tableofcontents

\section{ Introduction }
With few exceptions, virtually all long-lived states in nature are resonances,
so it is important to understand how these states behave under small perturbations, just as it is for bound states.

Recently, an example was discovered of a pre-existing resonance \cite{HR} of a Hamiltonian, 
which was not stable under an application of a small constant electric field:  none of the new resonances converge to the one pre-existing resonance when the field strength is taken to zero. 
This is surprising because it is known that discrete eigenvalues of atomic Hamiltonians move continuously into the complex plane and become 
resonances when a small  constant electric field is applied.  
 (See for example \cite{AvHe,CGH,CReAv,GGr1,GGr2,He,HS,H,YaTS}.)  
It might be expected that the same stability is true for atomic resonances near the real axis, 
for example the Auger states in helium, but this is not known.  

The example in \cite{HR} is a resonance created in a Friedrich's (or Weisskopf) model 
\[ H_F = 
   \left( \ba{cc} p^2  &  \mu\varphi \\  \mu\langle\varphi, \cdot \rangle  & 1 \ea  \right)    \]
acting in $L^2(\bbR) \oplus \bbC$.
For $\mu = 0 $, the model has an embedded eigenvalue in the continuous spectrum at 1, and
as $\mu \to 0$ the single resonance approaches  the embedded eigenvalue continuously. 
One might expect that perturbing $p^2 \to p^2 +fx$ and letting $f\to 0$ one would see the resonances approach those of $H_F$.
However, it is shown that this is not so.  None of the new resonances approach the resonance of $H_F$.  Instead an infinite set of new resonances is created and they all move to the real axis as $f \to 0$.

The natural question that arises then,  is whether this
unusual behavior is an artifact of the particular somewhat non-physical model above.
In response to these questions we consider the scenario of a shape resonance combined with
a small constant electric field. 

Thus consider the self-adjoint realization of the Hamiltonian

\be \label{H_eta}
H_{\kappa,\eta} =  p^2 + \tfrac{1}{\eta} \left(\delta_{\kappa} + \delta_{-\kappa}  \right)
\ee
in $L^2(\mathbb{R})$ with $p= -i\tfrac{d}{d x}$ and $\kappa > 0$. This is analogous to $H_F$ above when $\mu \ne 0$.  If we take $\eta \downarrow 0$ the monotone convergence theorem for forms \cite{RS} gives the operator 

\be\label{H_0}
H_0 =  H_{-}\oplus H_D \oplus H_{+} 
\ee
where $H_{-} =  -\frac{d^2}{dx^2}$ in $L^2((-\infty,-\kappa))$ with zero boundary conditions at $-\kappa$, $H_D =  -\frac{d^2}{dx^2}$ in $L^2((-\kappa,\kappa))$ with zero boundary conditions at $\pm \kappa$, and $H_{+} =  -\frac{d^2}{dx^2}$ in $L^2((\kappa, \infty))$ with zero boundary conditions at $\kappa$. This Hamiltonian is analogous to $H_F$ above with $\mu = 0$.  It has an infinite number of discrete eigenvalues embeddded in its continuous spectrum, namely the eigenvalues $(n\pi/2\kappa)^2, n \ge 1$ instead of the one embedded eigenvalue in the case of $H_F$ with $\mu = 0$. The Hamiltonian $H_{\kappa,\eta}$ has no eigenvalues for $\eta > 0$, but the embedded eigenvalues of $H_0$ become resonances which move continuously as a function of $\eta$.  These are the shape resonances.  Then as with $H_F$ we perturb with a small electric field by taking $p^2 \to p^2 + fx$ to obtain the self-adjoint realization of 
\be\label{Hss}
  H_{\kappa,\eta,f}  = p^2 + f x + \tfrac{1}{\eta} \left(\delta_{\kappa} + \delta_{-\kappa}  \right)
\ee
in $L^2(\mathbb{R})$.  This (and $H_{\kappa,\eta}$) will be our main objects of study.

The resonances are defined mathematically as the poles of the Green's functions, $ (H_{\kappa, \eta,f}-z)^{-1}(x,y)$, defined on the upper half plane upon being
extended below the positive real axis.  The resonance $z_n$ of $H_{\kappa, \eta}$ near the eigenvalue $E_n = (n\pi/2\kappa)^2$ (these are the  shape resonances) can easily be calculated perturbatively for small $\eta > 0$.  To second order we obtain (see Proposition \ref{zero2ndorder})  

\be \label{shape resonances}
\sqrt{z_n} = (n\pi/2\kappa)( 1 - \frac{\eta}{\kappa} + (\frac{\eta}{\kappa})^2 - i \frac{\eta^2}{\kappa} (n\pi/2\kappa)) +O(\eta^3).
\ee

On the other hand once the field is turned on, for small $f$ these resonances disappear without a trace: 

\bt\label{resonancedesc} 
           Let $\kappa,\eta > 0$,  $M > 20$
            and  $\beta \in (0,1)$.
            Then there is some $f_{\kappa,\eta,\beta} > 0$
             so that for any $f \in (0, f_{\kappa,\eta,\beta})$ 
            all resonances
            of $H_{\kappa, \eta, f}$  
                have absolute value between $f^{\frac{2}{3\beta}}$  and $f^{\beta/2}$  or are contained in the set 
\[  \{ z: |z|>f^{\beta /2}; |\arg{z} + 2\pi/3| < |z|^{-1}(1 + |z|^{-1/2}) M f\log(f^{-1} +|z|) 
                \tn{ or } |\arg{z}| <  |z|^{ - 3/2} M f\log(f^{-1} +|z|)  \}   \]  
\et

It is interesting to count the resonances just below the positive real axis for small $f$.  We have the following result:  Let $\mathcal{J}[a, b]$ be the number of resonances of $H_{\kappa, \eta, f}$ below the segment $[a,b]$ on the real axis counted with multiplicity.  Then 

\bt \label{resonancecountingtheorem} If $0 <a< b/2$, then
    $ 0 < \liminf_{f\to 0} \frac{\mathcal{J}[a, b]}{1/f} \le \limsup_{f\to 0} \frac{\mathcal{J}[a, b]}{1/f} < \infty$. \et
    
{The origin of these resonances is discussed briefly at the end of this section.}

\vspace{.1in}

The results of this paper and \cite{HR}
 stand in contrast to a theory of resonances under 
perturbations developed by Agmon in \cite{A98}.
In that work the resonances are  eigenvalues of an extension of the operator to a larger space.  These eigenvalues move analytically as a function of the
perturbation parameter. 
The results obtained in that paper do not apply to our models because of the singular nature of the electric field perturbation.

\vspace{.2in}

We now turn to the dynamics given by $e^{-itH_{\kappa, \eta, f}}$.  Specifically we are interested in the behavior of $e^{-itH_{\kappa, \eta, f}}\phi$ for small $f > 0$ where $\phi \in L^2(\mathbb{R})$ and where $\phi$ has bounded energy.  We enforce the latter restriction by replacing $\phi$ with $\chi(H_{\kappa,\eta, f})\phi$ with $\chi$ a continuous function of compact support.  Our first result does not mention resonances.  
            
            \begin{theorem} \label{small f approx}
            Suppose $T> 0$ and $\epsilon >0$ are given.  Then there is an $f_{\epsilon, T, \chi, \phi} > 0$ so that for $0 \le t \le T$, $0 < f < f_{\epsilon, T, \chi, \phi} $ and all $\eta > 0$
            \begin{equation}
            || e^{-itH_{\kappa, \eta, f}}\chi(H_{\kappa, \eta, f})\phi - e^{-itH_{\kappa, \eta}}\chi(H_{\kappa, \eta})\phi || < \epsilon 
            \end{equation}
            \end{theorem}
            Thus for arbitrarily long times the electric field can be chosen small 
            enough so that within a preassigned tolerance the particle acts as if there were no electric field . 
             This is certainly physically reasonable.
              In addition, it is certainly unavoidable that  $f_{\epsilon, T, \chi, \phi} $ cannot be chosen independent of
               $T$ for eventually a (classical) particle will be turned back by the electric field, no matter how small the
                field is as long as it is non-zero.  In fact this is true quantum mechanically as well.  Let $H_f = p^2 + fx$.  
                 Then one easily sees that the wave operators
                  $\lim_{t \to \pm \infty} e^{itH_{\kappa,\eta,f}}e^{-itH_f}$ exists and are unitary for $f>0$. 
                  This in turn follows 
                   (see \cite{Y})
                    from the fact that $(H_{\kappa,\eta,f} -i)^{-1} - (H_f- i)^{-1}$ is trace class 
                    (it is actually rank two) and that the spectra of both these operators are absolutely 
                    continuous (see Section \ref{resonancesection}).
            
            Our next result is the usual one making the connection between the lifetime 
            of states and the imaginary part of resonances (see for example \cite{H90}, \cite{KRW} or \cite{CGH}).
              Even though this kind of result has been proved in various settings, we have not seen 
              a proof which applies directly to our situation.  Since we are interested in the effect 
              of the shape resonances we take $\phi$ to be zero outside $[-\kappa, \kappa]$.  
              We also choose $\chi$ to be a continuous function with
               $1_{[a,b]} \le  \chi \le 1_{[a-\delta, b+\delta]}$ where $0 <a <b < \infty$.  
               We choose $a, b, \delta$ so that there are no eigenvalues of $H_D$ in $[a-\delta,a]\cup[b, b+\delta]$.
              We choose closed disjoint disks centered at the eigenvalues of $H_D$ in $[a,b]$
               with each disk $D_j$ containing exactly one eigenvalue, $E_j$.  
               For small enough $\eta$ there is exactly one resonance $z_j(\eta)$
                in each disk (see (\ref{shape resonances}) and Propositions \ref{Ftermzeroset} and \ref{singlesr}).  
                    
            \begin{theorem} \label{small eta approx}
            Given $\epsilon > 0$, $\phi$, and $\chi$ as above, denote the eigenvalues of $H_D$ in the support of $\chi$ by $E_j, j = 1, 2, \cdots,n$ and their corresponding normalized eigenfunctions by $\phi_j$. There is an $\eta_{\epsilon}> 0$ so that for all $t \ge 0$ and all $\eta$ with  $0 < \eta < \eta_{\epsilon}$
            \be 
           | \langle \phi, e^{-it H_{\kappa,\eta} } \chi(H_{\kappa,\eta})  \phi \rangle -  \sum_{j=1}^n| \langle \phi,\phi_j\rangle|^2 e^{-itz_j(\eta)}| < \epsilon. \ee 
           Here $z_j(\eta)$ is the resonance arising from the eigenvalue $E_j$.

            \end{theorem} 
            
           Combining the results of Theorems \ref{small f approx} and \ref{small eta approx} we see that the lack of resonances of $H_{\kappa, \eta, f}$ near the shape resonances of  $H_{\kappa, \eta}$ for small $\eta$ and $f$ does not contradict the expected physics (although it certainly contradicts our mathematical expectations).  We will come back to this shortly.
           
           \vspace{.2in}

The rest of this paper is organized as follows.  The  Green's function of the Hamiltonian perturbed by two delta functions  can be determined by
   standard methods and we describe this in Appendix \ref{greenbydelta}.
   The resonances are identified with the zeros of a certain determinant as described in that appendix.
   In Section \ref{resonancesection} we describe the resonances and determine
    regions in $\bbC$ where resonances do not exist,  In particular Theorem \ref{resonancedesc} is proved in this section. In Section \ref{dynamical work} we prove Theorems \ref{small f approx} and \ref{small eta approx} concerning the dynamics.
   In Section \ref{detgrowth} we use growth properties of the determinant to count the resonances 
   in certain domains of $\bbC$. 
   Theorem \ref{resonancedesc} follows from
    Proposition \ref{fddres} and Lemma \ref{originexpansion2}.

One may wonder if the instability in the model with Hamiltonian $H_{\kappa,\eta, f}$ as $f\to 0$ is due to the fact that the energy is unbounded below.  Note that when $ \eta \downarrow 0$ and $ f > 0$ a particle located to the right of $ \kappa$ is trapped no matter how small $f$ is.  The spectrum of the corresponding Hamiltonian in $L^2( (\kappa, \infty))$ with a Dirichlet boundary condition at $\kappa$ is pure point with discrete eigenvalues.  Thus we consider a secondary problem which has the advantage of not accessing an infinite energy reservoir but also does not allow escape to $+ \infty$.
  Let
  \be\label{Hds} 
    \tilde{H}_{\kappa,\eta,l}  = p^2  + \tfrac{1}{\eta} \left(\delta_{\kappa} + \delta_{-\kappa}  \right)
   \ee
  be an operator in $L^2(-\infty, l)$ with a Dirichlet condition at $l > \kappa$.  We think of $l \to \infty$ as analogous to $f \downarrow 0$.

This Hamiltonian, $\tilde{H}_{\kappa, \eta, l}$, has resonances similar to those of ${H}_{\kappa, \eta, f}$ at least in the right half plane. This bolsters our opinion that these resonances are remnants of the discrete eigenvalues of the two operators  $p^2$ with Dirichlet  boundary conditions at $\kappa$ and $l$            in $L^2(\kappa, l)$ and the operator $p^2 + fx$ with a Dirichlet boundary condition at $\kappa$ in $L^2(\kappa, \infty)$. The same instability occurs as $l\to \infty$ as when $f\downarrow 0$ and thus  it is unrelated to the unboundedness below of $H_{\kappa, \eta, f}$. We discuss this further in Section  \ref{resonancesection}.  

A trivial classical calculation shows that a particle moving to the right with kinetic energy $k^2$ will take a time of order of $2k/f$ to return after being pushed back by the electric field or a time $l/k$ to reflect off the wall at $l$ and return.  This might indicate  resonances near $k^2 - iNf/2k$ for small $f$ or $k^2 - iNk/l$  for large $l$ where $N$ is some measure of the number of times the particle comes back and hits the delta potential at $\kappa$ before escaping.  The former is in rough agreement with Theorem \ref{resonancedesc}.

\section{Acknowledgements}

We would like to thank Larry Thomas and Kiril Datchev for enlightening discussions.

\section{Description of resonances}\label{resonancesection}
   
    We now define the resonances we study in this paper.  We start with the Green's function which is analytic in the upper half plane as a function of the spectral parameter.  The resonances will be defined as poles in the lower half plane in the analytically continued Green's function.  To find the Green's functions we 
 remove the delta potentials from the Hamiltonian and use the second resolvent equation to add them back in.  This gives an easily solved equation for the Green's functions.  Details of this calculation are given in Appendix \ref{greenbydelta}.

   The  Hamiltonian (\ref{H_eta}) approaches the free Schr{\"o}dinger operator 
    in the $\eta  \to \infty $ limit.
    On the other hand the Hamiltonian (\ref{Hss}) 
     becomes an Airy Hamiltonian as $\eta \to \infty$.
     For now let $K_\infty$ be the Hamiltonian in (\ref{H_eta}) or (\ref{Hss})
     with $\eta = \infty$.  Then for $0 < \eta < \infty$ we may write
      \[K_\eta = K_\infty + \eta^{-1} (\delta_{\kappa}  + \delta_{-\kappa}) \]

   The analytic continuation of this function below the real axis 
   results in poles which we identify as the resonances of the model. 
        Let the square integrable solution at $-\infty$ of $K_\infty  u= z u $  for $\Im z> 0$ 
             be  $ u = \psi$ and the square integrable solution at $+\infty$   be $u = \phi$
              and denote  the Wronskian by $\cW = \psi'\phi - \phi'\psi$.  Then the 
               Green's function of the $K_\infty$ Hamiltonian is given by 
              $G_z^{(\infty)}(x,y) = \frac{1}{\cW} \psi(x) \phi (y)$ for $x< y$.

              Let $\alpha_\pm = \psi(\pm\kappa) \phi(\pm\kappa)$ and
              $\beta_{\pm} = \psi(\mp \kappa) \phi(\pm \kappa)$,
               notice $\alpha_+\alpha_- = \beta_+ \beta_-$.
               Let $g_\pm(x) = G^{(\infty)}_z(\pm \kappa , x)$, then
                the Green's function for $0 < \eta < \infty$ is given by
                \be\label{green1}
                     G^{(\eta)}_z(x,y) = G^{(\infty)}_z(x,y) - \frac{1}{\eta^2\cW  D}L_z(x,y)
                \ee     
                Where
                \[
                L_{z} (x,y) = 
                 \left( \ba{cc} g_+(x)& g_-(x) \ea \right)
                        \left[ \ba{cc} \eta\cW + \alpha_- & - \beta_+ \\ - \beta_+ &  \eta \cW  + \alpha_+\ea  \right]
                        \left( \ba{c} g_+(y) \\   g_-(y) \ea \right) \]
               and 
              \be\label{greendet}
                D(z) = 1 + \frac{\alpha_+ + \alpha_-}{\eta \cW} + \frac{\beta_+ (\beta_- - \beta_+)}{  \eta^2\cW^2  } 
                 \end{equation}
                Note that $\eta^2 \cW^2 D$ is the determinant of the above matrix.
                
              As we discuss in detail below, $D$ is an entire analytic function of $z$ 
              when $K_\infty$ is the $\eta = \infty$ case of (\ref{Hss}) 
               and an analytic function  of $\sqrt z$   (except for a pole at $0$) when $K_\infty$
                is the $\eta = \infty$ case of (\ref{H_eta}).
               Likewise the functions $\cW$, $\alpha_\pm$, $\beta_\pm$  and  $g_\pm(x)$  for fixed $x\in \bbR$ 
               are analytic in their corresponding variables ($z$ or $\sqrt z$).
               For fixed $x,y\in \bbR$, 
               the Green's functions
               $G^{(\eta)}_z(x,y)$ and  $G^{(\infty)}_z(x,y)$ are analytic in the upper half
               plane.

                We claim there are no zeros of $D$ in the upper half plane and zeros of 
                $D$ in the lower half plane lead to poles of $G^{(\eta)}$.
                To see there are no zeros of $D$ in the upper half plane notice that 
                 $D$ is independent of $x,y$. Therefore if $D(z) = 0 $ at some $z_0$ in the upper half plane
                 then $L_{z_0}(x,y) = 0$ for all $x,y\in \bbR$  so that $G^{(\eta)}$ remains analytic in the 
                 upper half plane as a bounded operator.  Making use of the fact that $\phi(x)\phi(y), \phi(x)\psi(y), \psi(x)\phi(y)$, and $ \psi(x)\psi(y)$ are linearly independent functions for $z_0$  in the upper half plane it is easily shown that $D(z_0)=0$ and $L_{z_0}=0$ cannot both be true.  A similar argument shows that a zero of $D$ in the lower half plane gives a pole of $G^{(\eta)}$.  We omit the details.

               Both (\ref{H_eta}) and (\ref{Hss})
                have no eigenfunctions, which is well known and
               follows from the fact that $K_\eta u = E u$
               has no square integrable solutions at $-\infty$ for $E \in \bbR$.  In this sense the two Hamiltonians are similar.
               On the other hand, $p^2 \to p^2 + fx$ is a large perturbation.  As discussed in the 
               introduction the spectra are vastly different with  
               $\sigma(p^2) = [0,\infty)$ while  $\sigma(p^2 + fx) = \bbR$.
               This indicates that we should treat the cases of the (\ref{H_eta})
                and (\ref{Hss}) Hamiltonians separately.
                Therefore, we begin in Section \ref{sectshaperes} with the 
                resonances of Hamiltonian (\ref{H_eta}) and continue in 
                Section \ref{mainproofs} with the analysis of the resonances of 
              Hamiltonian   (\ref{Hss}).  In particular Theorem \ref{resonancedesc}  is proved in  Section \ref{mainproofs} using Proposition \ref{fddres} with some help from Lemma \ref{originexpansion2}.

  \subsection{The Hamiltonian (\ref{H_eta}), the shape resonance}\label{sectshaperes}
   While (\ref{shape resonances}) gives an approximation of the resonances for small $\eta$,
   this approximation is not uniform in $n$.
   We now present uniform approximations of the shape resonances over the right half plane. Again
   proofs are contained in the appendix.
  
   We return to the free  problem   on $\bbR$ perturbed by two delta functions, equation (\ref{H_eta}),
   and refer to (\ref{greendet}) for how to calculate the resonances in the lower half plane.
 Then $\psi = e^{- i\sqrt z  x}$
 is the left solution,  $\phi = e^{i \sqrt z x}$
 is the right solution and the Wronskian is $\cW = \psi' \phi - \phi' \psi = - i 2\sqrt z$.
  In this case (\ref{greendet}) becomes
   \be\label{det0}  D_0 = 1 +  \frac{1}{ -  i\eta\sqrt z} + \frac{1- e^{ i4\kappa\sqrt z}  }{ - 4 \eta^2 z } = \frac{(2\eta \sqrt z + i)^2  + e^{4i\kappa \sqrt z}}{4\eta^2 z},\ee
  and the zeros $\cZ_0$ of this function are the resonances.


  Observe that $D_0$ may be factored as $D_0 = F_+(2\eta,2\kappa; \sqrt z)  F_-(2\eta,2\kappa; \sqrt z)$
  where
  \be\label{Fterm} F_\pm(s,t;w)   = 1 +i \frac{1 \pm  e^{itw}}{sw}  \ee
   The following holds for the  $F_{\pm}$ functions. Fix $t = 2\kappa > s = 2\eta > 0$ 
   and let $w = u+iv = \sqrt z$.
               \bp\label{Ftermzeroset} 
                 Suppose $t > s$.  Then
                      all zeros of $F_\pm(s,t;w)$ in the right half plane are contained in the set
                 \[\{w : u > 0;    -(2t)^{-1} \log (1 +(s u)^2)   < v <  - (4t)^{-1} \log ( 1 + (s u)^2)   \}\]
                Moreover,
              in the right half plane,
                 the zeros of $F_+$ are given as exactly one zero $w_n^+$ with real part in each interval $t^{-1}(n\pi, (n+1)\pi)$ for nonnegative even integers $n$. 
                 The zeros of $F_-$ in the right half plane are given exactly as
                   one zero $w_n^-$ with real part in each interval $t^{-1}(n\pi, (n+1)\pi)$ for nonnegative 
                   odd integers $n$.

            As for the zeros in the left half plane, these are exactly the reflections across the
          imaginary axis of the zeros in the right half plane.
             \ep

  This result is proved in  Appendix \ref{F+-}, in Propositions \ref{singlesr} and \ref{Fzeroset}. 

 \subsection{The shape resonance perturbed by an electric field (\ref{Hss}), proof of Theorem \ref{resonancedesc}} \label{mainproofs}
  Our goal in this section is first to find the function (\ref{greendet}) associated to (\ref{Hss}),
  then to locate the zeros of  the function. 
   It will be seen that the associated function is entire analytic in $z$.

  We refer to (\ref{greendet})  for the defnition of the function which
   determines  the resonances.
In this case, the Hamiltonian $K_\infty = H_{\kappa,\infty,f } = p^2 + fx$ 
 is unperturbed  by delta functions.
 The solutions $g$ to $(p^2 +fx - z)g = 0$ may be expressed in terms of Airy functions.
In particular, for $z$ in the upper half plane,
  the solutions of $(H_{\kappa,\infty,f} -z)g = 0$ which are $L^2$ on the left and right are respectively,
\[  \psi(x) = A\left( - \omega  \frac{z}{ f^{2/3}} (1 - \tfrac{xf}{z} )\right) 
         \textnormal{ and }     \phi(x) = A\left( -    \frac{z}{ f^{2/3}} (1 - \tfrac{xf}{z} )\right)\]
   where $\omega = e^{i2\pi / 3}$ and $A(x) = \frac{1}{2\pi}\int_{-\infty}^{\infty} e^{i(tx + t^3/3)} dt$.
    The Wronskian of these solutions is $  \cW =  \psi' \phi - \phi' \psi =   f^{1/3}e^{-i\pi/6}/2\pi $.  

With the notation  $\theta_\pm = 1 - \tfrac{\pm\kappa f}{z}$, we write  $\alpha$ and $\beta$ in (\ref{greendet}),
\be\label{alphaform}   \alpha_\pm = \psi(\pm\kappa)\phi(\pm\kappa) 
        = A\left( - \omega  \frac{z}{ f^{2/3}} \theta_{\pm} \right)
          A\left( -   \frac{z}{ f^{2/3}} \theta_{\pm}\right) \ee
 \be\label{betaform}      \beta_\pm = \psi(\mp\kappa)\phi(\pm\kappa) 
        =  A\left( - \omega  \frac{z}{ f^{2/3}} \theta_\mp  \right)
          A\left( -   \frac{z}{ f^{2/3}} \theta_\pm  \right) .\ee
          
We refer to $D$ in
  (\ref{greendet}) with these value of $\alpha_\pm$, $\beta_\pm$ and $\cW$ as $D_f$.
  The analyticity of $D_f$ follows immediately from the analyticity of the Airy function.

  The zeros of $D_f$ are exactly the resonances.  
  
   \subsubsection{Approximations of the function $D_f$}
     Although the function $D_f$ is entire, in
   some regions of the plane there may be highly oscillatory behavior, especially as $f\to 0$,
   so extracting useful estimates from the definition above of the Airy function is difficult.
   We rely on classical asymptotic expansions of the Airy functions which hold on
     limited sets of the plane and even then we must 
   be careful of sets where approximations are difficult.  
 Thus let $0 < a < 1/2$ and define for all $f > 0$
 \be     \label{originbound}
           \cR_0 = \{   z \in\bbC:  | z | > f^{a}  \}. \ee
  We obtain useful asymptotic expansions on $\cR_0$ for sufficiently small $f>0$.
    On the other hand, the expansions suggest the function $D_f$ is highly oscillatory
     in the vicinity of the rays $\theta = 0$ and $\theta = -2\pi /3$ 
    and thus in these regions it is difficult to obtain useful approximations.  
    However in the connected regions in the complement of these rays one does find useful approximations. 
         Write $z = r e^{i\theta}$,   and for given $ a \in (0,1/2)$ 
        define the sets for some $M >>1$ ($M > 20$ is sufficient)
     \[ \cR_1 = \left\{z = re^{i\theta}\in  \cR_0: -2\pi/3 + r^{-3/2} M  f\log(r + f^{-1}) 
             <  \theta <  - r^{-3/2} M  f\log(r + f^{-1})\right\} \] 
        and
          \[   \cR_2 = \left\{z\in  \cR_0 :   
                        -2\pi   \leq  \theta <   -2\pi/3 - r^{-1}(1 + r^{-1/2})M f \log(r + f^{-1})   \right\} . 
           \]
We will show that for small $f$, $D_f$ has no zeros in these regions.

  \bp \label{fddres}
    For $0 < a<1/2$ and    $ M  > 20$ there is an $ f_{a, M }>0$ so that for $0 < f< f_{a, M}$, 
    there are no zeros of $D_f$ in the sets $\cR_1$ or $\cR_2$.
  \ep

   \begin{remark} 
     Up to showing there are no zeros close to the origin, Theorem \ref{resonancedesc} follows from 
     Proposition \ref{fddres}.
    The separation of the zeros from the origin follows 
      from Lemma \ref{originexpansion2}.
   \end{remark}

   \br Proposition \ref{fddres} will follow from the asymptotic
    properties of $ D_f$ in $\cR_1$ and $\cR_2$.
    We will state the approximations and then prove the Proposition.
   Asymptotic approximations 
   over most of the complex plane
   are also needed in Section \ref{detgrowth}, 
   so in this section we state somewhat more details than are strictly necessary 
   for the proof of Proposition \ref{fddres}. 
   In particular, we state approximations over most of the upper half plane,
   where we have already established that $D_f$ has no zeros.
   \er

   It is convenient to define the pair of  functions
   \be\label{gb}   g_b(w)  = \sin(w) + b w \cos (w), \tn{ and } h_0  =    \left( {2\eta \sqrt z}   +  {i}\right)^2 + 
                          e^{i4\kappa \sqrt z} .\ee 
As stated in the remarks
 we will later require estimates  
 for $D_f$  over most of the complex plane. 
   Therefore, we  exchange $\cR_2$ for a set where asymptotic expansions can be obtained,
   \[  \wt\cR_2 = \{z\in \cR_0 :     4\pi/3 - (1 + r^{-1/2}) r^{-1}  fM \log (f^{-1} + r) 
    >   \theta > r^{-3/2} fM \log (f^{-1} + r)   \} . \]
    
   Now $D_f $ is approximated in $\cR_1$ and $\wt\cR_2$ in terms of the functions in (\ref{gb}).  This means that $\sqrt z = |z|^{1/2}e^{i\theta/2}$ where the ranges of $\theta$ are given in the definitions of $\cR_1$ and $\wt\cR_2$.
   $D_f$
    is approximated   in $\cR_1$    by 
\[F_1(z) = \frac{ g_{\eta/\kappa}(2\kappa \sqrt{z})    }{2z\eta^2}   
  e^{i\frac{4z^{3/2}}{3f}   +  i\frac{f\kappa^2}{2\sqrt z}      }   \]
   and $D_f$ is approximated in $\wt\cR_2$ by 
\[F_2(z) = \frac{1}{4z\eta^2}   h_0(z). \] 
        \bl  \label{fddapprx2}
    For $0 < a<1/2$ and    $ M  > 20$ there is an $ f_{a, M }>0$ and finite $ C$
    so that for $0 < f< f_{a, M}$, 
           and $z \in \cR_1$ we have,
          \[         
                  \left|\frac{D_f (z)}{F_1(z)} -1 \right|      \leq    C \frac1{\log(|z| + f^{-1}) } \]  
          and for $z \in \wt\cR_2$ we have 
                  the approximation
       \[ \left| {D_f (z)}  - {F_2(z)} \right|      \leq     Cf^{2}  |z|^{-5/2} \left( 1 + |z|^{-3/2} \right) . \]
               \el
               This lemma will follow from Lemma \ref{fddapprx} in Section \ref{DinRlemmaproof}.
           One more lemma is needed,  a simple lower bound for $g_b$.  
           An upper bound will be useful later, 
           in Section \ref{detgrowth}, so we state that here as well.
           The proof of the following  lemma is  the content of Appendix \ref{gbproof}. 
            \bl\label{gtermest}
              For $|w| >0 $ and $- \pi < \arg w < \pi$, let $w = u+ iv$.  Then
            \[              [b|w| +1]   \cosh|v|   >
                  |g_b(w)| >\sqrt{ (b|w|)^2 +1}     \sinh|v|. \]
        \el 
   \begin{proof}\textbf{of Proposition \ref{fddres} }
   
   The proof of Proposition \ref{fddres} follows from Lemmas \ref{fddapprx2} and \ref{gtermest}.
   Indeed, in $\cR_1$ the quantity  $F_1(z)$ cannot be zero so if $D_f(z) = 0$ then $1\le C/\log(r + f^{-1})$ which is impossible for small $f$.
   
   Since we have
   already eliminated the possibility of zeros in the upper half plane we can replace $\wt\cR_2$ by
   \[  \cR_2' =   \{z \in \cR_0 :  \pi  < \arg z < 4\pi/3 - r^{-1} (1 + r^{-1/2} ) M f \log(r + f^{-1})  \}    \]
   In this set we have $\Im\sqrt z = r^{1/2} \sin(\theta/2) \ge  r^{1/2}\sqrt3/2$ which gives $$|F_2(z)| = |h_0(z)/4z\eta^2| \ge [(\sqrt3 \eta r^{1/2} +1)^2 - 1]/4r\eta^2 > 3/4.$$
   On the other hand in $\cR_0$ we have the bound on the error term $ f^2 |z|^{-5/2} \left( 1 + |z|^{-3/2} \right) <2 f^{2 - 4a}$ for $0 < f < 1$ and thus for small $f$ we cannot have $D_f(z) = 0$ in $\cR_2'$ . \qed
\end{proof}

%

\subsubsection{Proof  of Lemma  \ref{fddapprx2}} \label{DinRlemmaproof}

 We approximate $D_f$ in three sectors of $\cR_0$.
  In particular,
  by 
    \beq\label{Deast}
      D^{(E)}
         &  :=&
            \frac{1}{4z\eta^2}  \left\{  h_0(z)+
                 2 e^{i\frac{4z^{3/2}}{3f}   +  i\frac{f\kappa^2}{2\sqrt z}     } g_{\eta/\kappa}(2\kappa \sqrt z)    \right\}  + fD^{(1)}  
            \eeq
   in the sector $\cR_0^E   =   \{z \in \cR_0:   |\arg z|  <  2\pi/3 -\epsilon\}$.
    Here the first order term is 
   \be\label{Dfo}
        D^{(1)}  =   \frac{- i }{2z^{3/2}\eta^2}
                    e^{i\frac{4z^{3/2}}{3f}   +  i\frac{f\kappa^2}{2\sqrt z}     } 
                      \left[   c_1\frac{ 3 }{ z }     g_{\eta/\kappa}(2\kappa \sqrt z)  
                          +  \kappa\eta   \sin(2\kappa \sqrt z)    \right] \ee
   where $c_1 = \frac{\Gamma(3+1/2)}{54\Gamma(1 +1/2) } = \frac{5}{72}$,
    although its precise value will not concern us.
   By $D^{(N)}(z) =       \frac{1}{4z\eta^2}        h_0 (z)$
      in the sector $\cR_0^N   =   \{z \in \cR_0:  \epsilon <    \arg z  <  4\pi/3 -\epsilon\}$,
   and finally 
 \beq  \label{Dsouth}  D^{(S)}(z)    & := & 
         \frac{1}{4z\eta^2} \left\{ \tilde h_0(z)
           +2 e^{i\frac{4}{3f}z^{3/2}   +  i\frac{f\kappa^2}{2\sqrt z}      } 
           g_{\eta/\kappa}(2\kappa \sqrt{z})  \right\}    + f D^{(1)}
     \eeq
   in the sector 
 $\cR_0^S   =   \{z \in \cR_0: - 4\pi/3  +  \epsilon <    \arg z  <  -\epsilon\}$.
  Here 
  $ \tilde h_0(z) = \left(  2\eta \sqrt z   - i\right)^2 + e^{-i 4\kappa \sqrt z}  $ where $\sqrt z = |z|^{1/2} e^{i\theta/2}$ and where the range of $\theta$ is given in the definition of  $\cR_0^S $.

   Finally, we introduce functions which act as error terms in $\bbC$,
     \[ \cE_1(z) = \frac{f^2}{|z|^{ 2}} \left|e^{i\frac{4z^{3/2}}{3f}    +   i\frac{f\kappa^2}{2\sqrt z}}\right|  
             e^{ 2\kappa|\Im  \sqrt z|}
               \left(1 + |z|^{-2}  \right)  \]
     and  
    \[ \cE_2(z) =   \frac{f^2}{|z|^{5/2}}  
         \left(   1 +  |z|^{-3/2}\right)  \left(  1 + \left|e^{  i\kappa 4\sqrt z}\right| \right)    
            ;\;\;     \cE_2'(z) =   \frac{f^2}{|z|^{5/2}}  
         \left(   1 +  |z|^{-3/2}\right)    \left(  1 + \left|e^{ - i\kappa 4\sqrt z}\right| \right)  .      \] 
     \bl\label{fddapprx}
            Fix $\epsilon>0 $ then,
         for small $f > 0$ we have in $\cR^E_0$,
     \[ D_f = D^{(E)} +  O(\cE_1) + O\left(\cE_2  \right)  \]
       in $\cR_0^N$ we have  
      \[
         D_f =
            D^{(N)}+ O\left(\cE_2 \right)    \]
       and in $\cR_0^S$ we have the approximation
 \[
    D_f  =  D^{(S)}  + O(\cE_1) + O\left(\cE'_2 \right)
  \]
        \el    
        
        \br
        Notice in $\cR^E_0$ and $\cR^S_0$ there are terms $\wt h_0 \neq h_0$
           in the overlapping set $\{ z \in \cR_0:  -2\pi/3+\epsilon < \arg z < \epsilon  \}$.
           However, in this set these terms are small compared to the error term $\cE_1$.
           The $h_0$ term becomes the leading term when $\epsilon < \arg z$ in $\cR^E$ and 
           $\wt h_0$ becomes the leading term in $\cR^S$ when $\arg z < -2\pi /3 - \epsilon$.
        \er

         \br
          Lemma \ref{fddapprx} follows from  asymptotic expansions of the Airy function as demonstrated in Appendix \ref{Dfapprx}. 
         \er

           We now show Lemma \ref{fddapprx2} follows from Lemma \ref{fddapprx} (and Lemma \ref{gtermest}).
            The method is simply to show all terms are small compared to the leading terms in the given regions.
         \bpf \textbf{of Lemma  \ref{fddapprx2}}
           We break the proof into roughly the following sectors,
           $ -\pi/3<  \theta <0; -2\pi/3< \theta < -\pi/3; -\pi < \theta < -2\pi/3;0< \theta < \pi/2$ and  $\pi/2 < \theta < \pi$.
\bi
         \item[1.]   First consider the  region
          \be \label{firstsixth}
            \left\{z: |z| > f^a; - M f\log(|z| + f^{-1})    |z|^{-3/2} > \arg z \geq  -\pi /3  \right\}
          \ee
           where according to Lemma \ref{fddapprx},
          $D_f$ is approximated by $D^{(E)}$ given in (\ref{Deast}).
          In turn $F_1$ is simply a term in $D^{(E)}$ which we 
        will show dominates the other terms.
         We first compare $F_1$ to the other non-error terms.
         Consider,
         \be\label{estimatequant1} \left|\frac{h_0/(4z\eta^2)}{F_1} \right|
          < C    \frac{|h_0|}{ |g_{\eta/\kappa}(2\kappa \sqrt z)|} 
           \left| e^{i\frac{4z^{3/2}}{3f}   +  i\frac{f\kappa^2}{2\sqrt z}      }\right|^{-1}   \ee
           we will show this is bounded by    $C (r + f^{-1})^{-(M-3/2)}$ for small $f$ 
           in the region given by (\ref{firstsixth}).
        Let $z = re^{i\theta}$,   of course 
          \[   |h_0| <    \left(2[4\eta^2  |z|  +1 ] + |  e^{  i  4\kappa  z^{1/2} } | \right)  
          < C  \left(   r  +  e^{    2\kappa r^{1/2} |\theta| }   \right)    \]
         for $C$ depending on $\eta$,  and 
         \[   
           \left| e^{i\frac{4z^{3/2}}{3f}   +  i\frac{f\kappa^2}{2\sqrt z}      }\right|^{-1} 
           = 
                   e^{  -\frac{4}{3f}r^{3/2} \sin|3\theta/2| + \frac{f\kappa^2}{2r^{1/2} }\sin|\theta/2| } 
                   < 
                   e^{  -\frac{1}{f}r^{3/2} |\theta| + \frac{f\kappa^2}{4r^{1/2} } |\theta| } 
                   <
                  C    e^{  -\frac{1}{f}r^{3/2} |\theta|  } 
              \]
      where in the first inequality we use $\sin|\theta/2| < |\theta/2|$ and $\sin|3\theta/2| > 3|\theta|/4$.
      The term $g$ is bounded below
        by Lemma \ref{gtermest}, that is for some $c>0$,
       \be\label{glowerbound}
          |g_{\eta/\kappa}(2\kappa \sqrt z)|  > \sqrt{ 4\eta^2 r + 1 } \sinh(2\kappa r^{1/2} \sin |\theta/2|)     
                >    c (1 + r)^{1/2}\sinh(2\kappa r^{1/2} \sin |\theta/2|)         \ee
        Thus we have for all $r > 0; - \pi /3 \leq  \theta \leq 0$ the bound (\ref{estimatequant1}) becomes
         \be\label{estquant1}
          \left|\frac{h_0/(4z\eta^2)}{F_1} \right| < 
          C \frac{r  +   e^{  2\kappa r^{1/2} |\theta|  } }{(1 + r )^{1/2}}
                 \frac{  e^{  -\frac{1}{f}r^{3/2} |\theta| } }{   \sinh(2\kappa r^{1/2} \sin |\theta/2|)     } 
         \ee
         Now, we evaluate the smallness of the right hand side of (\ref{estquant1})
         in the region described in (\ref{firstsixth}), and this is done by breaking the region into
          cases $r^{1/2 } |\theta |  \geq 1$ and $ r^{1/2} |\theta| \geq  1$.  
          If  $r^{1/2} |\theta| \ge 1$ 
           the hyperbolic sine term is bounded below
           and we have $r^{3/2}|\theta| > r  $ so that
        \[  
          \left|\frac{h_0/(4z\eta^2)}{F_1} \right| < 
          C \frac{r  +   e^{  2\kappa r^{1/2} |\theta|  } }{(1 + r )^{1/2}}  e^{- \frac{r }{f} } 
          < C (1  +  r)^{1/2}e^{- \frac{r}{f} + 2\kappa r^{1/2}}\]
          which is bounded by $e^{- r/(2f)}$ for small enough $f$.
       Now for $z= re^{i\theta}$ in the region $\{z: r^{1/2} \theta<1;   r^{3/2}|\theta| < M f \log(f^{-1} +r )\}$
        we have then that for some $c > 0$
        \be\label{sinhlwb}
           \sinh(2\kappa r^{1/2} \sin |\theta/2|)     >c r^{-1} M f \log(r + f^{-1}) 
        \ee
       so that (\ref{estquant1}) becomes
       \[
          \left|\frac{h_0/(4z\eta^2)}{F_1} \right| < 
          C  (1 + r )^{1/2} 
         \frac{  e^{  -\frac{1}{f}r^{3/2} |\theta| } }{    f \log (r + f^{-1})   }
             <   
          C  (1 + r )^{1/2} 
         \frac{   (r + f^{-1})^{-M}}{    f \log (r + f^{-1})   } < C (r + f^{-1})^{-(M-3/2)} 
       \]
%

         Now consider the first order term $f D^{(1)}$ in the same region.

         The sine term of (\ref{Dfo}) is bounded as
         \[  |\sin(2\kappa \sqrt z )| < 2 e^{ 2 \kappa r^{1/2} \sin ( |\theta|/2) } \]
         so for some finite $C$ depending on $\eta,\kappa$,  
         \be\label{estimated1}
          \left| \frac{ f D^{(1)}}{F_1} \right| <
             f  C  \left( \frac{1}{|z|^{3/2}}
                          + \frac{1}{ |z|^{1/2}}
                           \frac{  e^{ 2 \kappa r^{1/2} \sin  |\theta/2| }  }{   |g_{\eta/\kappa}(2\kappa \sqrt z) |   }     \right)  
                            \ee
       Recall that $|z| = r \ge f^{a} $ for $a < 1/2$, so the first term is of order $f^{1 - 3a/2}   $.
       Since the power series of $g_b$ around $0$ starts out as $(1+b)w \dots $  
       there exists $\epsilon_1 >0$ so that
        for $0 < |w| < \epsilon_1$ and any $b >0$,
        we have the lower bound 
       \be\label{gnearzero}
              |g_b(w)| > |w| .
       \ee
       Thus for $|z| = r < \epsilon_1^2/4\kappa = :\epsilon_2$
            \be\label{estimated12}
          \left| \frac{ f D^{(1)}}{F_1} \right| <
             f  C  \left( \frac{1}{|z|^{3/2}}
                          +  
                           \frac{  e^{ 2 \kappa r^{1/2} \sin  |\theta/2| }  }  { |z|^{3/2}}     \right)  
             <  C f |z|^{-3/2}  < f^{1/4}
                            \ee
            for sufficiently small $f$.
             Note that $|z| $ is small here so this is certainly of order $(\log (|z| + f^{-1}))^{-1}$. 
            
            Now we consider the second term on the right in (\ref{estimated1}) for $|z| \geq \epsilon_2$.
       First consider $r^{1/2}|\theta|  > 1$.  We apply (\ref{glowerbound})
       so that the second term on the right of (\ref{estimated1}) is bounded by
         \be\label{estimated13}
          \frac{Cf}{ r^{1/2}}
                           \frac{  e^{ 2 \kappa r^{1/2} \sin  |\theta/2| }  }{ (1 +r)^{1/2}   \sinh(2\kappa r^{1/2} \sin |\theta/2|)    }    
                            \ee
       so the exponential and hyperbolic sine terms are balanced and $r$ is bounded below so
       so that (\ref{estimated13})
      is of order $f (1 + |z|)^{-1} $.
      Finally consider $r^{1/2}|\theta| < 1$ and $r  \ge \epsilon_2$
       and apply (\ref{glowerbound}) and (\ref{sinhlwb}) to 
      (\ref{estimated1}) 
        so that
               \be\label{estimated14}
          \left| \frac{ f D^{(1)}}{F_1} \right| <
            C  f (1 + r)^{-3/2}
                          + 
                         C  \frac{ 1 }{   \log(r + f^{-1}) }    .
                            \ee

       We now address the error terms.  First consider  $|\cE_1/F_1|$,

              \be\label{estimated3}
               \left| \frac{\cE_1  }{F_1(z) }\right| <  
          \frac{  e^{  2\kappa  r^{1/2} \sin|\theta /2|   }  }{ | g_{\eta/\kappa}(2\kappa \sqrt z) |}    
           \left(
              \frac{f^2}{r}  
               \left(1 + r^{-2}  \right)\right)
            \ee
         First, for $|z| = r $ sufficiently small, ie, $f^a < r < \epsilon_2$,  and using (\ref{gnearzero}) we have
       \[      \left| \frac{\cE_1  }{F_1(z) }\right| <  
                   C f^{2 - 7a/2}.\] 
            Again, $a < 1/2$ and $|z|$ is small so this is of order $(\log(r + f^{-1}))^{-1}$.

          Now, assume $r \geq \epsilon_2 $. 
          If   $r^{1/2}|\theta| \geq 1$
           apply  (\ref{glowerbound}) to (\ref{estimated3}) to get
            \[
               \left| \frac{\cE_1  }{F_1(z) }\right| <   C
          \frac{  e^{  2\kappa  r^{1/2} \sin|\theta /2|   }  }{   (1 + r)^{1/2} \sinh(2\kappa r^{1/2}\sin|\theta/2|)   }    
           \left(
              \frac{f^2}{r}  
               \left(1 + r^{-2}  \right)\right)
           \]
           then as $r$ and $r^{1/2}|\theta|$ are bounded below and  the hyperbolic sine terms 
           and exponential terms are balanced for large $r^{1/2}|\theta|$  this is a term of order $(1+r)^{-7/2}f^2$.
          Finally, assume $r \geq \epsilon_2 $ and consider $r^{1/2}|\theta| < 1$.
            We use (\ref{sinhlwb})  on (\ref{estimated3})  so that,
                 \[
               \left| \frac{\cE_1  }{F_1(z) }\right| <  
          \frac{C  }{   M\log(r + f^{-1}) }    
           \left(
              \frac{f}{r}  
               \left(1 + r^{-3/2}  \right)\right)
               < C f \frac{1}{\log(r + f^{-1})}.
            \]

       To finish, we consider $\cE_2$,
       \be \label{estimatede2}
           |\cE_2/F_1| < 
              C  \frac{f^2}{r^{3/2}}(1 + r^{-3/2})  
              \frac{ e^{4\kappa r^{1/2 }\sin |\theta/2|}  }{  |g_{\eta/\kappa}(2\kappa \sqrt z)|   }
                   e^{  -\frac{4}{3f}r^{3/2} \sin(3 |\theta|/2 )}  
       \ee

       Now suppose   $\epsilon_2 > |z| > f^a$ and use (\ref{gnearzero}).
       We have the bound $    |\cE_2/F_1| < f^2 |z|^{-7/2}  $
       Now we may suppose that $|z| > \epsilon_2$,
        and use Lemma \ref{gtermest} to obtain,
       \be\label{estimated2}
           |\cE_2/F_1| < 
              C  \frac{f^2}{r^{3/2}} 
              \frac{ e^{4\kappa r^{1/2 }\sin |\theta/2|}  }{  (1 + r)^{1/2} \sinh(2\kappa r^{1/2}\sin|\theta/2|)   }
                   e^{  -\frac{1}{f}r^{3/2} |\theta|} .
           \ee
       Now $r > \epsilon_2$. First consider $r^{1/2} \theta \geq 1$.  Then $r^{3/2}|\theta | > r$ so we have,
      \be\label{estimated21}
           |\cE_2/F_1| < 
              C  \frac{f^2}{r^{3/2}} 
                e^{4\kappa r^{1/2 }\sin |\theta/2|}   
                   e^{  -\frac{1}{f}r} 
                   < C \frac{f^2}{r^{3/2}} e^{- r /(2f) } = O(\log(r + f^{-1}))^{-1}
       \ee
               for sufficiently small $f$.
        Finally suppose $r^{1/2}|\theta| < 1$ and $r > \epsilon_2 $,
      and apply the bound (\ref{sinhlwb}). Then (\ref{estimated2}) becomes
              \[    |\cE_2/F_1| < 
              C  \frac{f}{r} 
              \frac{ (r +f^{-1})^{-M}  }{ M  \log(r + f^{-1})} 
              <   C (r+f^{-1})^{-M}
           \]
%
%
       

         \item[2.]  The region
         \[\{z: |z| > f^a;      -2\pi /3 +  |z|^{-3/2}  M f\log(|z| +f^{-1} )  <  \arg z < -\pi /3 \}\]
          is handled exactly as the first part
                        with the exception that in this region we use the approximation 
                        (\ref{Dsouth}) instead of (\ref{Deast}).
                        
               Thus         
                        consider 
                        \[ \left| \frac{\tilde h_0(z)/ (4\eta^2 z)}{F_1(z)} \right|   \]
                        We simply have $\tilde h_0(z)$ is of order $ 1+ |z|$
                        so that this term is bounded as
                                 \be\label{estquant2}
          \left|\frac{\tilde h_0/(4z\eta^2)}{F_1} \right| < 
          C \frac{(1 + r)^{1/2}  e^{  -\frac{4}{3f}r^{3/2}\sin(3\psi/2)} }{   \sinh(\kappa r^{1/2} |\theta|/2)   } < C(r+1)^{1/2}r^{-1/2}e^{-r^{3/2}\psi/f}
         \ee
        where $\psi = 2\pi/3 - |\theta|$.  The worst case scenario is when $\psi$ is small where we have $e^{-r^{3/2}\psi/f} \le (r+f^{-1})^{-M}$ so using $(r+1)^{1/2}r^{-1/2} \le C f^{-a/2}$ we obtain
         
         \[    \left|\frac{\tilde h_0/(4z\eta^2)}{F_1} \right| <  
           C\left(r + f^{-1}\right)^{-(M - 1/4)}.  \]
           
           For the first order term $D^{(1)}$,
           we have exactly the bound (\ref{estimated1}) once more.
           For $|z|$ small we once more have  (\ref{estimated12})
            since these estimates did not depend on the argument of $z$.
            Returning to (\ref{estimated1}), for $|z| > \epsilon_2$
            the exponential terms are bounded, i.e.  
            $     \frac{  e^{ 2 \kappa r^{1/2} \sin  |\theta/2| }  }{   |g_{\eta/\kappa}(2\kappa \sqrt z) |   }  < C, $
           so the term in (\ref{estimated1}) is of order $f (1 + |z|)^{-1/2} \le (r + f^{-1})^{-1/2}$.

           We now turn to the error terms. When $r$ is large the exponential and hyperbolic sine terms are balanced in $|\cE_1/F_1|$ and we obtain the bound $|\cE_1/F_1| \le Cf^2(1+r)^{-7/2}$.  The worst scenario is when $r$ is small where again we get the estimate $|\cE_1/F_1| \le Cf^{2-7a/2}$.

           Finally consider the $\cE_2'$ error term. We have
           in place of (\ref{estimatede2}) 
       \be \label{estimatede22}
           |\cE_2'/F_1| < 
              C  \frac{f^2}{r^{3/2}}(1 + r^{-3/2})  
              \frac{       e^{  -\frac{4}{3f}r^{3/2} \sin(3 |\theta|/2 )}         }{  |g_{\eta/\kappa}(2\kappa \sqrt z)|   }
       \ee 
           Again, for small $|z|$ we apply (\ref{gnearzero})  and we have $|\cE_2' / F_1| $
           is of order $ f^2r^{-7/2}$. 
           On the other hand, where $z $ is not small, $|g|$ is bounded below so we have $|\cE_2' / F_1| $
           is of order $ f^2r^{-3/2}$. 
         \item[3.]  Now consider 
                         \[\{z: |z|>f^a: -\pi \leq \arg z < - 2\pi/3 -  |z|^{-1} (1 + |z|^{-1/2})
                 M f\log(|z| +f^{-1} )   \}\]
                         and we again 
                       use $D^{(S)}$ defined in (\ref{Dsouth}).
                       First note that aside from $\tilde h_0/(4z\eta^2)$  all other non-error terms
                       as well as    $\cE_1$    are of order  
             \be \label{estimate2}
             \left|e^{i\frac{4z^{3/2}}{3f}  + i2\kappa\sqrt z  +   i\frac{f\kappa^2}{2\sqrt z}}\right|
                \left(     1 + r^{-4} \right) 
                <
                C
             e^{ - \frac{ r^{3/2}}{f}  |\theta + 2\pi /3|   + \kappa r^{1/2} |\theta|  }
                   \left(     1 + r^{-4} \right) 
                  \ee
               Then since $r^{3/2}|\theta + 2\pi/3| \ge (1+r^{1/2})Mf\log(r+f^{-1})$ all these terms are of order $f^{-4a/2}(r+f^{-1})^{-M} <  (r+f^{-1})^{-M+1}$ for small enough $f$.
               
                
               Finally, for the remaining error term we have, 
               \be \label{estimate3}\cE_2' = O \left( \frac{f^2}{|z|^{5/2}} (1 + |z|^{-3/2}) \right).\ee
                 Using the bound $|z| > f^a$, we have this is of order $f^{2- 4 a}$.
               The approximation of $D_f$ in terms of $F_2$ i.e.  in terms of $h_0$
               in this set  follows from the observation that for the $z$ under consideration, $\tilde h_0(z ) = h_0 (z ).$ (See their definitions above.)

             \item[4.] In the section $ \{z: |z| > f^a; \pi/2 \leq \arg z \leq \pi\}    $,   we use $D^{(N)}$ where
                            the error term is $\cE_2$ . The bound on this  error  given in Lemma \ref{fddapprx}, namely  $\cE_2 \le C|z|^{-5/2}(1+ |z|^{-3/2})$ is exactly what is needed.
                \item[5.] Finally, consider 
                \[ \{z: |z| > f^a;   |z|^{-3/2}  M f\log(r + f^{-1})   < \arg z \leq \pi/2 \}.   \]
                 In this region, we again use approximation $D^{(E)}$.
                An approximation similar to (\ref{estimate2}) holds.  First note that aside from $h_0/(4z\eta^2)$  all other non-error terms
                       as well as    $\cE_1$    are of order

                             \be \label{estimate4}
             \left|e^{i\frac{4z^{3/2}}{3f}  - i2\kappa\sqrt z  +   i\frac{f\kappa^2}{2\sqrt z}}\right|
                \left(     1 + r^{-4} \right) 
                <
                C
             e^{ - \frac{ 3r^{3/2}}{5f}  |\theta |   + \kappa r^{1/2} |\theta|  }
                   \left(     1 + r^{-4} \right). 
                  \ee
                  From this inequality we obtain, 
                  with $r^{3/2}\theta > Mf\log(r + f^{-1}) $,
                  a bound of order
                  \[ 
             e^{ - \frac{3 r^{3/2}}{5f}  |\theta | (1-  5\kappa f r^{-1}/3)   }
                   \left(     1 + r^{-4} \right) 
                   <
                   C (r + f^{-1})^{-2M/5} f^{-4a} < \left(r + f^{-1}\right)^{-(\frac{2M}{5} -2)}
                  \]
                  where on the left hand side we used $(1 - 5\kappa f  r^{-1}/2) > 1 - O(f^{1/2}) > 2/3 $.  With $M \ge 20$ this gives the bound stated in Lemma \ref{fddapprx2}.
                Thus non-error terms other than $h_0/(4z\eta^2)$ are small, as well as $\cE_1$.
                The necessary smallness of $\cE_2$ again follows directly from its defintion.
         \ei 
        \epf

\section {Approximation of dynamics: Proofs of Theorems \ref{small f approx}, \ref{small eta approx}} \label{dynamical work}
              We first prove Theorem \ref{small f approx}.
                       
             \bpf 
             First we show that for small $f>0$
             \be \label {src}
              || ((\pm i - H_{\kappa,\eta, f})^{-1} - (\pm i -  H_{\kappa,\eta})^{-1})\phi||
                      \le 4f^{3/8}||\phi|| + 2 || 1_{\{|x| > f^{-1/4}\}} \phi || \ee
              for all $\eta > 0$.

              Let $z = \pm i$ and consider the difference:
            \[  ((z - H_{\kappa,\eta,f})^{-1} - (z - H_{\kappa,\eta})^{-1})\psi  
                       = f(z-H_{\kappa,\eta, f})^{-1} (z-H_{\kappa,\eta})^{-1}x\psi  + f(z-H_{\kappa,\eta,f})^{-1}
                          [x,(z - H_{\kappa,\eta})^{-1}]\psi,  \]
            then for the final term, find that
            \[   (z-H_{\kappa,\eta,f})^{-1}[x,(z - H_{\kappa,\eta})^{-1}]  
              =   i(z-H_{\kappa,\eta,f})^{-1}(z - H_{\kappa,\eta})^{-1}2p (z- H_{\kappa,\eta})^{-1}. \]
            The final factor is bounded by a constant,
            \[||p(z- H_{\kappa,\eta})^{-1}|| 
                \le C||p(1 + H_{\kappa,\eta})^{-1}|| 
                = C ||(1 + H_{\kappa,\eta})^{-1}p^2(1 + H_{\kappa,\eta})^{-1}||^{1/2} \le C,  \]
            where we have used $\eta >0$.

                  We thus have 
                  \[ || ((z - H_{\kappa,\eta,f})^{-1} - (z - H_{\kappa,\eta})^{-1})\psi || \le 2Cf||\psi|| + f||x\psi||\]
                  
                  If $\phi \in L^2(\mathbb{R})$  
                   then we take $\psi = (1+ i\epsilon x)^{-1}   \phi $ and obtain 
                  \begin{align}
                  || ((z - H_{\kappa,\eta,f})^{-1} - (z - H_{\kappa,\eta,0})^{-1})\phi ||
                  &   
                          \le 2Cf||\psi|| + f||x\psi|| + 2||\phi - \psi|| \\
                   & \le 2Cf ||\phi|| + \frac{f}{\epsilon}|| \phi|| +2||\phi - \psi||
                   \end{align}
                   We take $\epsilon = f^{5/8}$ so that 
                   
                   \be
                   ||\phi - \psi ||^2 = \int |\epsilon x/(1+i\epsilon x) \phi |^2 dx  \le   
                     f^{3/4} \int _{|x| < f^{-1/4}} |\phi |^2 dx + \int _{|x| > f^{-1/4}} |\phi |^2 dx 
                     \ee
                   This gives  (\ref{src}).  
                   
                   Suppose $ s \in [0,T]$.  Then given $\phi \in L^2$ there is an $f_{s,\epsilon, \chi, \phi} >0$ so that if $0 < f < f_{s,\epsilon, \chi, \phi}$
                   
                    $$||\chi(H_{\kappa,\eta, f}) e^{-isH_{\kappa,\eta, f}}\phi - \chi(H_{\kappa,\eta, 0}) e^{-isH_{\kappa,\eta}}\phi|| < \epsilon $$ for all $\eta > 0$.  This comes from the strong resolvent convergence uniform in $\eta$ reflected in (\ref{src}). 
                    The lemma then follows from the (norm) continuity of $\chi(H) e^{-itH}$ in the variable $t$ and the compactness of $[0,T]$.   
                    \qed
                    \epf   
                     
                   Now we prove  Theorem \ref{small eta approx}.
\bpf

Let $\delta_{\mu}(x) = \frac{\mu}{\pi}(x^2 + \mu^2)^{-1} = \frac{1}{2\pi i}((x -i\mu)^{-1} - (x +i\mu)^{-1})$.  We have by Stone's formula

\be
\langle \phi, \chi(H_{\kappa,\eta}) e^{-itH_{\kappa,\eta}} \phi \rangle =\lim _{\mu \downarrow 0}\int_{\mathbb{R}} \chi(\lambda) e^{-it\lambda }\langle \phi, \delta_{\mu}(H_{\kappa,\eta} -\lambda)\phi \rangle d\lambda.
\ee

We estimate
\be \label{endpoint errors}
|\lim_{\mu \downarrow 0} \int _{\mathbb{R}\setminus [a,b]}\chi(\lambda) e^{-it\lambda }\langle \phi, \delta_{\mu}(H_{\kappa,\eta} -\lambda)\phi \rangle d\lambda |\le \langle \phi, \tilde\chi(H_{\kappa,\eta})\phi\rangle.
\ee
where $0 \le \tilde\chi \le 1$ is a continuous function with support in $[a-\delta,a + \alpha] \cup [b- \alpha , b+\delta]$.  Here $\alpha $ is chosen so that  there are no eigenvalues of $H_D$ in the latter intervals.  By the monotone convergence theorem for forms (\cite{RS}, p.372), if $\eta$ is small enough the right side of (\ref{endpoint errors})$ \le \epsilon/2$ since  $\tilde\chi(H_D)\phi = 0$.

In the remaining integral we deform the path picking up the residues $e^{-itz_j(\eta)}\text{Res} \langle \phi, (H_{\kappa,\eta} - z)^{-1}_c \phi \rangle|_{z=z_j(\eta)}$ 
where $(H_{\kappa,\eta} - z)^{-1}_c$ is the meromorphic continuation of  $(H_{\eta} - z)^{-1}$ from the upper to the lower half plane.  (Note that the zeros of the determinant are simple.)  We are left with the integral over the path $\gamma = \gamma_1 + \gamma_2 + \gamma_3$  where $\gamma_1(t) = a -it ; \gamma_2(t) = a-i +t(b-a); \gamma_3(t) = b -i +it$ and the $t$ interval is $[0,1]$ in each case.  Thus 

\begin{align*}
\langle \phi, \chi(H_{\kappa,\eta}) e^{-itH_{\kappa,\eta}} \phi \rangle &= \sum_{j=1}^n  e^{-itz_j(\eta)}\langle \phi, (H_{\kappa,\eta} - z)^{-1}_c \phi \rangle|_{z=z_j(\eta)} + \int_{\gamma}e^{-itz}\langle \phi, (H_{\kappa,\eta} - z)^{-1}_c  - (H_{\kappa,\eta} - z)^{-1}\phi \rangle dz\\
&+ \mathcal{E}_1
\end{align*}
where $ |\mathcal{E}_1|\le \epsilon/2$.
\vspace{.2in}

Let $I=[-\kappa, \kappa]$.  By the monotone convergence theorem for forms \cite{RS},$ (z - H_{\kappa,\eta})^{-1}|_{L^2(I)}$ converges strongly to $(z - H_{D})^{-1}$ for $\Im z \ne 0$.  Consider  $(H_{\kappa,\eta} - z)^{-1}_c$ in the lower half plane as $\eta \to 0$.  The denominator $\eta^2\mathcal{W}D$ in the formula (\ref{green1}) is given by
  \[   \eta^2 \cW D = -i2 \sqrt z 
                \left( \eta -i \frac{ 1 + e^{i2\kappa \sqrt z}}{ 2 \sqrt z}   \right) 
                 \left( \eta -i \frac{ 1 - e^{i 2\kappa \sqrt z}}{2 \sqrt z}   \right). \]
             This converges uniformly on compact sets of the right half plane  to 
            \[ \frac{i}{2\sqrt{z}}(1- e^{4i\kappa \sqrt{z}}). \]
            
            The numerator of the continued Green's function is bounded uniformly on compacts of the right half plane for $(x,y)$ in compact subsets of $\mathbb{R} \times \mathbb{R}$.  Thus if we avoid small disks around the $n$ eigenvalues of $H_D$ in $[a,b]$, $\langle \phi, (H_{\kappa,\eta} - z)^{-1}_c \phi \rangle$ is uniformly bounded on compacts of a neighborhood $N$ of $\{z = x + iy:  a \le x \le b\}$, in other words on compacts of $N\setminus K$ where $K$ is the union of these small disks.  Since we have convergence in the upper half plane, Vitali's convergence theorem 
            (\cite{T}, Theorem 5.21) implies convergence in the lower half plane as well.  
            Thus $\langle \phi, (H_{\kappa,\eta} - z)^{-1}_c \phi \rangle \to \langle \phi, (H_{D} - z)^{-1} \phi \rangle$ uniformly for $z$ in $N$ but away from the union of any arbitrarily small disks centered at the eigenvalues $E_1, \dots, E_n, E_j \in (a,b)$.  Since the residues $\text{Res} \langle \phi, (H_{\kappa,\eta} - z)^{-1}_c \phi \rangle|_{z = z_j(\eta)} $ can be calculated by integrating around  small circles centered at $E_j$, we see that they converge to $|\langle\phi,\phi_j\rangle|^2$ as $\eta \to 0$.  We can find $\eta_{\epsilon} > 0$ so that the errors just discussed add up to at most $\epsilon/2$ if $0 < \eta < \eta_{\epsilon}$ which proves the theorem.
             \qed
\epf

 \section{The shape resonance perturbed by a distant  Dirichlet boundary  (\ref{Hds})}

   We consider (\ref{Hds}) on $L^2(-\infty , l)$ with a Dirichlet boundary at $l > \kappa$. 
    {The solution of $ (\tilde{H}_{\kappa,\eta,l} - z)g = 0$ which is in $L^2 $ on the left for $z$ 
    in the upper half plane is }$\psi_1(x) = e^{-i x \sqrt z}$
    and the { solution satisfying the boundary condition at $l$ is} 
     $\phi_1(x) = \sin((x-l) \sqrt z)$. 
     The Wronskian  $\psi_1' \phi_1 - \psi_1 \phi_1'$ is
   $ \cW^{(0)}  = \cW^{(0)} (l) = -\sqrt z e^{-i l \sqrt z}$.
   From (\ref{greendet}) the function whose zeros give the resonances is
  \beq   D^{(0)}   &=& \label{det1dirdda}
           1 + i \frac{1 - e^{i 2 l \sqrt z}\cos(2\kappa \sqrt z)  }{\eta\sqrt z}  
           -  \frac{ e^{i(\kappa + l)\sqrt z } \sin((\kappa - l)\sqrt z) \sin(2 \kappa \sqrt z) }{ \eta^2 z   }  
      \eeq 
    As we will see the shape resonances disappear as soon as the Dirichlet condition appears near infinity 
    and an infinite number of other resonances appear just as for the Hamiltonian $H_{\kappa, \eta, f}$.  The latter can be seen for example from (\ref{alt}) below and the Weierstrass factorization theorem.  In Proposition \ref{zeroDirprop} below we find the possible locations for this infinite set of resonances.
    
    Again we must first verify that all the poles  of $G^{(\eta)}$ are exactly the zeros of $D^{(0)}$. It is easy to see that if $\phi(\kappa) \ne 0$ then the four functions $g_{\pm}(x)g_{\pm}(y)$ and $ g_{\mp}(x)g_{\pm}(y)$ are linearly independent (for example take $-\kappa < x < y < \kappa$) and thus if      
    $L_z(x,y) = 0$,  $\beta_+ = 0, \eta \mathcal{W} + \alpha_+ = \eta \mathcal{W} + \alpha_{-} = 0$.  This and $D^{(0)}(z) = 0$ imply $\eta \mathcal{W} = 0$, a contradiction. If $\phi(\kappa) = 0$ then $\alpha_+ =0$,   but $z \ne 0$, and  $L_z(x,y) = 0$ imply $\eta \mathcal{W} +  \alpha_+ = 0$, again a contradiction.
      The point $z=0$ is a branch point of the resolvent so cannot be a pole.  We can see that $L_z(x,y)/\sqrt z$ as well as $G_z^{(\infty)}(x,y)$ are entire functions of $\sqrt z$ so that the only singularities in this variable in the resolvent $G_z^{\eta}(x,y)$ come from the zeros of $D^{(0)}$.  But $\eta^2 D^{(0) } (0) = \eta^2 + 2l\eta + 2\kappa(l-\kappa) > 0$ and thus  $G_z^{\eta}(x,y)$ is analytic in $\sqrt z$ in a neighborhood of the origin.  
       Thus the zeros of $D^{(0)}$ are exactly the resonances of (\ref{Hds}).
       Moreover, $D^{(0)}$ is analytic in $\sqrt z$, so zeros of $D^{(0)}$ (resonances) may exist
      only  in the set $\{ |z|> 0; -3\pi < \arg z < 0\}$.

       \bp\label{zeroDirprop}
      Let $\eta,\kappa > 0$ and $c =3\pi/4$. Then there exists  $l_{\eta,\kappa} > \kappa$
       so that if $l > l_{\eta,\kappa}$  then all zeros of 
       $D^{(0)}$ with $-2\pi \le \arg z < 2\pi$
        are contained in the set
  \be\label{zerosetddDir}   
            \left\{ z:  0 >     \arg z > - c l^{-1}|z|^{-1/2} \log(l + |z|) 
                       \tn{ or }   -2\pi  \le   \arg z  < -2\pi  + c l^{-1}|z|^{-1/2} \log(l + |z|)     \right\}   
          \ee

      \ep

       It will be helpful to rewrite $D^{(0)}$ in an alternative form 
        as a sum of two functions   $D^{(0)} = D_1^{(0)}  + D_2^{(0)} $ where
      \beq   \label{alt} D_1^{(0)}  =     -i \frac{e^{i2l\sqrt z}}{2\eta^2 z}g_b(2\kappa \sqrt z)
         \tn{ and }
           D_2^{(0)}  =    
           \frac{  \left(   2\eta\sqrt z  +i \right)^2      +        e^{i4\kappa \sqrt z}}{4\eta^2 z}   
  \eeq
   for  $g_b(\omega ) =  \sin\omega + b\omega \cos\omega$  where $b = \eta /\kappa$.
   See Appendix \ref{dirichletdeterminantsect} for more details.
   In the following lemma we show that in the two different regions we can 
    approximate $D^{(0)}$ with $D^{(0)}_1$ and $D^{(0)}_2$ respectively.  Proposition \ref{zeroDirprop} will follow directly.

  \bl\label{ddDirapprx}
      Let $\eta,\kappa > 0$
      and $c =3\pi/2$.
       Then there exists $l_{\eta,\kappa} > \kappa$
       so that if  $l > l_{\eta,\kappa}$  then in the set
  \be   \label{setrestriction45}
            \left\{  z  = r e^{i\theta}:    -2\pi < \theta < 0,
           - 2\pi  +  c l^{-1}r^{-1/2} \log(l + r)    <   \theta < - c l^{-1}r^{-1/2} \log(l + r)   \right\}   
          \ee  
          we have 
    \be  \label{firstappsect4} \left| \frac{D^{(0)} }{D_1^{(0)} } - 1 \right|  <  C/ \log(l+r)
            \ee
            while in the set 
            \be\label{sec4set2}
          \left\{  z  = r e^{i\theta}:   
           - 3\pi  \le \theta \le -2\pi - c l^{-1}r^{-1/2} \log(l + r)   \right\}  
        \ee
        we have 
            \[   \left| \frac{D^{(0)} }{D_2^{(0)} }-1\right|  <   C/ \log(l+r). 
            \]
       Here the constant $C = C_{\eta,\kappa}$ depends only on $\eta$ and $\kappa$.

  \el

  Proposition \ref{zeroDirprop} follows immediately from Proposition \ref{ddDirapprx}.
  Indeed we have 
  \[ |D^{(0)} | > |D_1^{(0)} | [ 1 - C/\log(l+r)] \]
   in the first region and similarly in the second region.
   We need only show $D_1^{(0)}$ and $D_2^{(0)}$ do not vanish in the respective sets where
   they approximate $D^{(0)}$. 
    But  $D_1^{(0)}$ does not vanish in the lower half plane of $\sqrt z$, indeed 
   the exponential prefactor does not vanish, and as for $g_b$ 
   see  Lemma \ref{gtermest} for a demonstration 
   of a bound away from 0.  
   Finally in the set defined in (\ref{sec4set2}) we have $\Im z > 0$ thus $|(2\eta \sqrt z + i)^2| 
   > ( 1 +2\eta \Im \sqrt z)^2$ and $|e^{i4\kappa \sqrt z } |< 1$,
   therefore $ | 4\eta z D^{(0)}_2 | \geq  ( 1 +2\eta \Im \sqrt z)^2 -1 > 0$.
   This concludes the proof of Proposition \ref{zeroDirprop}.
   
      Now we prove Lemma \ref{ddDirapprx}. 
  \bpf
   We note for $0 < \arg z < 2\pi$ the resolvent is analytic and therefore $D^{(0)}$ has no zeros in this sector.
 
  We begin in the sector $-\pi \leq \arg z < 0$.
  Write $z = re^{i\theta}$ then  we have,
  \[  \left| \frac{D_2^{(0)}}{D_1^{(0)}}  \right| 
     = \frac{ \left|  (2 \eta \sqrt z + i)^2 + e^{i4\kappa\sqrt z}  \right|    }
                   {2 \left| e^{i2l\sqrt z} g_b(2\kappa \sqrt z)\right|}  . \]
   Then
   we bound $g$ below by Lemma \ref{gtermest}, so that
   for some $C > 0$ depending on $\eta$ and $\kappa$,
    \[   \left| \frac{  (2 \eta \sqrt z + i)^2    }{D_1^{(0)}}  \right| 
         <  C\frac{ r +1    }
         {  e^{2l r^{1/2} \sin(|\theta|/2)} ( r  + 1)^{1/2} \sinh  (  2\kappa r^{1/2} \sin(|\theta|/2)  )  }.
                       \]    
     If $|\theta| \le \pi$ we have $\sin(|\theta|/2) \ge |\theta|/\pi$ and thus  $ 3(4 l)^{-1} \log(l+r)  <  \sin( |\theta|/2) r^{1/2} $ and we have 
         \[   \sinh  (  2\kappa r^{1/2} \sin(|\theta|/2)  )  
          >     2\kappa r^{1/2} \sin(|\theta|/2) >3 \kappa  (2l)^{-1} \log(l+r)   \]
          and 
       $e^{2l r^{1/2} \sin(|\theta|/2)} \geq (l + r)^{3/2}  $ and thus
     \[   \left| \frac{  (2 \eta \sqrt z + i)^2    }{D_1^{(0)}}  \right| 
         <
           C\frac{  (r+1)^{1/2}}{ (l + r)^{3/2} (   l^{-1}  \log(l + r)  ) } 
           \le
           \frac{C}{\log(l + r)   }.
                       \]                      
  Now on the other hand, for $C$ depending only on $\eta$ and $\kappa$,
   \[  \left| \frac{ e^{i4\kappa\sqrt z}  }{D_1^{(0)}}  \right|   
   <  
       C  \frac{ e^{4\kappa  r^{1/2} \sin(|\theta|/2)}   }{ (1 + r )^{1/2} e^{2l r^{1/2} \sin(|\theta|/2)}\sinh(2\kappa r^{1/2} \sin(|\theta|/2)) } \le C \frac{e^{-2(l-2\kappa)r^{1/2}\sin(|\theta|/2)}}{(1+r)^{1/2} 2\kappa r^{1/2}\sin(|\theta|/2)}
   .  \]
  Thus \be 
    \left| \frac{ e^{i4\kappa\sqrt z}  }{D_1^{(0)}}\right|  \le C \frac{e^{-(3/2)(1-2\kappa/l)\log(l+r)}}{\kappa l^{-1}\log(l + r)} \le C (l+r)^{-(1-6\kappa/l)/2)}/\log(l+r) \le C/\log(l+r)
  \ee
  for large $l$. 
 
  Therefore we have $|D_2^{(0)}/ D_1^{(0)}| < C/ \log (l+r)$ for large enough $l$ 
  which implies (\ref{firstappsect4}).
    A similar argument works for $\pi \le |\theta| \le 2\pi$.
%
     \qed
   \epf

    \section{Counting of resonances, proof of Theorem \ref{resonancecountingtheorem}}\label{detgrowth}
            In this section we will investigate the number of zeros of $D_f(z)$.
            We have already established that the zeros are restricted to neighborhoods of the origin or of the
            rays $\arg z =  0$ and $\arg z = - 2\pi/3$, however,
            we now find estimates on the number of zeros, in particular near the real axis.
            For $\lambda$ an analytic function on $\bbC$, let us define $n_\lambda(r)$ to be the number
            of zeros (counted with multiplicity) 
            of $\lambda$ with modulus less than $r$, and let $M_\lambda(r)$ be
            defined as $\sup\{ |\lambda(z)| : |z| < r  \}$.
            
%
%
%
             
               We first note that one can conclude there are an infinite number of zeros of 
               $D_f$ in the plane for fixed $f>0$.
               This follows
              simply by referring to the Hadamard factorization theorem 
              to find
              \[   D_f = z^m  e^{a z +b} \prod_i   \left(1 - \frac{z}{z_i} \right)e^{{z}/{z_i} } .  \]
             Since $\log M_{D_f}(r) \sim \frac{4}{3f}r^{3/2}$ (see Lemma \ref{fddapprx2}).
 and the representation above contains only  $e^{z}$ terms, there must be sufficiently many 
              zeros to obtain the growth of the stated order and type.  
              (See Korotyaev \cite{K} for an analysis of the asymptotics of the number of resonances in
               $\{z: |z| <R\}$ as $R \to \infty$ of a similar model for a fixed $f$.)
               Here we are more interested in obtaining asymptotics for the number of resonances in bounded regions as $ f \to 0$.  We will show 
              that these grow in proportion to $1/f$. 
               (cf. Theorem  \ref{resonancecountingtheorem}).

              In particular, the approach to zero counting we use is estimating integrals
               which count the number of zeros.
              One could in principle count zeros by integrating $\frac{D_f'}{D_f}$ over contours,
               however this would involve 
              new difficulties of finding useful approximations of this new function. 
              Instead we bypass this and use the well known integrals  of Jensen and Carleman
               for  estimating the number of zeros.
              The advantage here is that we will only need the approximations 
              of $D_f$ which are stated in Lemma \ref{fddapprx2}. 

            We apply these theorems to the normalized functions $B_f(z) = \frac{D_f(z)}{D_f(0)}$
             and $\wt B_f(z) = B_f(\omega^{-1}  z)$,
               for $\omega = e^{ i 2\pi/3}$,   to count the zeros in 
             the right half plane and the \tql southwest\tqr\ half plane respectively (by definition the southwest half plane is the half plane centered at $\theta = -2\pi/3$).
                 Let $R_l < R_u$ and let $T_{R_l,R_u} = \{z\in \bbC:  R_l \le |z| < R_u\}$.
                We estimate the number of zeros in $T_{R_l,R_u}$ in the  right and southwest half planes.
                We may allow $R_l$ and $R_u$ to scale with $f$ with the restrictions that
                  \[   R_u /2 >    R_l > f^{1/100}  . \]

                  Let   $\cN_{R_l,R_u}[B_f]$ be the number of zeros of $B_f$ in $ T_{R_l,R_u} $ with positive real part.
            \bt  \label{zerocountLR}
               Given $\epsilon > 0$, there is some $f_{\eta,\kappa,\epsilon} >0 $
                so that for  $f_{\eta,\kappa,\epsilon} > f>0 $ 
               \be   
                 \label{rightcountlower}
                  \frac{1+\epsilon }{2\pi f}[\frac{5}{3}  R_u^{3/2} - R_l^{3/2}]   >  \cN_{R_l,R_u}[B_f] >  
               \frac{1-\epsilon }{2\pi f}[  R_u^{3/2} - \frac{5}{3} R_l^{3/2}].
                         \ee
            \et
            Theorem \ref{resonancecountingtheorem} follows immediately from  the above theorem.

             \br
                From the usual $2\pi/3$ rotational symmetry of the Laplacian plus electric field, one
                  expects the number of zeros localized about the ray $\theta = - 2\pi / 3$  to 
                 grow 
                  at the same rate as the zeros about the ray $\theta = 0$.
                 Indeed, our estimates are symmetric  about these two rays so a similar growth
                  of the number of zeros holds  in the left half plane.
             \er
             
              \subsection{Zero counting theorems of Carleman and Jensen
                          and proof of Theorem \ref{zerocountLR}}\label{JCintegrals}

             As stated above, Theorem \ref{zerocountLR} follows from 
              two classical theorems relating certain integrals of a function
               to sums of over zeros of that function.
               We state these theorems here and the evaluation of these integrals for our function $B_f$.
                We then prove Theorem \ref{zerocountLR} from the combination of these results.

             First we state Jensen's theorem, which will allow us to obtain upper bounds of the 
             number of zeros in bounded regions.
           \bt[Jensen]\label{jensen}
             Let $\lambda$ be a function analytic in $\bbD_R$, and suppose $\lambda(0) = 1$,
              and $\{z_1,\ldots,z_k\}$ 
             are the zeros of $\lambda$ in $\bbD_R$ counted with multiplicity, then
             \[  \int_0^R  \frac{ n_\lambda(t)}{t} dt =  \sum_{i=1}^k \log\frac{R}{|z_i|}  = I_J[\lambda;R]  := 
                 \frac{1}{2\pi} \int_0^{2\pi} \log |\lambda(Re^{it})| dt .\]
            \et  
              It follows, if $\lambda$ is analytic in $\bbD_{v R}$ for $v > 1$, that 
              \[    I_{J}[\lambda ;vr] - I_{J}[ \lambda; r]  =  \int_{r}^{vr}  \frac{ n_\lambda(t)}{t} dt 
                        =  \int_{r}^{vr}  \frac{ n_\lambda(t)  -  n_\lambda(r)  }{t} dt  +  n_\lambda(r)  \log v 
                                    \geq n_\lambda(r)  \log v \]
              So that we have the upper bound,  
              \be \label{jenub}
               n_\lambda(r) \leq  \frac{I_{J}[\lambda;  v r] - I_{J}[\lambda;r]  }{\log v} .
              \ee
              
              Now we introduce Carleman's theorem 
              which we will use to obtain lower bounds of the number of zeros in bounded regions in the 
              half planes.
              Let us introduce the notation 
              \[      A_ {R_L,R_\rho} = \{ z\in\bbC :  R_\rho < |z| < R_L ; \Re z > 0   \}.    \]
              and for given entire function $\lambda$ let $\{z_1,...,z_k\}$ 
              be the enumeration with multiplicity
              of the zeros of $\lambda$ in 
              $  A_ {R_L,R_\rho}$.
                              \bt[Carleman]\label{carleman}
                     Let $\lambda(0) = 1$ and suppose $\lambda(z) $  is not zero for $|z| = R_\rho$.
                      Then we have, 
                       \beq\label{carlemanint}
                         \sum_{i = 1}^k  \left(   \frac{1}{ |z_i|}  -  \frac{|z_i|}{R_L^2}    \right)\cos (\arg z_i) 
                            &=&   I_C[\lambda;R_\rho , R_L] :=
                            I_1[\lambda]  + I_2[\lambda]  ,
                            \eeq 
           where the $I_1,I_2$ are given by
            \[ I_1[\lambda] =     \int_{R_\rho}^{R_L}\log(|\lambda(is)|\cdot| \lambda(-is)| )
                              \left( \frac{1}{s^2}  - \frac{1}{R_L^2}\right)\frac{ds}{2\pi}   \]
            and, 
             \[
                I_2[\lambda] =      \int_{-\pi/2}^{\pi/2} \left[
                     \log \frac{   |\lambda(R_Le^{i\theta})|^2   }
                                        {   |\lambda(R_\rho e^{i\theta})|^{  (\alpha + \alpha^{-1} )   }   } 
                               \cos\theta
                            + [  \arg\lambda(R_\rho e^{i\theta})  ]
                             \left( \alpha - \alpha^{-1} \right) \sin\theta
                          \right]    \frac{d\theta}{R_L 2\pi} 
             \]
            with $\alpha = R_L/R_\rho$.
            \et

            Let $a,b$ be chosen so that $ \frac12 - \frac1{100} <   a <  \frac12$ 
            and $  \frac23 < b < \frac23 + \frac{1}{100}. $
           Notice we then have 
           \be\label{parameterdiff} 
                1/18 < 3a/2 - b < 1/12.\ee
        Let $R_L > 2 f^a$ and suppose $C> 0$ is given and 
        $R_\rho$ is chosen so that  $C^{-1} f^b < R_\rho <C f^b $.
             The Carleman integrals for $B_f$ and $\wt B_f$ are as follows:   
            \bp\label{carlemanhalfintegrals}
               For  small enough $f$ there is an $ r$ in  $((1 - f )R_L , R_L)$ so that for $B = B_f, \tilde B_f$
                \[ I_C[B; R_\rho,r]   =  \frac{8(3 + \sqrt 2)}{15 \pi }\frac{r^{1/2}}{f}  +O( f^{-17/18} + \log(1+r) \log f^{-1}). 
                       \] 
            \ep
       
                The Jensen integral for $B_f$ is as follows,  
            \bp\label{jensenintegral}
               Let $a < 1/2$ and $2f^a < R_L$.
               Then there is an $ r$ in  $((1 - f )R_L , R_L)$ so that
               \be \label{jensenvalue}  
                   I_J[B_f;r] = \frac{8}{9\pi}  \frac{r^{3/2}}{f}  +    O( (1+ r^{1/2})\log^2(f^{-1} + r)    ).
                   \ee
           \ep

             Propositions \ref{jensenintegral} and \ref{carlemanhalfintegrals} are demonstrated in 
              Section \ref{integralcalcs}.
              In particular Proposition  \ref{jensenintegral}  is proved in Section \ref{jensproof}
               and Proposition \ref{carlemanhalfintegrals}  is proved in Section \ref{carlsproof}
             as a combination of Lemmas \ref{axisintlemma} and \ref{arcintlemma}.
              For the remainder of this section we combine  Propositions \ref{jensenintegral} and \ref{carlemanhalfintegrals} 
              with Theorems \ref{jensen} and \ref{carleman} to prove  Theorem \ref{zerocountLR}.
                       \bpf  \textbf{of Theorem \ref{zerocountLR} }

              Let $R >4 f^a$ and let   $ s \in (1/2, 1)$ be a constant 
               (independent of $f$) to be fixed later.
                   For $|\nu | = 1$ let $N(\nu ,r) $ be the number of zeros in the set  $\{z: |z| < r;   |\arg z - \arg \nu| < \pi /2\}$.

                   For a given lower radius $R_0 = O(f^b)$ and $R $ as above we will
                   evaluate the difference of the Carleman integrals
                    with lower bound $R_0$ and upper bounds near $sR$ and $R$.                    That is, let
                    $R_1$ be chosen as the \tql$r$\tqr\ in Proposition \ref{carlemanhalfintegrals}
                     so that $(1 - f) sR < R_1 < sR$ and $R_2$ so that $ (1-f) R < R_2 < R$. 
                     We assume $f$ has been chosen small enough so that $R_2 > R_1$.
                   Label the zeros in the set                     
                   $\{z:  |z| < R_1 ; |\arg z|  < \pi/2\}$ as
                   $(r_i,\theta_i)$,
                   and label the zeros in the set 
                   $\{z:  R_1 \le |z| < R_2 ; |\arg z|  < \pi/2\}$ as
                   $(r'_i,\theta_i')$.   
                   Note that from Theorem \ref{resonancedesc}, 
                   for small enough $f$ there are no resonances with $|z| \le R_0$.

                From    (\ref{carleman}) evaluated at $R_0$ and $R_1,R_2$
                   \be\label{ICdiff1}       I_C[B_f;R_0,R_2] - I_C[B_f;R_0,R_1] 
                                  =    \sum_i  \left(\frac{1}{ R_1^2} - \frac{1}{R_2^2} \right) 
                                         r_i \cos \theta_i   
                                          + \sum_j  \left( \frac{1}{r_j'}   -  \frac{r_j' }{R_2^2} \right) \cos\theta_j' . \ee
               We estimate from above, noting that $r^{-1} - r R^{-2}$ is decreasing in $r$ for $0< r < R$.
               \be  \label{ICdiff2}            I_C[B_f;R_0,R_2] - I_C[B_f;R_0,R_1]    \leq
                                            N(1,R_1)  R_1  \left( \frac{1}{R_1^2} - \frac{1}{R_2^2} \right)   + 
                                             (N(1,R_2) - N(1,R_1) )  \left( \frac{1}{R_1} - \frac{R_1}{R_2^2}\right)  
                                                     \ee 
                  The right hand side of (\ref{ICdiff2})
                is simply $       (N(1,R_2)R_1  )  \left( \frac{1}{R_1^2} - \frac{1}{R_2^2}\right)  $. 
                 Now we apply the calculation of Proposition \ref{carlemanhalfintegrals} to obtain,
               \[                  \frac{8(3 + \sqrt 2)}{f 15 \pi} \left( R_2^{1/2} - R_1^{1/2} \right) + O(f^{-17/18} )   
                       \leq   (N(1,R_2)   R_1  )  \left( \frac{1}{R_1^2} - \frac{1}{R_2^2}\right)    \]  
                We may rewrite this as,
                 \be \label{n1lower1}
                   N(1,R_2) \geq        \frac{8  (3 + \sqrt 2)}{f 15 \pi}
                  \left( R_2^{1/2} - R_1^{1/2} \right)
                       R_1^{-1}    \left( \frac{1}{R_1^2} - \frac{1}{R_2^2}\right)^{-1}  
                  +   R_1^{-1}  \left( \frac{1}{R_1^2} - \frac{1}{R_2^2}\right)^{-1}  
                  O(f^{-17/18} ).      
                  \ee
                  For the first term, we have
                  \beq
                  \left( R_2^{1/2} - R_1^{1/2} \right) 
                       R_1^{-1}    \left( \frac{1}{R_1^2} - \frac{1}{R_2^2}\right)^{-1}  
                       = \frac{R_1R_2^2}{(R_2^{1/2} +R_1^{1/2})(R_2 + R_1)} \ge \frac{s}{4}((1-f)R)^{3/2}
                    %
                  \eeq              
                  and 
                  \[   R_1^{-1}  \left( \frac{1}{R_1^2} - \frac{1}{R_2^2}\right)^{-1}  
                  O(f^{-17/18} )    =    
                    R \ O( f^{-17/18})
                  \]
                 Thus (\ref{n1lower1}) becomes,
                 \be\label{lowerboundonRHP}
                     N(1 , R)  \geq   N(1,R_2) 
                 \geq     \frac{ ((1-f)R)^{3/2} }{ f}    
                             \left[    \frac{s}{ 2 \pi}  \right]   + O \left( R f^{-17/18} \right)     \ee
                          The error can be larger than the main term if $R \sim f^a$ as is allowed. 
                         Here we remind the reader of the assumption of the theorem which 
                         says $ R_l > f^{1/100}$ which guarantees that the error term above is smaller than the main term.
                A similar
                argument gives a similar  lower bound  for $ N(\omega^{-1},R_2)$: 
                 \be\label{lowerboundonSWP} N(\omega^{-1},R) \geq      
                                \frac{ ((1-f)R)^{3/2} }{ f} \left[   \frac{s}{ 2 \pi}   
                             \right]   
                              + O \left( R f^{-17/18} \right).    \ee
                                    
                 Now we develop upper bounds on  $n_{B_f}(\cdot)$ using Jensen's theorem. 
                 Let $v > 1$  be a constant to be  fixed later. 
            Given $R' > 2(1-f)f^a$ first set $R_L = R'(1-f)^{-1}$ and then find $R_1' \in ((1-f)R_L, R_L) = (R', (1-f)^{-1}R')$ so that we can use Proposition \ref{jensenintegral} with $r = R_1'$.  Similarly we choose $R_L = vR'(1-f)^{-1}$ and find $R_2' \in ((1-f)R_L, R_L) = (vR', ((1-f)^{-1}vR')$ and again use Proposition \ref{jensenintegral} with $r = R_2'$.  We then apply the inequality (\ref{jenub})  
                 to find
               \[  n_{B_f}(R'_1) \leq \frac{  I_J[B_f; R'_2]  - I_J[B_f;R'_1]}{\log(R'_2/R'_1)}.  \]
                 Thus we have  a bound
                \begin{align}
                    n_{B_f}(R')  \leq n_{B_f}(R'_1)  &\leq  \nonumber
                          \frac{8}{9\pi}\frac{R'^{3/2}}{f}\frac{v^{3/2}(1 -f)^{-3/2} - 1 }{\log [v(1 - f)]} 
                                  +          O(   f^{-1/4} + R'\log^2(f^{-1} + R')    ).\\
                     \intertext{It is not hard to see the $(1-f) $ factors amount to a shift of order $R'^{3/2}$ so that}
                        \label{jenub2}    
                              n_{B_f}(R') &
                                 \leq  \frac{8}{9\pi}\frac{R'^{3/2}}{f}\frac{v^{3/2}  - 1  }{\log v } 
                                  +         O(  R'^{3/2}  + f^{-1/4} + R'\log^2(f^{-1} + R' )    ).
                  \end{align}
                 In particular, for $R' = 2f^{a}$, with $a = .4$,
                 we have $   n_{B_f}(R') = O(f^{-2/5})$. The remaining zeros are outside this disc, ie in the set $\cR_0$ (which follows by using another $a$ slightly larger than $.4$).
                 But by Proposition \ref{fddres} there are no zeros in $\cR_1$ or $\cR_2$,
                 so $N(-1,R)$ and $N(\omega^{-1},R)$ only differ by the zeros with absolute value less than $f^a$.
                  Therefore, (\ref{lowerboundonSWP}) is a lower  bound
                 on the number of zeros in the left half plane, 
                 i.e.  for $ R > 4f^a$
                 \be\label{lowerboundonLHP} N( - 1 ,R) \geq     
                   \frac{ ((1-f)R)^{3/2} }{ f} \left[    \frac{s}{ 2 \pi} \right] 
                               + O( R f^{-17/18}) \ee 
                 
                 Of course  for any $R  > 0, N(-1,R) + N(1,R) = n_{B_f} (R) + O(f^{-2/5})$ (the $O(f^{-2/5})$ term comes from adding in the possible zeros on the negative imaginary axis).
                  Thus, combining (\ref{jenub2}) 
                 with $R' = R$ and  (\ref{lowerboundonLHP})
                 we have
               \be \label{ubRHP}   N(1,R) \leq   \frac{1}{\pi}\frac{R^{3/2}}{f}
                              \left[ \frac{8(v^{3/2}  - 1 ) }{9\log v } - \frac{s}{2} \right] 
                                  +         O(  R^{3/2} + R f^{-17/18}+ R\log^2(f^{-1} + R)    ) . \ee
                 Moreover, we have $  \cN_{R_l,R_u}[B_f] =  N(1,R_u)  - N(1,R_l)  $, 
                  combining (\ref{lowerboundonRHP}) with $R = R_u$  
                  and (\ref{ubRHP}) with $R = R_l$ it follows that                  \[   \cN_{R_l,R_u}[B_f] \geq  
                  \frac{s }{2\pi f}[  R_u^{3/2} + R_l^{3/2} -  (16/9s)
                           ( \frac{v^{3/2}  - 1  }{ \log v }) R_l^{3/2}] +        
                            O(    R_u f^{-17/18} + R_u^{3/2}+ R_l\log^2(f^{-1} + R_l)    ) \].
                                     
               Setting $s=1$ and using $\lim \frac{v^{3/2}  - 1  }{ \log v } = 3/2$
                the first term becomes $\frac{1}{2 \pi f}[R_u^{3/2} -  \frac{5}{3}R_l^{3/2}]$. 
                 Since we are not allowed to do this we step back a bit to obtain for $R_u > 2 R_l$ and small enough $f$
              \[   \cN_{R_l,R_u}[B_f] \geq  
               \frac{1-\epsilon }{2\pi f}[  R_u^{3/2} - \frac{5}{3} R_l^{3/2}]. \]

                Here dropping the error term is permitted since $ R_u > f^{1/100} $.

           On the other hand,
                 for  (\ref{ubRHP}) with $R = R_u$    
                  and (\ref{lowerboundonRHP}) with $R = R_l$  
                and again setting $s = 1$ and $(v^{3/2} -1)/\log v = 3/2$ for the first term in the upper bound,
                we find an upper bound of
                   $\cN_{R_l,R_u}[B_f] $ at $\frac{1}{\pi f}( \frac{5}{6} R_u^{3/2} -\frac{1}{2} R_l^{3/2})$.  
                   Thus stepping back and again dropping the error term we obtain
               \[   \cN_{R_l,R_u}[B_f] \leq  \frac{1+\epsilon }{2\pi f}[ \frac{5}{3} R_u^{3/2} - R_l^{3/2}].  
                       \qed \]
              \epf

                \subsection{Calculations of Jensen's and Carleman's integrals for $D_f$}\label{integralcalcs}
   
               The first order of business for both $I_J$ and $I_C$ is to find a radius
                   $r > 0$ so that the function $B_f$ does not vanish on the circle of radius $r$.       
             \bl   \label{cartanperimeter}  
               Suppose  $R> f^{a}$, $c>1$ , and $\delta > 0$.  Then there exists $f_{R,c} > 0$ such that for $f \in (0, f_{R,c})$ there is an $r$ with  $(1-\delta) R <  r < R$
                so that for all $z$ with $|z| = r$, we have $|B_f(z)| \geq  (e^{- 7} \delta)^{c \frac{4(2e R)^{3/2}}{3f}}$.
             \el
             This follows from Cartan's lemma, which formulates the principle that
             ``an analytic function cannot vanish faster than it grows".  
              We prove Lemma \ref{cartanperimeter} from the statement of 
              Cartan's lemma.  Further discussion  of Cartan's lemma is left  to Appendix \ref{cartansection}.

             For a function $\lambda$ which is analytic in $\{|z| < R\}$   let $M_\lambda(R) = \sup\{ |\lambda(z)|: |z| < R \}$.

                       \bt[Cartan]\label{cartanapplied}
                         Let $\lambda$ be a function analytic in the disk $\{z: |z| < 2eR\}$ with $\lambda(0) = 1$.  Thus
                         $|\lambda(z)| \le M_\lambda: = M_\lambda(2eR) $  for
                         $\{z: |z| < 2eR\}$.
                         Given $\delta > 0$  there is a collection of disks $(C_j)$
                      with sum of radii
                       \[          \sum r_j <   e^6 R\delta^{\frac{1}{\log M_\lambda }}  \]
                      so that  $\{z: |z| < R, \  |\lambda(z)| < \delta \} \subset  \cup_j C_j $
                       \et 
              To use this lemma we note: 
              \bl\label{maxD}
                  For any $c>1$ and sufficiently small $f> 0$ we have for $r > f^{a}$,
                 \[ M_{D_f}(r) \leq \frac{1}{2r \eta^2} e^{ c \frac{4}{3 f}  r^{3/2}}. \]
             \el
             This lemma follows immediately from the approximations in Lemma \ref{fddapprx}.
              We also need some control on $D_f$ in the vicinity of the origin.
              \bl \label{originexpansion}
                For any $\epsilon > 0$ the following approximation holds in $|z| < f^{2/3 + \epsilon}$:  
               \[  D_f(z)  =   f^{-1/3}[c_1 + c_2zf^{-2/3} + c_3f^{1/3} + O\left((|z|f^{-2/3} + f^{1/3})^2 \right)]
                     \]
                where $c_i$ depends on $\eta $ and $\kappa$ and all $|c_j| > 0$.
             \el            
              This approximation is calculated  in Section \ref{originDf}.
              \br
                 It follows immediately that $B_f(z) = \frac{D_f(z)}{D_f(0)}$ has an expansion 
                 $B_f (z) = 1 + \frac{c_2}{c_1}zf^{-2/3} + O((|z|f^{-2/3} + f^{1/3})^2)$ when $|z| < f^{2/3 + \epsilon}$ for some $\epsilon >0$.
               \er
             We now prove Lemma \ref{cartanperimeter}.  
             \bpf 
                    From Lemmas \ref{originexpansion} and \ref{maxD}, we have, for $R\geq f^a$,
                     for any $c > 1$ and small $f$
                    \[   \log  M_{B_f}(2eR)    
                           < c \frac{4}{3 f } (2eR)^{3/2}.  \]
                     Consider the circles $C_r= \{ z\in\bbR : |z | = r \}$. We determine how many of these may pass through 
                     a set where $B_f$ is small.
                     It follows from Theorem \ref{cartanapplied} that
                       $  S_R =   \left\{ z:|z| < R;  |B_f(z)| <  e^{ - k  c \frac{4 (2eR)^{3/2} }{3f}}   \right\}   $
                        is contained in a collection of disks
                      with total sum of radii $e^{6- k}R$.
                    Thus, for $k > 0$
                    \[  |\{ 0<r<R:   C_r \cap S_R \neq \emptyset    \}| 
                                   < 2  e^{6 - k}  R  \]
                   Given $\delta$ (small) choose $k >0$ so that  $\delta =   e^{7 - k} (> 2e^{6-k})$.  Then there is some $r$ with $(1 - \delta)R <  r < R$ 
                   so that $C_r \cap S_R = \emptyset$.  \qed
            \epf
          
              \subsubsection{Approximation of $\log B_f$}
              
              As discussed above, $D_f$ and therefore $B_f(z) = \frac{D_f(z)}{D_f(0)}$ 
              are entire.  We find approximations of $B_f$ 
             in the approximate sectors $0 < \theta < 4\pi/3$ and $ -2\pi/3 < \theta < 0$.
             This will cover (almost) the entire plane.
             
              Now in $\wt \cR_2$ we approximate $D_f$ from 
             Lemma \ref{fddapprx2}  and 
             find $D_f(0) = c_3+ c_1f^{-1/3} + O(f^{1/3})$ from Lemma \ref{originexpansion}. Thus
                \[ B_f(z)  = (D_f(0))^{-1} \left[ \left(i(2\eta\sqrt z)^{-1} + 1\right)^2 
                            + (4\eta^2 z )^{-1} e^{i4\kappa \sqrt z} \right] +f^{1/3} O(f^{2(1-2 a)}).  \]
              Therefore, for small $f> 0$, in $\wt\cR_2$ 
               \be \label{logB2}
                     \log|B_f(z)| = O(|\log f|).
               \ee

              On the other hand, in  $\cR_1$,
              from Lemma \ref{fddapprx2}  and Lemma \ref{originexpansion} 
                   we have for small enough $f$ that 
                 \[      \left|  \frac{F_1   }{ D_f(0)}\right| 
                             \left(1 -  C \frac{1}{\log(|z| + f^{-1})} \right)
                            < |B_f(z)| < 
                            \left|  \frac{F_1}{ D_f(0) }\right| 
                            \left(1+    C\frac{1}{\log(|z|+f^{-1})}  \right).\]
                  Now we approximate $\log|F_1|$.  We have
                  from Lemma \ref{gtermest} and the definition of $\mathcal{R}_1$ that
                  
                  \[\log |g_{\eta/\kappa} (2\kappa \sqrt z) | =O(|z|^{1/2} + \log f^{-1})\]
                in $\mathcal{R}_1$.
                 Thus, we have in $\mathcal{R}_1$
                 \[  \log|F_1| = - \frac{4   |z|^{3/2}}{ 3f}  \sin\left[\frac{3}{2} \arg z \right]+
                                                O(  |z|^{1/2} + \log|z|^{-1} + \log f^{-1})   \]
                 and the bound $|z| > f^a$ implies for 
                $z   \in \cR_1$ and small $f$
                 \be\label{logB1}
                  \log| B_f( z)|  =  - \frac{4   |z|^{3/2}}{ 3f}  \sin\left[\frac{3}{2} \arg z \right]
                                                   +     O( |z|^{1/2} + |\log f|)   . \ee

             \subsubsection{Proof of Proposition \ref{jensenintegral}}\label{jensproof}
              Here we estimate the Jensen integral $I_J$  of $B_f$.
              Recall that         
               $a$ satisfies $    a <  \frac12$   and we have 
             $R_L > 2 f^a$.
            \bpf   
                We will use approximations $F_1$ and $F_2$ in sectors covered by $\cR_1$ and $\cR_2$ respectively.
               In the remaining sets we use Cartan's lemma to control where $B_f$ may vanish.
               First, apply Lemma \ref{cartanperimeter} to $B_f$ with 
                $R = R_L$.  Then there exists $R_L (1-f) < r < R_L$ so that,
                 \be \label{loglb5}  
                 \log|B_f(z)| >  c \frac{4  (2e  R_L)^{3/2} }{3 f} \log(f/e^7) \ee
               for all $z$ so that $|z| = r$.
                  We now apply approximations   for
                  $\log|B_f| $ found above to the circle $\{z : |z| =  r\}$.

              Let $\epsilon' = r^{-3/2} f M \log(f^{-1} + r) $,  
               and $\epsilon'' =   r^{-1}(1 + r^{-1/2}) f M \log(f^{-1} + r)$.
              For 
              for $  \epsilon'< \theta < 4\pi/3 -\epsilon''$
              we have, $z = r  e^{i\theta}$ is in $\wt\cR_2$
              thus, by (\ref{logB2}) we have 
             \[   \int_{\epsilon'}^{4\pi/3 - \epsilon''} \log|B_f( r e^{i\theta})| \frac{d\theta}{2\pi}   
             =   O(|\log f|) .   \]

           For $-\frac{2\pi}{3}  +\epsilon'< \theta <- \epsilon'$,  
           $z = re^{i\theta} $ is in $\cR_1$.  Thus
           using (\ref{logB1})
                  \beq \nonumber
                         \int_{-2\pi/3 + \epsilon'}^{-\epsilon'}   \log|B(re^{i\theta})|  \frac{d\theta}{ 2\pi }
                           &= &    \int_{-2\pi/3 + \epsilon'}^{-\epsilon'}     
                            - \frac{4}{3} \frac{ r^{3/2}}{f} \sin\left(\frac{3}{2}\theta \right)  \frac{d\theta}{2\pi} 
                                   + O(|\log f| +   r^{1/2})\\
                          &=&\nonumber
                        \int_{-2\pi/3 }^{0}  
                         - \frac{4}{3} \frac{r^{3/2}}{f} \sin\left(\frac{3}{2}\theta \right)  \frac{d\theta}{2\pi} 
                         +   O(|\log f| +   r^{1/2})
                        \eeq
               The last integral is exactly  $\frac{8}{9\pi}  r^{3/2} f^{-1} $. 
               Here extending the integral from $[  {-2\pi/3 + \epsilon'},{-\epsilon'}   ]$ to   $  [  {-2\pi/3 },{0}    ] $
                gives an error of order $[\epsilon']^2 f^{-1}r^{3/2}$
               which is of order $ f   r^{-3/2} \log^2(r + f^{-1})  $.  But $r^{-3/2}  < f^{ - 3a/2}$,
               thus this amounts to an error of order $f^{1/4} \log^2(f^{-1})$ which is negligible.

               Finally, applying the  bound  (\ref{loglb5})
               over the remaining intervals
                $- \epsilon' < \theta < \epsilon'$ and $ - 2\pi /3 - \epsilon'' < \theta < - 2\pi /3 + \epsilon'  $,
              we obtain an additional error term of order $(\epsilon''  + \epsilon') r^{3/2} f^{-1} |\log f| $
               which is of order $ (1+ r^{1/2})\log^2 (f^{-1}+r)  $.\qed 
              \epf

             \subsubsection{Proof of Proposition \ref{carlemanhalfintegrals}} \label{carlsproof}
             We prove the proposition in two parts.
              That is,  in the next lemma we calculate $I_1$, and in the lemma following we calculate $I_2$.
             We include the calculations of both $B_f$ and $\wt B_f$ in these lemmas.
           We recall that         
               $a,b$ satisfy $ \frac12 - \frac1{100} <   a <  \frac12$ 
            and $  \frac23 < b < \frac23 + \frac{1}{100} $
             so that (\ref{parameterdiff}) holds and we have 
             $R_L > 2 f^a$ and $R_\rho$ so that $ C^{-1}f^b  <  R_\rho   < Cf^b$.

            \bl\label{axisintlemma} 
              For any    $R_L > r > R_L (1-f) $ and $ B =B_f,\wt B_f $ we have, for small enough $f$
              \[      I_1[B] =  \int_{R_\rho}^{r} \left(\frac{1}{s^2} - \frac{1}{r^2}\right) \log(|B ( i  s)B (-i  s)|)  \frac{ds}{2\pi}  
                            =  \frac{16\sqrt 2 }{15\pi  f}  r^{1/2} +      O(  f^{ - 17/18})   \] 
            \el

             \bpf        
             The proofs for both functions $B_f, \wt B_f$ are similar. 
             (This follows from the symmetry of the approximations for $B_f$ and $\wt B_f$.)
             We can therefore carry out the proof with  $B = B_f$  and  
             conclude a similar result for  $\wt B_f$.
             
             The integral $I_1$ will be split into 2 parts. 
             For $\{ |z | > f^a \}$ we may use approximations of $\log |B_f|$ from  (\ref{logB2}) and (\ref{logB1}).
              For $\{|z| < f^a\} $ these approximations   do not apply. 
             Instead we use bounds on the vanishing of the function $B_f$
             to control the integral.

             From Lemmas \ref{maxD} and \ref{originexpansion} we have, for any $c > 1$ 
             and small $f$ and  $R \geq f^a$, the upper bound,
             \be \label{logub}
               \log|M_{B_f} (2eR) |  \leq   c\frac{4}{3f}[2eR]^{3/2}  =: \overline M  
               \ee
              Now we apply Cartan to $B_f$ for $R = f^a$ 
              let $ A_k =   \{|s| <   f^{a}:    |B_f(i   s)|  <  f^{k \overline M }\}  $ we have,
                $|A_k| < e^7 f^{a + k}$.
                 Use H\"older's   inequality to bound the  integral
              \[   \int_{ R_\rho  }^{f^{a}} \left(\frac{1}{s^2} - \frac{1}{r^2 }\right)    \log|B_f(is) B_f(-is)|  ds .\]
              We will use exponents $p = 1.01 $ and $\frac{p}{p-1} = 101 $ for H\"older's inequality.
            For the first factor in the integrand, for small enough $f$, for any $p > 1$,
            (recall $C^{-1}f^b < R_\rho < Cf^b$)
             \be \label{axis2} |  \int_{R_l }^{f^{a}}   |\frac{1}{s^2} - \frac{1}{r^2}|^p  ds |^{1/p} 
                      \leq  | \int_{R_\rho }^{f^{a}}    \frac{1}{s^{2p}}   ds|^{1/p}  
                       < C'   f^{ -  b( 2  -p^{-1})  }   .\ee
              
              For the second integral, we partition the interval inside and outside the following set
                $E = \{ s:  f^{b}< |s| < f^a   ;|  B(is )| < 1\} = \cup_{k=0}^{\infty} A_k \setminus A_{k+1}$.
               Thus   we have
             \[   \left|  \int_{R_\rho }^{f^{a}}  {\bf 1}_E(s)    \log^{\frac{p}{p-1}}|B(i   s)B_f(-i   s)|  ds \right|<
                   \sum_{k\geq 0}   |\log^{p/(p-1)} f^{2( k+1) \overline M}|  |A_{k }| \]
              It follows, using $ a > 49/100$ and $p =1.01$, for  small enough $f$ that, 
                        
              \begin{align}  \label{axis1}       & \left|  \int_{R_l }^{f^{a}}   {\bf 1}_E(s)      \log^{\frac{p}{p-1} }  |B(i  s)B(-i  s)|  ds \right|^{(p-1)/p} <[ f^a  e^7   \left|2  \overline M \log  f \right|^{\frac{p}{p-1} }\sum_{k=0}^\infty   (k+1)^{\frac{p}{p-1} } f^{k} ]^{(p-1)/p} \nonumber \\
                   & \le f^ {-1 + 3a/2 + a(p-1)/p } \log f^{-1}
                      \end{align}
            
            Finally, consider the integral from $f^b$ to $f^a$ on the set $E^c$.  
             We use (\ref{logub}) to find $   \log  |B(is)B(-is)|  = O(f^{3a/2-1})$ 
             for $s \in [0,f^a] \backslash E $, then for small enough $f$
             \be \label{axis3}       | \int_{R_l }^{f^{a}}  
                         {\bf 1}_{\bbR \backslash E} (s)  \log^{\frac{p}{p-1}}  |B(is)B(-is)|   ds |^{(p-1)/p}
                   <        C f^{ -1 +  3a/2 + a(p-1)/p}
            \ee
            and (\ref{axis1}) dominates (\ref{axis3}).
             Now combining (\ref{axis2}) - (\ref{axis3}) we have 
            \be\label{axis4}
                  \int_{R_l }^{f^{a}}    \left(\frac{1}{s^2} - \frac{1}{r^2 }\right)    \log|B(is) B(-is)|  ds  
                      =  O( f^{-17/18} ). 
                   \ee


            The remaining part of the integral, along  $f^a$ to $r$ is calculated
             from the approximations  (\ref{logB2}) and (\ref{logB1}),
            thus we have  
           \[   \int_{f^{a}}^r \left(\frac{1}{s^2} - \frac{1}{r^2}\right)  \log |  B(is) B(-is)| ds  
                       =       \int_{f^{a}}^r \left(\frac{1}{s^2} - \frac{1}{r^2}\right)  
                                        \left( \frac{2\sqrt 2 s^{3/2}}{3f} + O(s^{1/2} + |\log f| )  \right)ds.
          \]  
          Integrating we obtain,
            \[      \int_{f^{a}}^r \left(\frac{1}{s^2} - \frac{1}{r^2}\right)  \log |  B(is) B(-is)| ds  
                       =\left.\frac{2\sqrt 2}{3f}(2s^{1/2}  - \tfrac{2}{5}r^{-2} s^{5/2} \right|_{f^a}^r  
                   + O\left( \cE_3  \right)
            \]
           where
           \[  \cE_3 =  \left| s^{-1/2}|^r_{f^a} \right| +   \left| r^{-2}s^{ 3/2}|^r_{f^a} \right|  
                             + |\log f| \left(   \left| s^{-1} |^r_{f^a} \right| + r^{-2} \left| s  |^r_{f^a} \right|  \right)    \]
          as $1/2 >  a$ for small enough  $f$ we have $O\left( \cE_3  \right) = O(f^{-1/2})   $.
          On the other hand,
           \[  \left.\frac{2\sqrt 2}{3f}(2s^{1/2}  - \tfrac{2}{5}r^{-2} s^{5/2} \right|_{f^a}^r
                                                       = \frac{16\sqrt 2}{15f}r^{1/2}  + O(f^{\frac{1}{2} a - 1})  \]
          Since $a > \frac{1}{2} - \frac{1}{100}$, $f^{\frac{1}{2 } a - 1 } = O(f^{-17/18})$,
            so combining this equation with (\ref{axis4}) completes the lemma.
          \qed 
           
           \epf

              We calculate $I_2[B]$ for upper radius
               $R_L > 2 f^{a }$    and lower radius $R_\rho$.
             \bl\label{arcintlemma} 
                Let $B = B_f,$ or $\wt B_f$.
                 For small enough $f$, there is an $r$   so that $R_L (1 - f) < r < R_L $ and 
                         \begin{align}
                              &\int_{-\pi/2}^{\pi/2} \left[
                     \log \frac{   |B(re^{i\theta})|^2   }
                                        {   |B(R_\rho e^{i\theta})|^{  (\alpha + \alpha^{-1} )   } } \cos\theta
                                       + [  \arg B(R_\rho e^{i\theta})  ]
                             \left( \alpha - \alpha^{-1} \right) \sin\theta
                          \right]    \frac{d\theta}{r2\pi} 
                          =
                      &+O(f^{-17/18} + \log (1+r) \log f^{-1}) 
                       \end{align}
                       where
                       $\alpha = r/R_\rho$.
              \el
                \bpf
               We break the integral into parts covered by (\ref{logB2}) and (\ref{logB1}), 
                 first we use Theorem \ref{cartanperimeter} to find a suitable radius.
                  Thus,  there exists $R_L (1-f) < r < R_L$ so that,
                 \be \label{loglb6}
                 \log|B_f(z)| >  c\frac{4  (2e R_L)^{3/2} }{3 f} \log(f/e
^7) \ee
               for all $z$ so that $|z| = r$.
                
                     Using (\ref{logB1}) we have,  for $\epsilon' = M  r^{-3/2}  f \log (f^{-1}+r )$
                  \be\label{seint1}
                         \int_{-\pi/2}^{- \epsilon'}   \log|B_f(re^{i\theta})|^2  \cos\theta \frac{d\theta}{r2\pi }
                          = \int_{-\pi/2}^{-\epsilon'} 
                      -2\left[\frac{4}{3} \frac{r^{3/2}}{f} \sin\left(\frac{3}{2}\theta \right) \right] 
                         \cos \theta \frac{d\theta}{r2\pi}  + O(  r^{ - 3/2} + r^{-1}|\log f|)
                       \ee
                 We will make the replacement
                   $2\sin(3\theta/2) \cos(\theta) = \sin(5\theta/2) + \sin(\theta/2)$ and
                   extend the  integral to $-\pi/2 < \theta  < 0$.
                   The extension of the interval creates a new error term, namely 
                   the integral over the set   $ - \epsilon'< \theta< 0$, which  is 
                  of order  $[\epsilon']^2r^{1/2}f^{-1}=  O(f^{-17/18} )$
                and thus
          \be\label{seint11}
                         \int_{-\pi/2}^{-\epsilon'}   \log|B_f(re^{i\theta})|^2  \cos\theta \frac{d\theta}{r2\pi }
                          =  \frac{8(3- \sqrt 2  )}{15\pi}  \frac{r^{1/2}}{f}
                        +  O( f^{-17 /18} ).
                        \ee
                        This gives the main contribution to the result for $B = B_f$.
                         The remaining part of the proof
                        for $B = B_f$ is to show the remaining terms are small.
                        
                  The integral in the upper interval $\epsilon' < \theta < \pi/2$ 
                  is seen to be of order $ |\log f| $ directly from  (\ref{logB2}).

                 For the remaining part we have, by  (\ref{loglb6}),
                       \be\label{seint2}
                             \left| \int_{-\epsilon'}^{\epsilon'}\log|B_f(re^{i\theta})|^2  \cos\theta \frac{d\theta}{r2\pi }  \right|
                                        <  4 c \epsilon'    \frac{4  (2e R_L)^{3/2} }{6\pi r f} |\log(f/e^7)| =    r^{-1}O(\log^2(f^{-1} +r ))             
                                         \ee
               which of course is of order $f^{-17/18}$.

               The radius of the inner circle $R_\rho$ is of order $ f^{b}$.
                Thus we have,  by Lemma \ref{originexpansion},   
               \be \label{originB}   B_f(R_\rho e^{i\theta}) =  1 + O(f^{b - 2/3}),\ee
                so that 
               $ \left|\log|B(R_\rho e^{i\theta})| \right|  =  O(f^{b - 2/3}) $.
               Moreover, $\alpha^{-1} + \alpha = O(rf^{- b})$, therefore,
                \be \label{inner1}  
                        \int_{-\pi/2}^{\pi/2} \log|B_f(R_\rho e^{i\theta})|^{(\alpha^{-1} + \alpha)} 
                        \cos\theta \frac{d\theta}{r 2\pi}  
                        = O(  f^{-2/3}).  \ee
                Finally,  we again use (\ref{originB}) to see the argument of $B_f$ along the inner arc is of order $ f^{b - 2/3} $
                 \be   \label{inner2}  
                      \left|\int_{-\pi/2}^{\pi/2}  \left( \alpha - \alpha^{-1} \right)   
                      \arg B_f(R_\rho e^{i\theta})  \sin\theta  \frac{d\theta}{r2\pi} \right| 
                          =   O(f^{-2/3}). \ee
                          This completes the proof for $B = B_f$.

            For the southwest half plane, ie integrating $\wt B_f (z) = B_f(\omega^{-1} z)$,
             the strategy is entirely the same, albeit with the change that
            we use (\ref{logB1})  in  $ \epsilon' < \arg(\omega^{-1} z)  <   \pi/2 $
             and  (\ref{logB2}) in $   -\pi/2 < \arg(\omega^{-1} z) < - \epsilon'' $ where
              $\epsilon'' =   r^{-1/2}(1 + r^{-1}) f M \log(f^{-1} + r)$.

              First we integrate $\wt B_f$ over $ \epsilon'$ to $\pi/2$ and as before we extend the interval to
               $0$ to $\pi/2$ which creates an error of order $f^{-17/18}$ .
              We thus obtain an equation similar to (\ref{seint11})
          \be\label{swint3}
                         \int_{\epsilon'}^{\pi/2}   \log|\wt B_f(re^{i\theta})|^2  \cos\theta \frac{d\theta}{r2\pi }
                          =  \frac{8(3- \sqrt 2  )}{15\pi}  \frac{r^{1/2}}{f}
                        +  O( f^{-17/18}).
                        \ee
              From (\ref{logB2}) we see  the integral over $-\pi/2$ to $-\epsilon''$ creates a term of
              order $|\log f|$.
              Finally, in this integral the gap between $\cR_1$ and $\wt\cR_2$ is of width $(\epsilon' + \epsilon'')$
              thus (\ref{seint2}) becomes
                       \be\label{swint4}
                             \left| \int_{-\epsilon''}^{\epsilon'}\log|\wt B_f(re^{i\theta})|^2  
                                 \cos\theta \frac{d\theta}{r2\pi }  \right|
                                        <   (\epsilon'+ \epsilon'') 2 c  \frac{4  (2e R_L)^{3/2} }{6\pi rf} |\log(f/e^7)| = 
                                           O\left( (1 + r^{-1})\log(f^{-1} +r )\log f^{-1}\right).
                                         \ee
               This completes the proof of Lemma \ref{arcintlemma}.                \qed
            \epf 

           We complete the section by proving 
            Proposition \ref{carlemanhalfintegrals}
             from Lemmas  \ref{axisintlemma}  and     \ref{arcintlemma}.
             The proof is similar for cases $B = B_f$  and $B = \wt B_f$.
             For small $f$, let $r$ satisfying $(1- f) R_L< r < R_L$ be chosen according to Lemma \ref{arcintlemma},
             then combine the integral with the result in Lemma \ref{axisintlemma}. 
            Thus we have 
            
            \begin{align}
            I_C (B; R_\rho ,  r) &= \frac{16\sqrt 2}{15\pi} \frac{r^{1/2}}{f } 
                                       +  \frac{8(3-\sqrt 2  )}{15\pi}  \frac{r^{1/2}}{f}  + O(f^{-17/18 } + \log(1+r) \log f^{-1}) \nonumber \\
                                 & =  \frac{r^{1/2}}{f}   \frac{8}{15\pi}(3+\sqrt 2)   +O( f^{-17/18} + \log(1+r) \log f^{-1}). 
                                  \qed  \end{align}

\appendix

\section{Green's functions under perturbation by delta functions}\label{greenbydelta}

  We review relations for the Green's functions of the Hamiltonian 
   $\fh_0 = p^2 + V$ on   $L^2(\mathbb{R})$ perturbed by two delta functions. 
  Let $\fh_1$ be a Hamiltonian on $L^2(\mathbb{R})$, so that
\[    \fh_1  =   \fh_0 + \tfrac{1}{\eta} \left(\delta_\kappa  +  \delta_{-\kappa}\right).  \]
We treat $\fh_1$ as a perturbation of a better understood operator $\fh_0$, 
although the $\eta > 0$ parameter is taken to be small.

  Let $G^{(i)}(x,y)$ be the integral kernel of the resolvent $(\fh_i - z)^{-1}$.  We suppose $\Im z > 0$.
  Let $\psi$ be the left solution (the solution which is square integrable near $-\infty$), and $\phi$ the right solution, of $(\fh_0 - z)\zeta= 0$. Then
 for $x < y$
  \[  G^{(0)}_z(x,y) = \cW^{-1} \psi(x) \phi(y)   \]
  where $\cW= \phi\psi' - \phi'\psi$.
 Let $\Delta = \eta^{-1}(\delta_{-\kappa} + \delta_{\kappa})$.
  From  the resolvent equation 
\[  (\fh_0-z)^{-1} - (\fh_i - z)^{-1} =  (\fh_0 - z)^{-1}\Delta(\fh_i - z)^{-1}  \]
 which implies for the Green's functions
\be \label{greenresolvent}
      G^{(0)}_z(x,y) = G^{(1)}_z(x,y) + \tfrac1\eta \left[     G_z^{(0)}(x,\kappa)G_z^{(1)}(\kappa,y) 
                        + G^{(0)}_z(x,-\kappa) G^{(1)}_z(-\kappa,y)       \right]  .\ee

Substituting $x = \pm \kappa$ into (\ref{greenresolvent}) we have the matrix equation,
\[    \left(\ba{c} G^{(0)}_z(\kappa,y) \\ G^{(0)}_z(-\kappa,y)  \ea  \right) 
         =   \left(  \ba{cc} 1 + \frac{1}{\eta} G^{(0)}_z(\kappa,\kappa) & \frac{1}{\eta} G^{(0)}_z(\kappa,-\kappa)  \\
                   \frac1\eta G^{(0)}_z(-\kappa,\kappa)  &   1 + \frac{1}{\eta} G^{(0)}_z(-\kappa,-\kappa)    \ea   \right)   
                     \left(\ba{c} G^{(1)}_z(\kappa,y) \\ G^{(1)}_z(-\kappa,y)  \ea  \right)   \]
Inverting this we have,
\be \label{greenrelation}            \left(\ba{c} G^{(1)}_z(\kappa,y) \\ G^{(1)}_z(-\kappa,y)  \ea  \right)   
                 =     \frac{1}{D(z)} 
                  \left(  \ba{cc}  1 + \frac{1}{\eta} G^{(0)}_z(-\kappa,-\kappa) & - \frac{1}{\eta} G^{(0)}_z(\kappa,-\kappa)  \\
                  - \frac1\eta G^{(0)}_z(-\kappa,\kappa)  &   1 + \frac{1}{\eta} G^{(0)}_z(\kappa,\kappa)    \ea   \right)  
                     \left(\ba{c} G^{(0)}_z(\kappa,y) \\ G^{(0)}_z(-\kappa,y)  \ea  \right)     \ee
 where, the determinant is
 \[  D(z) = 1  +  \frac{  G^{(0)}_z(\kappa,\kappa)  +G^{(0)}_z(-\kappa,-\kappa)     }{ \eta} +
                      \frac{   G^{(0)}_z(\kappa,\kappa)G^{(0)}_z(-\kappa,-\kappa)    
                      -    G^{(0)}_z(-\kappa,\kappa)  G^{(0)}_z(\kappa,-\kappa)   }{\eta^2} .  \]                    
 Now substituting (\ref{greenrelation}) into (\ref{greenresolvent}) we have a formula for
  the Green's function $G^{(1)}_z(x,y)$.

              
              Let $\alpha_\pm = \psi(\pm\kappa) \phi(\pm\kappa)$ and
              $\beta_{\pm} = \psi(\mp \kappa) \phi(\pm \kappa)$;
               notice $\alpha_+\alpha_- = \beta_+ \beta_-$.
                     With this notation, the determinant becomes,
              \be\label{determinant}
                D(z) = 1 + \frac{\alpha_+ + \alpha_-}{\eta \cW} + \frac{\beta_+ (\beta_- - \beta_+)}{  \eta^2\cW^2  }  
                \ee
                Let $g_\pm(x) = G^{(0)}_z(\pm \kappa , x)$.  Combining the above we have the formula
                \be\label{green}
                     G^{(1)}_z(x,y) = G^{(0)}_z(x,y) - \frac{1}{\eta^2\cW  D}
                        \left( \ba{cc} g_+(x)& g_-(x) \ea \right)
                        \left[ \ba{cc} \eta\cW + \alpha_- & - \beta_+ \\ - \beta_+ &  \eta \cW  + \alpha_+\ea  \right]
                        \left( \ba{c} g_+(y) \\   g_-(y) \ea \right).
                \ee     
                     
              \section{Analysis of relevant functions} \label{functions}
               Here we analyze the auxiliary functions arising from the shape resonance and the 
               shape resonance under the perturbation of an electric field.
              \subsection{The functions $F_\pm(s,t;w)  =   1 + i \frac{1 \pm e^{itw}}{sw}   $}\label{F+-}

               For  $s < t\in\bbR^+$, we consider the  zeros of 
                  \[ F_\pm(s,t;w)  =   1 + i \frac{1 \pm e^{itw}}{sw} . \]
                We first locate the real parts of the zeros within the intervals $[n\pi,(n+1)\pi]$, 
               this allows us an identification of the zeros of $F_\pm$ to $ 0 = 1 \pm e^{itw} $ at $ s \to 0$.
                Let $I_n^\pm = \pm[ n\pi   + (0,\pi)]$ for $n \geq 0$, we remark that the $\pm$ symbol
                 in the $I$ term is unrelated to that in the $F$ term.
                 Let $\bbI_n^\pm$ be the subset of $z \in \bbC$ so that
                 the real part of $z$ is in $I_n^\pm$.
                 For $n\geq 0$, we will associate the function $w^\pm_{2n}(s,t)$
                   to the function $F_+ $ and the interval $I^\pm_{2n}$;
                 respectively, associate  $w^\pm_{2n+1}(s,t)$ to the function $F_-$ and the interval $I^\pm_{2n+1}$.
                 These functions $(w_n^\pm)$ have the following property.
                \bp\label{singlesr}
                 Suppose $0< s, t$
                  then for all $n  \geq 0$,   there are functions
                   $w_{2n}^\pm:(s,t) \to t^{-1}\bbI^\pm_{2n}$ 
                   and $w_{2n+1}^\pm:(s,t) \to  t^{-1}\bbI^\pm_{2n+1} $
                  so that $0 = F_+(s,t; w_{2n}^\pm(s,t)) =  F_-(s,t; w_{2n+1}^\pm(s,t))  $.
             Moreover, these are all the zeros in the  plane.
                 \ep

                \bpf
                   First consider the behavior at  the origin.   $F_+(s,t;w)$ has a pole of order 1 at $w =0$ 
                and $F_-(s,t;0) = 1 + t/s$ at $w = 0$, as $s,t > 0$ there is no solution at the origin.

                 We simplify matters by multiplying by $itw$, 
                 which adds a zero at the origin for $F_-$ and removes the pole for $F_+$.
                 Let $q =\  ^t/_s$ and set $\zeta = q ( i ws -1)$ we have $F_\pm = 0$  
                 is equivalent to 
                \be\label{trig1}  \zeta e^{-  \zeta}      =   \pm  e^q. \ee
                 Let  $\zeta = x + iy$. The imaginary part
                  of the equation obtains $x$ in terms of $y$,
                 $x =   y \cot  y$.
                    On the other hand the real part of (\ref{trig1})
                    is $e^{-x}(x\cos y + y \sin y) = \pm e^q$ and replacing $x = y \cot y$ we find,
                    $y = \pm  e^q  (\sin y)  e^{ y \cot y}$.
                    In the last expression we multiplied by $ e^{y\cot y} $,
                    notice the replacement $x \to y \cot y$ creates nonremoveable singularities 
                    at $ y = \pi n$ for $n \in \bbZ$.
                    
                    First we will focus on $|y| > \pi$.
                    We determine  the image of $I_n^\pm$,  for $n \geq  1$,
                     under $   h(y) =  (\sin y)  e^{ y \cot y} $.
                    Find the derivative,   $h'(y) = - \frac{y - \sin 2y  }{\sin y} e^{y\cot y}$.
                     For $|y| > \pi$ we clearly have  $|y| >  |\sin 2y|$ so $h(y)$ is monotonic on $I^\pm_n$ for $n =1,2,..$
                     From the asymptotic form of the components of $h$ it is clear that 
                     $h(   [n\pi - 0]) = 0 $, and $h(   [ n\pi + 0]) =   (-1)^n \infty$.
                     
                     From these behaviors we have that for any $q > 0$,
                      $y = e^q h(y)$ has a unique  solution on $I_{2n}^\pm$ for all $n \geq 1$
                     and  $y =  - qe^q h(y)$ has a unique solution on $I_{2n+1}^\pm$ for all $n\geq 0$.
                      Of course,
                      $y = e^q h(y)$ has no  solution on $I_{2n+1}^\pm$ for any $n \geq 0$
                     and  $y =  - e^q h(y)$ has no solution on $I_{2n}^\pm$ for any $n\geq 1$.
                     
                       Now we consider $I_0^+$,   restate equation $y = \pm  e^q (\sin y)  e^{ y \cot y}$
                      as $ h_{(i)}  = \frac{y}{\sin y} e^{ - \frac{y}{\sin y} \cos y}  = e^q$, the $-$ branch is neglected 
                      as it clearly it has no solution on $I_0^+$ or $I_0^-$.
                      Notice $h_{(i)}(0) = e^{-1}$, and $h_{(i)} (\pi - 0) = \infty$.
                      But $h'_{(i)} =   [ \sin^2y - y\sin 2y + y^2  ]   (\sin^{-3} y) e^{- y \cot y}$, 
                      and writing $h_{(ii)}(y) = \sin^2 y - y\sin 2y + y^2 $ we note  
                        $h_{(ii)}'(y) = 2y(1 - \cos 2y) > 0$ for $y \in (0,\pi)$ 
                        so $h_{(ii)}(y) > 0$ for $y \in (0,\pi)$,
                       therefore $h'_{(i)} > 0$ for  $y\in (0,\pi)$
                        which shows there is only one solution in $I_0^+$. 
                        By the even-ness of $h_{(i)}$, we see that 
                        there is exactly one reflected solution in $y \in (-\pi ,0)$.
                        
                        Thus zeros are $\zeta = x + iy$ 
                        where $y$ is found above and $x$ is found as a function of $y$.
                        To complete  the proof, return to the orinal coordinates, i.e.
                        $\Re w = y/t$, and $\Im w = - (x + q)/t$. \qed
%
                 \epf

                 \bp\label{zero2ndorder}
                    For fixed $t>0$ given $n =0,1,2..$
                    and small enough $0 < s < s_n$ the approximation of $w_n(s,t)^\pm$  
                    to second order is, 
                     \[  w_{n}^\pm (s,t) = 
                           w_{  n}^\pm 
                            \left[1 - \frac st  + \left(\frac{s}{t}\right)^2 - i \frac { s^2}{2t} w_{  n}^\pm  \right]+ O_n(s^3)    \]
                 where, for $n \geq 0$,    $w_{ n}^\pm    = \pm(1+ n)\pi/t $.
                 \ep
                 
                   \bpf
                     From Proposition \ref{singlesr} the zeros of $F_\pm$ are in one to one
                      correspondence to the zeros of $ 1 \pm e^{itw}$ for small $s$, less the zero at $w = 0$.
                      Analyticity of $ w_{\nu n}^\pm (s)$ is clear from the requirement to solve  
                       $0 = F_+(s,t; w_{2n}^\pm(s,t)) =  F_-(s,t; w_{2n+1}^\pm(s,t))  $.
                       We can therefore consider each zero $w_n^\pm$ as an analytic function in a neighborhood of $s = 0$.
                      Let us define
                       $\wt F_\pm = -isw F_\pm  = - isw + 1 \pm e^{ i t w}$ and set $\wt F_\pm = 0$,
                      (this inserts  a root at zero for $F_-$ and removes a singularity for $F_+$,
                       we may ignore the behavior at zero)
                        then 
                      at $s = 0 $ we have the above solutions $w_{ n}^\pm = \pm (1+n)\pi/t$. 
                        Find  the first 2 derivatives, which must be identically 0,
                         \[   \tilde F'_\pm  =  - i sw' - iw \pm ( e^{itw} )iw't =0  \]
                         and 
                         \[   \tilde F''_\pm = -isw'' - i2w' \mp (e^{itw})(w't)^2 \pm (e^{itw})iw'' t = 0.\]
                        For both equations we use  the solution $1 \pm e^{ i t w} = 0 $, for $w = w_n^\pm$ at $s = 0$. 
                         So that the first two derivatives of  $w$ are $(w_n^\pm)' = -t^{-1} w_n^\pm$ 
                          and $(w_n^\pm)'' = 2t^{-2} (w_n^\pm) - i t^{-1} (w_n^\pm)^2$.
                          \qed
                   \epf

                  Independent of the previous estimates, we now find bounds 
                    for the zeros set of $F(s,t,w)$ for fixed $s$ and $t$, and $w$ in the right half plane.
                 In the following, let $w = u+ iv$.
 
                 \bp\label{Fzeroset} If $t >  s$
                      all zeros of $F_\pm$ for $u > 0$ are contained in the set
                 \[\{w : u > 0;    -(2t)^{-1} \log(1 +2[s u]^2)   < v <  - (4t)^{-1} \log ( 1 + [s u]^2)   \}\]
                   \ep
                \bpf Consider $  isw F_\pm = isw -1 \mp  e^{itw}$.
                 Let $w = u+iv$. Then $|isw - 1  |^2 = (su)^2  +  (1 + sv)^2$.
                   On the other hand $|e^{itw}|^2 = e^{-2tv}$.
                   
                   Clearly there are no zeros for $v>0$ since for all $\{w: v > 0\}$,
                    \[                                |isw - 1  |^2 = (su)^2  +  (1 + sv)^2
                                          >     1  \geq   e^{-2tv} =  |e^{itw}|^2 .   \]
                  Next consider $ \{w :  u> 0; v \leq - u\}.$ Let us fix $u = u_0$ and let $|v| = r$.
                 Consider the functions, for $v < 0$, 
                  \[\fz_1(r) = |e^{itw}|^2  = e^{2t r}  \tn{ and } \fz_2(r) = |isw - 1  |^2 = (su_0)^2 + (1 - sr)^2. \]
                   We also define $ \fz_3(r ) = (sr)^2 + (1 - sr)^2$.  Notice, for $r > u_0$, $\fz_3(r) > \fz_2(r)$.
                   Next observe that $\fz_1(r) > \fz_3(r)$.  This follows by noting that $\fz_1(0) = \fz_3(0) = 1$ 
                    and $\fz_1'(r) = 2te^{2tr} > 4s^2 r - 2s = \fz_3'(r)$.
                     Thus it remains only to consider the set $\{w : u > 0; 0\geq v > u\}$.

                     Now we establish the lower bound of the set.
                    Again fix $u = u_0>0$, and consider $r $ so that $(2t)^{-1} \log(1 + 2[su_0]^2) < r < u_0$,
                     then
                     \[ \fz_1(r) = e^{2t r }    > 1 + 2[su_0]^2\]
                       but 
                       \[  \fz_2(r) = [su_0]^2 + (1 - sr)^2  =  
                                                [su_0]^2  + [sr]^2 - 2sr + 1 < 2[su_0]^2 + 1 -2sr   \]
                    thus we clearly have $\fz_1(r) > \fz_2(r)$ in this set and therefore there are no
                    zeros of $F_{\pm}$ in the set $\{w : u > 0;   -  (2t)^{-1} \log(1 + 2[su]^2) > v > - u  \}$.
                    
                    Finally we demonstrate the upper bound of the set. Fix $u = u_0 > 0$ and 
                      consider $r$ so that $0 \leq r <(4t)^{-1} \log(1 + [su_0]^2)  $.
                      For such $r$ we have
                      \[\fz_1(r) < \left(1 + [su_0]^2\right)^{1/2 }   < 1 + [su_0]^2/2.\]
                    On the other hand, in this set $2sr < 2tr < \frac{1}{2} \log(1 + [su_0]^2) <  [su_0]^2/2$ and
                    \[  \fz_2(r) > [su_0]^2+1 - 2sr . \]
                     Thus $\fz_2 > \fz_1$ in this region.
                 \qed
                \epf  
                   For the final claim of Proposition \ref{Ftermzeroset}, it is trivial to check 
                   that  $F_{\pm} (s,t,w) = 0$ 
                    implies $F_{\pm} (s,t, - \overline{w}) = 0$.  
                    Therefore, the zeros in the left half plane are exactly the 
                   zeros in the right half plane reflected across the imaginary axis.
                  This completes the proof of Proposition \ref{Ftermzeroset}

              \

          \subsection{The function $g_b(w)  =  \sin w + bw\cos(w)  $}\label{gbproof}
          We fix  $b > 0$ and show $g_b$ has the following growth away from the real axis:
         \bl
              For $|w| >0 $ and $- \pi < \arg w < \pi$ , let $w = Re^{i\theta} = u + i v$.  Then
                \[              [bR +1]   \cosh|v|   >
                  |g_b(w)| >\sqrt{ (bR)^2 +1}     \sinh|v|. \]
        \el
        \bpf
         \bi
         \item[1.]
          Let $w = R e^{i\theta}$ then
            \[  g_b(w) =
                \tfrac{bR e^{i\theta} + i }{2} e^{-iRe^{i\theta}} + \tfrac{bR e^{i\theta} - i }{2} e^{ iRe^{i\theta}} 
                 =  
                  e^{iRe^{i\theta}} \frac{bRe^{i\theta} -i}{2} 
                   \left(1 + \frac{bRe^{i\theta} +i}{bRe^{i\theta } -i }    e^{- i2 Re^{i\theta}}   \right) . \]
                For all $Rb > 0$ and $ -  {\pi} < \theta < 0$
                   we have,  $ |bRe^{i\theta} +i|  < \sqrt{ (bR)^2 +1}<  |bRe^{i\theta } -i|$
                so
                \[        1 + e^{2R\sin\theta}>\left| 1 + \frac{bRe^{i\theta} +i}{bRe^{i\theta } -i }    e^{- i2 Re^{i\theta}}   \right|  
                           > 1 - e^{2R\sin\theta} .  \]
                Thus for all $  -  \pi  <  \theta < 0$ we have,
                \[        \tfrac{1}{2}     [  1 + e^{2R\sin\theta}]    [bR +1] e^{-R\sin\theta}
                 >    |g_b(w)| >\tfrac{\sqrt{ (bR)^2 +1}}{2} e^{-R\sin\theta} (1 - e^{2R\sin\theta}). \]
                
                \item[2.]
                    For $\pi > \theta >  0$ we use $|g_b(\bar w)| = |g_b(w)|$ and the result for $ -  {\pi} < \theta < 0$.
             \ei
        \epf

                  \section{Analytic functions} 
                  
                  We discuss here the variant of Carleman's Theorem
                  that is used in Section \ref{JCintegrals}.
                  \bt
                      Suppose $\lambda$ is an entire function with $\lambda(0) = 1$ and
                        $S= \{ z\in\bbC :  R_1 < |z| < R_2 ; \Re z > 0   \}$ . 
                         Let $\{z_j = r_j e^{i\theta_j}:  j = 1, ...\}$ be the zeros of
                          $\lambda$ in $S$.  Suppose $\lambda(z)$ is non-zero for $|z| = R_1$.
                      Then we have, for $\alpha = \frac{R_2}{ R_1}$
                       \beq \label{carlformula}
                         \sum_n \left(      \frac{1}{r_n} - \frac{r_n}{R^2_2}    \right)\cos \theta_n 
                            &=&  
                            \int_{R_1}^{R_2}\log(|\lambda(is)|\cdot| \lambda(-is)| )
                              \left( \frac{1}{s^2}  - \frac{1}{R^2_2}\right)\frac{ds}{2\pi} \\
                           & &\nonumber   +
                               \int_{-\pi/2}^{\pi/2} \left[
                     \log \frac{   |\lambda(R_2e^{i\theta})|^2   }
                                        {   |\lambda( R_1 e^{i\theta})|^{  (\alpha + \alpha^{-1} )   }   } 
                               \cos\theta
                            + [  \arg\lambda( R_1 e^{i\theta})  ]
                             \left( \alpha - \alpha^{-1} \right) \sin\theta
                          \right]    \frac{d\theta}{R_2 2\pi} 
                       \eeq
                  \et

                 \begin{remark}
                  Note that the zeros on the boundary of $S$ can be included in the statement of the theorem with the exception of those zeros $z$ with $|z| = R_1$ because the left side of (\ref{carlformula}) does not change if these zeros are included.
                  \end{remark}

                  \bpf
                  The key ingredient of the theorem is the integral 
                  \[    I_\gamma = \frac1{2\pi i} \int_\gamma \log \lambda (z) \left( \frac{1}{z^2} + \frac{1}{R^2_2}\right) dz  \]
                  for a positively oriented simple closed curve $\gamma$ with certain properties. We assume that $\lambda$ is non-zero on $\gamma$, that $\gamma(0) = \gamma(1)$ is nonzero and pure imaginary.  We assume that with the exception of the points $t = 0,1$, $\gamma(t) \in S$. We also assume that all of the zeros of $\lambda$ in $S$ are inside $\gamma$.
                  
                  Integrating by parts one obtains,
                  \[ I_\gamma = \frac{1}{ R_2 2\pi i}
                       \left( {\log \lambda} (\frac{z}{R_2} - \frac{R_2}{z}) \right) _{z = \gamma(0)}^{z = \gamma(1)} 
                      - \frac{1}{ R_2 2\pi i}  
                         \int_\gamma \frac{\lambda'}{ \lambda} \left( \frac{z}{R_2} - \frac{R_2}{z} \right) dz.   \]
                   The first term is purely imaginary and the real part of the second term is, in terms of the 
                   zeros $z_n =  r_n e^{i\theta_n}$ of $\lambda$,
                   \[     \Re I_\gamma =   \Re \frac1{R_2} \sum_n\frac{R_2}{z_n} -\frac{z_n}{R_2}  
                     = \frac{1}{R_2}\sum_n \left(\frac{R_2}{r_n}  -  \frac{r_n}{R_2 }\right)\cos \theta_n,   \]
                     which is valid for  the above described  curves $\gamma$.  Thus  
                     \[ \sum_n \left(\frac{1}{r_n}  -  \frac{r_n}{R^2_2}\right)\cos \theta_n = 
                    \Re   \int_\gamma \log \lambda (z) \left( \frac{1}{z^2} + \frac{1}{R^2_2}\right) \frac{dz}{2\pi i} .  \]
                     It is easy to see that we can let $\gamma$ approach the boundary of $S$ so that $\Re I_\gamma$ is continuous. The left side of the above equality does not change in this limit.  
                      The (real part of the) integral over the imaginary axis becomes,
                      \[  I_1 =  \int_{R_1}^{R_2}\log(|\lambda(is)|\cdot| \lambda(-is)| )
                              \left( \frac{1}{s^2}  - \frac{1}{R_2^2}\right)\frac{ds}{2\pi}    \]
                       The real part of the
                        integral over semicircles $\gamma_r= (re^{i\theta} : -\frac{\pi}{2} < \theta < \frac{\pi}{2})$ is
                       \beq 
                          \Re I_{\gamma_r}
                          &=& \nonumber
                           \int_{-\pi/2}^{\pi/2}   
                        \Re\left[  \log\lambda(z)  \left( \frac{1}{z^2} + \frac{1}{R_2^2} \right) \frac{z}{2\pi} \right]  d\theta \\
                          &=&\nonumber 
                             \int_{-\pi/2}^{\pi/2} \left[
                               \log|\lambda(re^{i\theta})| \left(\frac{R_2}{r} + \frac{r}{R_2} \right) \cos\theta
                            +   \arg\lambda(re^{i\theta}) \left( \frac{R_2}{r} - \frac{r}{R_2} \right) \sin\theta
                          \right]    \frac{d\theta}{R_2 2\pi}   \eeq
                       Of course
                        the second term on the right vanishes for $r = R_2$ so the only term involving the argument
                        is the integral along the arc at radius $r = R_1$.  It is important to note that it is independent of which (continuous) branch of $\arg z$ we choose.
                      Thus the real part of the integral over the curved edges of the horseshoe is
                      \beq     I_2   & = &\nonumber
                                  \Re( I_{\gamma_{R_2}} - I_{\gamma_{R_1}} )   \\
                             & = &\nonumber      \int_{-\pi/2}^{\pi/2} \left[
                     \log \frac{   |\lambda(R_2e^{i\theta})|^2   }
                                        {   |\lambda(R_1 e^{i\theta})|^{  (\alpha + \alpha^{-1} )   }   } 
                               \cos\theta
                            + [  \arg\lambda(R_1 e^{i\theta})  ]
                             \left( \alpha - \alpha^{-1} \right) \sin\theta
                          \right]    \frac{d\theta}{R_2 2\pi}.\qed  \eeq
                      \epf
                      \subsection{Cartan estimate}\label{cartansection}
                      We recall here the Cartan estimate.
                      \bt{(Cartan)}
                       Let $\lambda$ be a function analytic in the disk
                        $\{z: |z| < 2eR\}$ with $|\lambda(0)| =1$.
                      Suppose $a > 0$.  Then there is a collection of disks $(C_j)$
                      with sum of radii $\sum r_j < a R$, so that in the
                       set $\{z: |z| < R\}\backslash \left(\cup_j C_j\right) $ 
                       the estimate 
                       \[ \log|\lambda(z)| > - H_a\log M_\lambda(2eR) \]
                       holds with  $H_a = \log\frac{15e^3}{a}$.
                       \et
                       
                       From this we can derive the (essentially) equivalent statement:
                       \bt\label{cartanapplied1}
                         Let $\lambda$ be a function analytic in the disk $\{z: |z| < 2eR\}$ with $\lambda(0) = 1$ and 
                         with  $|\lambda(z)| <M_\lambda $.  
                      
                         Given $\delta > 0$  there is a collection of disks $(C_j)$
                      with sum of radii
                       \[          \sum r_j <   e^6 R\delta^{\frac{1}{\log M_\lambda }}  \]
                      so that  $\{z: |z| < R;  |\lambda(z)| < \delta \} \subset  \cup_j C_j $
                       \et
                        
                       \section{Details of resonances and associated functions}

  \subsection{The associated function for resonances of (\ref{Hds})} \label{dirichletdeterminantsect}
    Consider  $\phi\in L^2(-\infty , l)$ 
    so that $\tilde H_{\kappa,\eta,l} \phi = z \phi$ for $z$ in the upper half plane.
   The 
    solution to the left, which is square integrable at $-\infty$,
   of (\ref{Hds}) is
    $\psi_1(x) = e^{-i x \sqrt z}$ and the solution at the right,
    which satisfies a Dirichlet condition at $l$ is $\phi_1(x) = \sin((x-a) \sqrt z)$.
   The Wronskian becomes
   \[ \cW^{(0)} =    \phi\psi' - \phi'\psi = -\sqrt z e^{-i l \sqrt z} \]
   Then from the definitions of $\alpha_\pm$ and $\beta_\pm$ in Appendix \ref{greenbydelta},
   we have
   \[   \alpha_\pm = - e^{\mp i \kappa \sqrt z} \sin((a \mp \kappa)\sqrt z) \;  
   \tn{  and }
   \;  \beta_\pm = - e^{\pm i \kappa \sqrt z} \sin((l \mp \kappa)\sqrt z)   . \]
    So we have
   \[   \beta_- - \beta_+ = - e^{- li\sqrt z} \sin(2\kappa \sqrt z)  \]
   and thus,
   \[     \beta_+ (\beta_- - \beta_+ ) = e^{i (\kappa - l) \sqrt z } \sin((l - \kappa )\sqrt z) \sin(2 \kappa \sqrt z). \]
  On the other hand,
   \[  \alpha_+ + \alpha_- = i \left( e^{i l \sqrt z} \cos ( 2 \kappa  \sqrt z)  - e^{- i l \sqrt z}\right). \]
   We therefore have the function $D^{(0)}  = D$  as defined in (\ref{determinant}) 
  \begin{align}
     D^{(0)}  &= \label{detdirdd}
           1 + i \frac{1 - e^{i 2 l \sqrt z}\cos(2\kappa \sqrt z)  }{\eta\sqrt z}  
           -  \frac{ e^{i(\kappa + l)\sqrt z } \sin((\kappa - l)\sqrt z) \sin(2 \kappa \sqrt z) }{ \eta^2 z   } 
            \\
            \intertext{which can be expanded as}
            & =  \nonumber
            1 + \frac{i}{\eta \sqrt z} - \frac{i}{\eta\sqrt z}e^{i2l\sqrt z}\cos(2\kappa \sqrt z)
                    - \frac{1}{4\eta^2 z}\left( e^{i2(l+\kappa)\sqrt z} + 1 - e^{i4 \kappa \sqrt z} 
                    - e^{i2(l-\kappa)\sqrt z} \right).
      \end{align} 
      Alternately we may rewrite $D^{(0)}$ in the form:
      \beq
         D^{(0)}
           & = & \label{detdirdd2}
           \frac{  \left(   2\eta\sqrt z  +i \right)^2      +        e^{i4\kappa \sqrt z}}{4\eta^2 z}  
               -i \frac{e^{i2l\sqrt z}}{2\eta^2 z}g(2\kappa \sqrt z)
  \eeq
   for  $g_b(w ) =  \sin w + bw \cos w$  where $b = \eta /\kappa$.

   \subsection{The associated function for resonances of (\ref{Hss})} \label{appendixshapestark}

 In this appendix we discuss the determinant $D(z)$ for the Stark Hamiltonian  $H_{\kappa,\eta, f} $ for $f>0$.
 
 First we look at the Green's function for the Hamiltonian $H_f = H_{\kappa, \infty,f}$ for $\Im z > 0$.  This function is constructed from 
 the left and right
   square integrable solutions of the differential equation $-a''(x) + fxa(x) = z a(x)$.
  They can both be written in terms of the Airy function defined as $A(x) = \frac{1}{2\pi}\int_{-\infty}^{\infty} e^{i(tx + t^3/3)} dt$.  They are
  respectively,
\[  \psi(x) = A\left( - \omega  \frac{z}{ f^{2/3}} (1 - \tfrac{xf}{z} )\right) 
         \textnormal{ and }     \phi(x) = A\left( -    \frac{z}{ f^{2/3}} (1 - \tfrac{xf}{z} )\right)\]
where $\omega = e^{i2\pi/3}$.
  From  \cite{AS}, $A(0) = [3^{2/3}\Gamma(2/3)]^{-1}$ and
 $A'(0) = - [3^{1/3}\Gamma(1/3)]^{-1}$.  Using $\Gamma(1/3)\Gamma(2/3) = \frac{\pi}{\sin{\pi/3}}$
 we can compute the Wronskian:
 \[  \cW =  \psi' \phi  - \phi'\psi =   f^{1/3}\frac{\sin\tfrac{\pi}{3}}{3\pi}  (1 - \omega  )  =     f^{1/3}e^{-i\pi/6}/2\pi.\]
Thus it is an easy matter to write down the kernel of the resolvent of $H_f$.  Namely for $\Im z > 0$ and 
 $x < y$
\[    G^{(0)}_z(x,y) =  \cW^{-1} \psi (x) \phi(y) .  \]
As one might expect this is analytic everywhere. For a detailed analysis of this operator see for example \cite{liu}.

      \subsubsection{Behavior at the origin of $D_f$ for the shape resonance plus electric field}      \label{originDf}
            
             \bl \label{originexpansion2}
             Given $\epsilon > 0$, consider $D_f(z)$ where $|z|f^{-2/3}< f^{ \epsilon}$.  Then for small $f > 0$
             \[D_f(z) =  f^{-1/3}\left [c_1 + c_2( zf^{-2/3})  + c_3f^{1/3} + O((|z|f^{-2/3} + f^{1/3})^2) \right]\]
             with the $c_j$ depending only on $\eta$ and $\kappa$ and with $c_1$ nonzero.  
             \el

           \bpf

            The value of the Airy function and its derivative at zero are, (\cite{AS} 10.4.4; 10.4.5)
            \be\label{Airyatorigin}
            a_0 =  A(0) = \frac{1}{3^{2/3} \Gamma(2/3)} \textnormal{ and }
            a_1 =  A'(0) = \frac{-1}{3^{1/3} \Gamma(1/3)}.
            \ee
            Then the Maclaurin series is derived from the basic equation $A'' - z A = 0$. So $A''(0) = 0$ and
            if $A(z) = \sum a_k z^k$ we have $a_{k+3} = \frac{1}{(k+3)(k+2)} a_{k}$.  Thus
           \[  A(z) =   a_0\left(1 + \frac{z^3}{3!} + \frac{4 z^6}{6!} + \cdots \right)  + 
                       a_1 \left( z + \frac{2z^4}{4!} + \frac{2\cdot 5 z^{7}}{7!} + \cdots  \right)  = a_0 g_0(z) + a_1 g_1(z)  \]
             
             Let $\zeta_\pm = -zf^{-2/3} \pm \kappa f^{1/3}$, 
             then $\alpha_\pm = A(\omega \zeta_\pm)A(\zeta_\pm) $
             and $\beta_\pm = A(\omega \zeta_\mp) A(\zeta_\pm)$.
            We compute
             \begin{align}\alpha_\pm & = (a_0g_0(\zeta_\pm))^2 + \omega (a_1g_1(\zeta_\pm))^2 + a_0a_1(1+\omega) g_0(\zeta_\pm)g_1(\zeta_\pm),  \nonumber \\
            \beta_\pm & =   a_0^2g_0(\zeta_-) g_0(\zeta_+)  +a_1^2\omega g_1(\zeta_-)   g_1(\zeta_+)  + a_0a_1(g_0(\zeta_\mp)g_1(\zeta_\pm) + \omega g_0(\zeta_\pm)g_1(\zeta_\mp) ), \nonumber \\
            \beta_{-} - \beta_{+} & =  a_0a_1(\omega - 1)\left[ g_0(\zeta_-)g_1(\zeta_+) -  g_0(\zeta_+)g_1(\zeta_-)\right].
            \end{align}
            
            To compute $D_f(z)$ for small $z$ and small $f$ we introduce the symbol $|\zeta| = |z|f^{-2/3} +f^{1/3}$.  Using the Taylor series above we see that 
            \be \label{Wronskian}
            g_0(\zeta_-)g_1(\zeta_+) -  g_0(\zeta_+)g_1(\zeta_-) = (\zeta_{+} - \zeta_{-}) (g_1'(\zeta_-) g_0(\zeta_-) -  g_1(\zeta_-) g_0'(\zeta_-) ) + (\zeta_{+} - \zeta_{-}) ^2 O(|\zeta|^2).
            \ee
            Noticing that (\ref{Wronskian}) contains a Wronskian, we evaluate it at $0$ instead of $\zeta_{-}$ and find 
            \be
            \beta_{-} - \beta_{+}  = - a_0a_1(1-\omega)2\kappa f^{1/3} + O(f^{2/3}|\zeta|^2).
            \ee
            
            We calculate 
            
             \begin{align}
            \alpha_{+} + \alpha_{-} &= 2a_0^2 - 2a_0a_1(1+\omega) zf^{-2/3} + O(|\zeta|^2) \nonumber \\
            \beta_{+} & = a_0^2 - a_0a_1((1+\omega) zf^{-2/3} - (1-\omega) \kappa f^{1/3}) + O(|\zeta|^2) \nonumber \\
             \beta_{+}(\beta_{-} - \beta_{+}) &  =  a_0^3a_1(\omega -1) 2\kappa f^{1/3} - 2(a_0a_1)^2 \left [(\omega^2 - 1) 
                   \kappa (zf^{-2/3}) f^{1/3} + \kappa^2 (1-\omega)^2 f^{2/3} \right ]+ O(f^{1/3}|\zeta|^2). 
             \end{align}
            
            Thus using the value of $a_0a_1$
            
           \begin{align}
           D_f(z)& = 1 + \beta_{+}(\beta_{-} - \beta_{+})/(\eta \cW)^2 + (\alpha_{+} +\alpha_{-})/\eta \cW \nonumber \\
           &=  4\pi a_0^2 e^{i\pi/6} \eta^{-2}(\eta + \kappa) f^{-1/3}  + \frac{2i}{\sqrt 3}\eta^{-2} ( \kappa + \eta)(zf^{-2/3}) f^{-1/3} + 1 - 2 ( \frac{\kappa}{\eta})^2 +  O(f^{-1/3}|\zeta|^2)).
           \end{align} \qed
           \epf

\subsubsection{Properties of the Airy function}
We collect here some useful formulas for Airy functions from \cite{AS}. 
The first, for any $z\in \bbC$, is the identity of rotations of parameters of the 
 Airy function, (\cite{AS}, line 10.4.7),
\be\label{ar}
   A(z )  + \omega A(\omega z) + \bar\omega A(\bar\omega z) =0.
\ee
The following is the asymptotic expansion of the Airy function.
For $|\arg z| < \pi$, we have, (\cite{AS} 10.4.59)
\be\label{as59}
    A(z) \sim \frac1{2 \sqrt \pi   z^{1/4}}  e^{-\frac23 z^{3/2} }
             \sum_{k \geq 0} (-1)^k c_k \left[\frac3{2z^{3/2}}\right]^{k}   
\ee
where $c_0 = 1$ and $c_k = \frac{\Gamma\left(3k+\frac12\right)}{54^k k! \Gamma\left(k+\frac12\right)}$.
 The following two asymptotic formulas are derived from (\ref{as59}).  Let $\omega^{1/2} = e^{i\pi/3}$ and $\theta_\pm = 1 - \frac{\pm \kappa f}{z}$.
The asymptotic form for $z$ rotated by $\omega$ and $\bar \omega$ become,
for $f>0$ and $f^{-2/3}|z|$ large, in the region 
  $-\frac{2}{3}\pi  + \epsilon < \arg z < \frac{4}{3} \pi -  \epsilon $
\beq  \label{Ai5}
        A\left(-\omega \frac{ z  }{f^{2/3} }   \theta_\pm\right) 
        &=&  A\left(  \frac{z    }{f^{2/3} }  [\omega^{-1/2} \theta_\pm]\right) 
        \sim
          \frac{e^{i\pi/12} f^{1/6} }{(2 \sqrt \pi )z^{1/4} \theta_\pm^{1/4}} 
                e^{ i\frac{2}{3} \frac{[z\theta_\pm]^{3/2}}{f} }  S(\omega^{ - 1/2}\theta_\pm z)   
          \eeq
and in $- \frac{4}{3} \pi  + \epsilon < \arg z < \frac{2}{3 } \pi - \epsilon$
\beq  \label{Ai6}
        A\left(-\bar \omega \frac{ z }{f^{2/3} }   \theta_\pm\right)
        &=&  A\left(  \frac{z   }{f^{2/3} } [\omega^{1/2}  \theta_\pm]\right)  
         \sim 
          \frac{e^{-i\pi/12} f^{1/6} }{(2 \sqrt \pi )z^{1/4} \theta_\pm^{1/4}} 
                e^{ -i\frac{2}{3} \frac{[z\theta_\pm]^{3/2}}{f} }  S(\omega^{ 1/2}\theta_\pm z)   
\eeq
where 
\be \label{asymterm}  S(\theta z) =   \sum (-1)^k c_k \left(\frac{3f}{2 [z\theta]^{3/2}}\right)^{k} \ee
having required $\left| \arg( -\omega' z  )\right| < \pi - \epsilon$, 
  for $\omega'\in \{\omega,\bar\omega\}$.  Here by writing = we do not imply convergence of the series.
  
  Similarly for $-2\pi +\epsilon  <\arg z < 0 - \epsilon $
   
  \be\label{Ai7}
     A\left( -\frac{z \theta_\pm }{f^{2/3}}\right) =    A\left( e^{i\pi}\frac{z \theta_\pm }{f^{2/3}}\right)
        \sim 
        \frac{e^{ - i \pi/4 } f^{1/6} }{2\sqrt{\pi} z^{1/4} \theta^{1/4}_{\pm} } e^{i \frac{2[z\theta_\pm]^{3/2}}{3f}} 
        S( e^{i\pi} \theta_\pm z)
  \ee
  and for $2\pi  - \epsilon  > \arg z  >  0 +  \epsilon$
 
  \be\label{Ai8}
     A\left( -\frac{z \theta_\pm }{f^{2/3}}\right) =    A\left( e^{-i\pi}\frac{z \theta_\pm }{f^{2/3}}\right)
     \sim
     \frac{ e^{i\pi/4} f^{1/6} }{ 2\sqrt\pi   z^{1/4} \theta_\pm^{1/4}} e^{-i  \frac{2[z\theta_\pm]^{3/2} }{3f}}
        S(e^{-i\pi} \theta_\pm z).
  \ee
\subsubsection{Approximation of $D_f$ in the region $\cR_0 = \{|z| > f^{a}\} $.}\label{Dfapprx}
In this section we derive approximations for the function $D_f$ in the set $\cR_0$ from the 
asymptotic expansions stated above.
To determine the behavior of the determinant in this  region we will have to break it into 3
\tql patches\tqr\  where there are adequate asymptotic expansions for all terms.
First we state some elementary approximations which will be used frequently.
         First we expand powers of $\theta_\pm$ defined in Section \ref{mainproofs} in orders of $f$,
           and similarly powers of the product $\theta_+ \theta_- = 1 - (\kappa f/ z)^2$. 
           We have
        \be
          \label{thetaexpand}
        \theta_\pm^{p}   =  
            \sum_{k=0}^{\infty}  \frac{(p)_k}{k!} \left(\mp \frac{\kappa f}{z} \right)^k;\;
             \tn{ and }\;
            (\theta_+\theta_-)^p = \sum_{k\geq 0 }  \frac{ (-1)^k (p)_k}{k!} \left(\frac{\kappa f }{z}\right)^{2k}
            \ee
           where $(p)_k = p(p-1)\cdots (p - k + 1)$.          
          We apply the $\theta$ approximation to the exponential terms which gives
         \begin{align}      \label{exptheta} 
            e^{i\frac{2}{3f}z^{3/2}\theta_\pm^{3/2}} 
               &  =  e^{i\frac{2}{3f}z^{3/2} \mp i     \kappa z^{1/2} +i \frac{ f\kappa^2}{4}z^{-1/2}   } 
                  \left(1 + O(z^{-3/2}f^2) \right).  
         \end{align}

            The 4 expressions for the  \tql$S$-terms\tqr\ 
            in (\ref{Ai5}) - (\ref{Ai8})  result in two expansions, which are easily derived. 
            First notice that for small $f$
            \[\theta_\pm^p  = 1  + O \left( \frac{f}{|z|} \right)  =  1 + O(f^{1-a}) = O(1) \]
            as we are in the region $\cR_0$.  We also obtain
            \begin{align}
                S(e^{-i\pi}\theta_\pm z) 
                = S(\omega^{1/2} \theta_\pm z)  =& \nonumber \\
                \sum_k  i^k  c_k\left( \frac{3f}{2z^{3/2}\theta_\pm^{3/2}} \right)^k
                =& 1 + i c_1\frac{ 3f }{2z^{3/2} \theta_{\pm}^{3/2}}  + O\left(  f^{2}z^{-3} \right). \\
                \end{align}
                The last term follows since
                 $\theta_\pm = O(1)$ 
                   so we can replace all higher order terms with an error of the lowest order term 
                   which is $O (f^2 z^{-3}) = O(f^{2 - 3a})$.   Then applying the approximation
                    of $\theta_\pm = 1 + O(f/|z|)$
                   we have
                   \begin{align} \label{Si}
                S(e^{-i\pi}\theta_\pm z) 
                &= S(\omega^{1/2} \theta_\pm z) = 
                  1 + i c_1\frac{ 3f }{2z^{3/2} }     +f^2 z^{-5/2}  O\left( 1 + |z|^{-1/2}\right),  
                  \end{align}
                  \begin{align}
                S(e^{i\pi}\theta_\pm z) 
               & = S(\omega^{-1/2} \theta_\pm z)  \label{Sii}
                =\sum_k(-i)^k  c_k\left( \frac{3f}{2z^{3/2}\theta_\pm^{3/2}} \right)^k 
                = 1 - i c_1\frac{ 3f }{2z^{3/2} }   +f^2 z^{-5/2}  O\left( 1 + |z|^{-1/2}\right).
           \end{align}

              \paragraph{Eastern  patch}
              First we cover a region  given by the sector  
              $- 2\pi / 3 + \epsilon < \arg z < 2\pi / 3 - \epsilon$. 
              We will approximate $D_f $ to the first order in $f$ (after the exponential)
              by the function,
         \[  D^{(E)}
           :=
            \frac{1}{4z\eta^2}  \left\{  h_0(z)+
                 2 e^{i\frac{4z^{3/2}}{3f}   +  i\frac{f\kappa^2}{2\sqrt z}     }   
                    \left[ \left(   1 - i c_1\frac{ 3f }{ z^{3/2} }  \right)   g_{\eta/\kappa}(2\kappa \sqrt z)  
                          - i \frac{  \kappa\eta f}{ z^{1/2}}  \sin(2\kappa \sqrt z)    \right]
                               \right\} 
            \]
      where  
     $$h_0(z)  = \nonumber    \left( {2\eta \sqrt z}   + {i}\right)^2 + 
                          e^{i4\kappa \sqrt z} .$$
                          
                             We also use the following two error terms: 
     \[ \cE_1 = \frac{f^2}{z^{ 2}}
                 \left(1 +  |z|^{-2}  \right) 
              e^{i\frac{4z^{3/2}}{3f}    +   i\frac{f\kappa^2}{2\sqrt z} + 2\kappa |\Im \sqrt z|} .  
             \]
    \[ \cE_2 =   \frac{f^2}{z^{5/2}}  
           \left(   1  +   |z|^{-3/2}\right)         \left(  1  +  e^{ -  \kappa 4 \Im \sqrt z}\right) .             \]
      \bl
        For fixed $\epsilon > 0$, $a  < 1/2$, and sufficiently small $f>0$ we have the approximation 
        \[  D_f = D^{(E)} +  
                           O(\cE_1) + O(\cE_2)   \]
        in the set 
       \[     \{ z: |z| > f^{a}; |\arg z| < 2\pi / 3 - \epsilon   \} .   \]
       \el

\bpf
We apply (\ref{ar}) to the  $\beta_\pm $,
\be\label{betapmnes}
       \beta_\pm = A(-\omega \tfrac{z}{ f^{2/3}} \theta_\mp) A(-  \tfrac{z}{ f^{2/3}} \theta_\pm)
         =   A(-\omega \tfrac{z}{ f^{2/3}} \theta_\mp)
               [ - \omega  A(-\omega \tfrac{z}{ f^{2/3}} \theta_\pm)- 
                             \bar\omega  A(-\bar\omega \tfrac{z}{ f^{2/3}} \theta_\pm)].  \ee
The first term on the right is invariant under $+\to -$ so
\be\label{betadiff}  \beta_- - \beta_+ 
    =\bar\omega\left[  A(-\omega \tfrac{z}{ f^{2/3}} \theta_-) A(-\bar\omega \tfrac{z}{ f^{2/3}} \theta_+)]
         -
          A(-\omega \tfrac{z}{ f^{2/3}} \theta_+) A(-\bar\omega \tfrac{z}{ f^{2/3}} \theta_-) \right].\ee
              We apply formulas  (\ref{Ai5}) and (\ref{Ai6})  to  (\ref{betadiff})  
        to find (using $\theta_\pm - \theta_\mp = \pm (\theta_+ - \theta_-)$)
    \beq  \bar\omega  A\left(-\omega\frac{z}{f^{2/3}}  \theta_\pm\right) 
         A\left(-\bar\omega \frac{z}{f^{2/3}}  \theta_\mp \right)  
          & \sim & \nonumber
            \bar\omega \frac{f^{1/3} S(\omega^{-1/2}\theta_\pm z)
            S(\omega^{1/2}\theta_\mp) }{4\pi z^{1/2} [\theta_+\theta_-]^{1/4}}
                       e^{ \pm i \frac{2z^{3/2}}{3f} (\theta_+^{3/2}  - \theta_-^{3/2}) } .
     \eeq
      The ratio of the $S$ terms to the $\theta$ terms is 
      approximated with (\ref{thetaexpand}), (\ref{Si}) and (\ref{Sii}),
     \[    \frac{  S(\omega^{-1/2}\theta_\pm z)
            S(\omega^{1/2}\theta_\mp)   }{   [\theta_+\theta_-]^{1/4}  }  
            =   1 + O( f^{2}|z|^{-2})  +f^2 |z|^{-5/2}  O\left( 1 + |z|^{-1/2}\right)
             = 1 + f^2 |z|^{-2}  O\left( 1 + |z|^{-1}\right).  \] 
     We then apply (\ref{exptheta}) to obtain,  
     \beq  
        \bar\omega  A\left(-        \omega \frac{z}{f^{2/3}}  \theta_\pm\right) 
          A\left(-\bar\omega \frac{z}{f^{2/3}}  \theta_\mp \right)  
                       & = &\label{betanes1}
                   \bar\omega  \frac{f^{1/3} }{4\pi z^{1/2} }
                       e^{ \mp i 2 \kappa \sqrt z  }
                         \left( 1 +  O\left(  \frac{f^2}{ |z|^{3/2}}  \left( 1 + |z|^{-3/2}\right) \right) \right).
      \eeq
              Substituting into (\ref{betadiff}) gives,      
                         \beq
         \beta_- - \beta_+
           & = &\label{betanesdiff}
             i\bar\omega \frac{   f^{1/3}  }{2\pi \sqrt z }   \left[\sin\left( 2 \kappa \sqrt z \right) 
                        + O\left( f^2 |z|^{-3/2}  \left( 1+ |z|^{-3/2}\right)   e^{2\kappa |\Im\sqrt z| }  \right)
                          \right].
           \eeq
           We prepare to multiply by $\beta_+$; turning to   (\ref{betapmnes})
            we find we need to approximate the first term  on the right in (\ref{betapmnes}).
   The ratio of the $S$ term to the $\theta$ term is
   \[
          \frac{    S( \omega^{-1/2} \theta_+  )S(\omega^{-1/2} \theta_-)      }{  [\theta_+\theta_-]^{1/4}}
            =
           1 - i c_1 \frac{3 f}{z^{3/2}} + O\left( \frac{f^2}{z^2} (1 + |z|^{-1}) \right).
    \]
   Furthermore, applying (\ref{exptheta}) to the exponential terms we have   
   \[    
         -\omega A\left(-\omega\frac{z}{f^{2/3}}  \theta_+\right) 
          A\left(-\omega \frac{z}{f^{2/3}}  \theta_- \right)  
          = 
                 \frac{ e^{-i\pi/6} f^{1/3} }{4\pi \sqrt z }
                     e^{  i \frac{4 z^{3/2} }{3  f } + i\frac{f\kappa^2}{2 \sqrt z} }
                        \left(1       - i c_1 \frac{3 f}{z^{3/2}}  +
                         O\left( f^2 |z|^{-3/2}  \left( 1  + |z|^{-3/2}\right)  \right) \right) .
    \]

    We combine the term above  and (\ref{betanes1})  
    into (\ref{betapmnes}) to obtain,
    \begin{align}
           \beta_+ 
        =
                          \frac{ e^{-i\pi/6} f^{1/3} }{4\pi z^{1/2} }&\left[
                  \left(   e^{  i \frac{4 z^{3/2} }{3  f } + i\frac{f\kappa^2}{2\sqrt z}   } 
                              \left(1       - i c_1 \frac{3 f}{z^{3/2}}    \right)
                          +i e^{i2\kappa\sqrt z} \right)  \right.  \\
         &  \nonumber
             \left.   
                  \ \       +O  \left( \frac{f^2}{ z^{3/2}}    \left(1  +  |z|^{-3/2}  \right)  
                       \left(  \left| e^{  i \frac{4 z^{3/2} }{3  f }+ i\frac{f\kappa^2}{2\sqrt z}  } \right|
                            +  e^{ - 2\kappa \Im\sqrt z}\right)   \right)
                           \right] 
    \end{align}
    We combine the above and  (\ref{betanesdiff})  to obtain 
    \begin{align}
       \beta_+ (\beta_- - \beta_+) 
       & = \nonumber
                 e^{-i\pi /3}   \frac{f^{2/3}}{8\pi^2 z}     
                            \left(   e^{  i \frac{4 z^{3/2} }{3  f } + i\frac{f\kappa^2}{2\sqrt z}   } 
                              \left(1       - i c_1 \frac{3 f}{z^{3/2}}    \right)
                          +i e^{i2\kappa\sqrt z} \right) 
                   \sin\left( 2 \kappa \sqrt z \right)   + f^{2/3} \cE_\beta , \\
     \intertext{where the error term $\cE_\beta$ is,
    bearing in mind that $f|z|^{-3/2}= O(f^{1 - 3a/2}) = O(f^{1/4})$ in $\cR_0$,}
           \cE_\beta     &  =  \nonumber
                                      f^2 |z|^{-5/2}  \left( 1 + |z|^{-3/2}\right) \left[ 
                             e^{  i \frac{4 z^{3/2} }{3  f }+ i\frac{f\kappa^2}{2\sqrt z}  } 
                                        e^{ 2\kappa |\Im \sqrt z| }   
                                    +   1  + e^{ - 4\kappa\Im \sqrt z}        \right].
    \end{align}
    We divide by $(\cW \eta )^2$ to obtain,
    \beq
      D^{(\beta)} = \frac{\beta_+ (\beta_- - \beta_+)}{(\cW \eta)^2}
       & =&\nonumber
                        \frac{1  }{2\eta^2 z}                     
                    \left(   e^{  i \frac{4 z^{3/2} }{3  f }+ i\frac{f\kappa^2}{2\sqrt z}    }  \left(1   - i c_1 \frac{3 f}{z^{3/2}}  \right) +i e^{i2\kappa\sqrt z} \right)
                   \sin\left( 2 \kappa \sqrt z \right)   + \cE_\beta .  
    \eeq

    Now we turn to the term $\alpha_\pm$.
     We again apply (\ref{ar}) to the $\phi$ term to obtain
    \be\label{alphaeast}
    \alpha_\pm =A \left(  \frac{-  z \theta_\pm}{f^{2/3}} \right)  A \left(  \frac{-\omega z \theta_\pm}{f^{2/3}} \right) 
                        =  -\omega A^2\left(  \frac{-\omega z \theta_\pm}{f^{2/3}} \right) 
                            -\bar\omega  A\left(  \frac{-\omega z \theta_\pm}{f^{2/3}} \right)   
                                                         A\left(  \frac{-\bar\omega z \theta_\pm}{f^{2/3}} \right). 
    \ee
    After applying  (\ref{Ai5})  the first term on the right becomes 
    \[
      -\omega A^2\left(  \frac{-\omega z \theta_\pm}{f^{2/3}} \right) 
                \sim 
             \frac{e^{-i\pi / 6} f^{1/3} }{4\pi\sqrt z}
              \frac{S^2\left( \omega^{ - 1/2} \theta_\pm \right)}{\theta_\pm^{1/2}}
              e^{i\frac{4z^{3/2}}{3f} \theta_\pm^{3/2}  }.  
     \]
     The ratio of $S$ to $\theta$ terms is 
      \[     \frac{S^2\left( \omega^{ - 1/2} \theta_\pm \right)}{\theta_\pm^{1/2}} =
                1 \pm  \frac{\kappa f}{2 z}   - i c_1\frac{ 3f }{ z^{3/2} } + \frac{f^2}{z^{2}} O\left(1 + |z|^{-1}  \right).  \]
      Then applying (\ref{exptheta}) to the exponential term we have 
      \be \label{alphaeast1}
            -\omega A^2\left(  \frac{-\omega z \theta_\pm}{f^{2/3}} \right) 
               =
                        \frac{e^{-i\pi / 6} f^{1/3} }{4\pi\sqrt z}
              e^{i\frac{4z^{3/2}}{3f}   \mp i 2\kappa \sqrt z+ i\frac{f\kappa^2}{2\sqrt z}}
              \left(   1 \pm  \frac{\kappa f}{2 z}  - i c_1\frac{ 3f }{ z^{3/2} } + \frac{f^2}{z^{3/2}} O\left(1 +|z|^{-3/2}  \right)\right). 
      \ee
        Next we apply (\ref{Ai5}) and (\ref{Ai6}) to the second term on the right hand side of (\ref{alphaeast})
        and we immediately see the exponential terms cancel
      \beq
                 -\bar\omega A\left( - \omega \frac{z\theta_\pm}{f^{2/3}} \right)
                             A\left( -\bar  \omega \frac{z\theta_\pm}{f^{2/3}} \right)
                              & \sim &
                     \frac{  e^{i\pi/3}  f^{1/3}  }{  4\pi z^{1/2} }
                     \frac{S( \omega^{1/2}\theta_\pm z)  S(\omega^{-1/2}\theta_\pm z) }{  \theta_\pm^{1/2}} .
       \eeq
        On the other hand, the ratio of the $S$ terms to the $\theta$ terms is  constant to order 2 in $f$
        \[    \frac{S( \omega^{1/2}\theta_\pm z)  S(\omega^{-1/2}\theta_\pm z) }{  \theta_\pm^{1/2}}  =  1  + \frac{f^2}{z^{2}} O\left(1 + |z|^{-1}  \right)   \]
       Therefore we have,
       \be\nonumber
       \alpha_+ + \alpha_- 
                  =           
                     \frac{  e^{i\pi/3}  f^{1/3}  }{  2\pi \sqrt z }
                       \left(1 +  e^{i\frac{4z^{3/2}}{3f}  +  i\frac{f\kappa^2}{2\sqrt z}   }  \left[-i \cos(2 \kappa\sqrt z) 
                      \left(   1 - i c_1\frac{ 3f }{ z^{3/2} } \right) - \frac{\kappa f}{2z}  \sin(2\kappa \sqrt z)  \right]
                        \right) 
                       +  f^{1/3} \cE_\alpha
       \ee
         where 
        \[   \cE_\alpha =         \frac{f^2}{z^{5/2}} O\left(1 + |z|^{-1}  \right)   +
             \frac{f^2}{z^{ 2}}    O\left(1 + |z|^{-3/2}  \right)
             e^{i\frac{4z^{3/2}}{3f}    +   i\frac{f\kappa^2}{2\sqrt z}}  e^{2\kappa |\Im z|}.
             \]
             Thus

       \be\nonumber
       D^{(\alpha)} = 
       \frac{  \alpha_+ + \alpha_- }{ \cW\eta }      
                  =         
                  \frac{  i }{  \eta \sqrt z }
                      \left(1 +  e^{i\frac{4z^{3/2}}{3f}  +  i\frac{f\kappa^2}{2\sqrt z}   }  \left[-i \cos(2 \kappa\sqrt z) 
                      \left(   1 - i c_1\frac{ 3f }{ z^{3/2} } \right) - \frac{\kappa f}{2z}  \sin(2\kappa \sqrt z)  \right]
                        \right) 
                       +   \cE_\alpha.
     \ee
     So 
      \be
         D_f 
           =
           \frac{1}{4z\eta^2}  \left\{  h_0(z)+
                 2 e^{i\frac{4z^{3/2}}{3f}   +  i\frac{f\kappa^2}{2\sqrt z}     }   
                    \left[ \left(   1 - i c_1\frac{ 3f }{ z^{3/2} }  \right)   g_{\eta/\kappa}(2\kappa \sqrt z)  
                          - i \frac{  \kappa\eta f}{ z^{1/2}}  \sin(2\kappa \sqrt z)    \right]
                               \right\} + \cE_1 + \cE_2
      \ee
      where
      \beq
          h_0(z) & = &\nonumber    \left( {2\eta \sqrt z}   + {i}\right)^2 + 
                          e^{i4\kappa \sqrt z}   
      \eeq
     and the error terms $\cE_\alpha + \cE_\beta$ are of order $\cE_1 + \cE_2$.
      \qed
      \epf

   \paragraph{The Southern patch}
   The second patch covers the sector $- \frac{4}{3}\pi + \epsilon < \arg z < -\epsilon$.
   Define the function 
   \[     D^{(S)}(z)   =
             \frac{1}{4z\eta^2} \left\{\tilde h_0(z)
           +2 e^{i\frac{4}{3f}z^{3/2}  +  i\frac{f\kappa^2}{2\sqrt z}      }
          \left[ \left(  1   -  i c_1\frac{3f}{  z^{3/2}  }     \right)    g_{\eta/\kappa}(2 \kappa \sqrt z)
            - i \frac{  \kappa\eta f}{ z^{1/2}}  \sin(2\kappa \sqrt z)    \right]  \right\}  
             \]  
             
             with $ \tilde h_0(z) = \left(  2\eta \sqrt z   - i\right)^2 + e^{-i 4\kappa \sqrt z}  $ .
    \bl 
       For fixed $\epsilon > 0$, $a  < 1/2$, and sufficiently small $f>0$ we have the approximation 
               \[  D_f = D^{(S)}  +  O( \cE_1)  + O( \cE_2' )   \]
        in the set 
       \[     \{ z: |z| > f^{a};    -4\pi/3 + \epsilon  < \arg z <  - \epsilon   \} .   \]
    \el
   Here  
    \[ \cE_2' =   \frac{f^2}{z^{5/2}}   \left(   1  +  |z|^{-3/2}\right)   
          \left(  1  + e^{ - i\kappa 4\sqrt z}\right) .                   \]

    \bpf
   First  the $\beta_\pm$ are  approximated by applying $\bar \omega$ 
   times the rotation identity (\ref{ar}) to the $\psi$ term in (\ref{betaform})
   so we can write
   \be 
        \label{betasw}
      \beta_\pm = - A\left( - \frac{z}{f^{2/3} } \theta_\pm \right) 
                              \left[\bar \omega  A\left( -\frac{z}{f^{2/3}} \theta_\mp \right)  + 
                                  \omega A \left( - \bar \omega  \frac{z}{f^{2/3} } \theta_\mp \right)\right] .
    \ee
   A similar operation can be carried out for the $\alpha_\pm$ terms,
   \be
     \label{alphasw}
      \alpha_\pm = - A\left( - \frac{z}{f^{2/3} } \theta_\pm \right) 
                              \left[\bar \omega  A\left( -\frac{z}{f^{2/3}} \theta_\pm \right)  + 
                                  \omega A \left( - \bar \omega  \frac{z}{f^{2/3} } \theta_\pm \right)\right] .
   \ee
   The first term of (\ref{betasw}) is invariant under $+ \to -$ and therefore the difference of the $\beta$ terms   is   
   \[  \beta_ -  - \beta_+ = \omega  \left[
       A\left(- \bar\omega \frac{z}{f^{2/3}}\theta_- \right)  A\left(- \frac{z}{f^{2/3}}\theta_+ \right) 
           -       A\left(- \bar\omega \frac{z}{f^{2/3}}\theta_+ \right)  A\left(- \frac{z}{f^{2/3}}\theta_- \right) \right].  \]
   We apply the approximations (\ref{Ai6}) and (\ref{Ai7}) to the above formula giving
   \be \label{betasw2}
       \omega
      A\left( - \frac{z \theta_\pm}{f^{2/3}} \right)  
    A\left( -\bar\omega\frac{z \theta_\mp}{f^{2/3}} \right)  
        =    e^{ i\frac{\pi}{3}}\frac{   f^{1/3} }{4\pi \sqrt z }
      \frac{  S(\omega^{1/2}  \theta_\mp ) S(e^{i\pi} \theta_\pm ) }{(\theta_+\theta_-)^{1/4}} 
        e^{ \pm i \frac{2 z^{3/2} }{ 3f } \left(\theta_+^{3/2} - \theta_-^{3/2} \right)}.
        \ee
   We apply (\ref{thetaexpand}), (\ref{Si}) and (\ref{Sii}) to the   $\theta$ and    $S$ terms to 
    obtain the approximation
   \be\label{Sovertheta}
  \frac{  S(\omega^{1/2}  \theta_\mp ) S(e^{i\pi} \theta_\pm ) }{(\theta_+\theta_-)^{1/4}} 
      = 1  +    O\left( \frac{f^2}{z^2}\left(1 +   |z|^{-1}\right) \right). \ee
     Then  applying (\ref{exptheta}) to the exponential terms
   leads to
   \beq
   \beta_- - \beta_+ 
         &=&\label{betadiffsouth2}
         \frac{  e^{  - i\pi /6} f^{1/3}  }{2\pi \sqrt z   }  
         \left[ \sin(2\kappa \sqrt z) 
          +
         O\left(
          \frac{f^2}{|z|^{3/2}}     e^{  \kappa 2 |\Im\sqrt z| } 
          \left(  1 +   |z|^{-3/2} \right) \right)
         \right].
      \eeq
     Now we have from (\ref{Ai7})
     \be\label{betasw3}
         - \bar \omega  A\left(  -  \frac{z}{f^{2/3}} \theta_+  \right)A\left(  -  \frac{z}{f^{2/3}} \theta_-  \right)
                =
        e^{-i\pi/6} \frac{f^{1/3}}{4\pi \sqrt z} 
          \frac{  S(e^{i\pi}\theta_+) S(e^{i\pi} \theta_-)  }{ (\theta_+\theta_-)^{1/4}}
        e^{  i  \frac{2z^{3/2} }{3f}(\theta_+^{3/2} + \theta_-^{3/2})  }.
     \ee
    Using (\ref{asymterm}) and (\ref{thetaexpand})  we have the following approximation for the product of $S$ terms divided by
     $(\theta_+\theta_-)^{-1/4}$,     
      \be\label{Sovertheta2}
        \frac{S(e^{i\pi}\theta_+  )   S(e^{i\pi} \theta_-)}{(\theta_+\theta_-)^{1/4}} =
         1 - ic_1\frac{3f}{z^{3/2}} + O\left(   \frac{f^2}{z^{2}}  \left( |z|^{-1}+  1 \right) \right).
     \ee
     To complete the approximation, we apply (\ref{exptheta}) to the exponential term finding
     
     \[ 
         - \bar \omega  A\left(  -  \frac{z}{f^{2/3}} \theta_+  \right)A\left(  -  \frac{z}{f^{2/3}} \theta_-  \right)
         =
        e^{-i\pi/6} \frac{f^{1/3}}{4\pi \sqrt z} e^{i \frac{4 z^{3/2}}{3f}   +  i\frac{f\kappa^2}{2\sqrt z}       }
        \left[ 1 - ic_1\frac{3f}{z^{3/2}} + O\left(   \frac{f^2}{z^{2}}  \left( |z|^{-1}+  1 \right) \right) \right].
            \]
    We have the approximation, for $\beta_+$ of (\ref{betasw})
    combining  (\ref{betasw2}) and approximations (\ref{Sovertheta}) and (\ref{exptheta}),
    and   the approximation of (\ref{betasw3}),
    \begin{align}
    \beta_+ 
             = \nonumber
             \frac{e^{-i\pi/6}  f^{1/3} }{4\pi \sqrt z} &   
              \left[-i e^{ - i  2\kappa \sqrt z}
              +  e^{i \frac{4 z^{3/2}}{3f}   +  i\frac{f\kappa^2}{2\sqrt z}       }
               \left(         1 - ic_1\frac{3f}{z^{3/2}}\right) \right.\\
          & \nonumber
               \left.  + O\left( \frac{f^2}{z^2} 
                 \left( 
                 e^{  2\kappa \Im \sqrt z}
              +  \left| e^{i \frac{4 z^{3/2}}{3f}   +  i\frac{f\kappa^2}{2\sqrt z}       } \right|
                 \right)
                    \left(1+  |z|^{-1}\right) \right)  
               \right].
    \end{align}
    Thus, using the above and (\ref{betadiffsouth2}) we have the approximation
   \beq
      \beta_+ (\beta_- - \beta_+)  
         & = & \nonumber
          \frac{ e^{-i\pi/3}  f^{2/3}}{  16 \pi^2 z } 
            \left[   - \left( 1- e^{-i 4\kappa \sqrt z} \right)   +
             2\sin(2\kappa \sqrt z)   e^{i \frac{4 z^{3/2}}{3f} +  i\frac{f\kappa^2}{2\sqrt z}    }
              \left(         1 - ic_1\frac{3f}{z^{3/2}}\right) 
            \right] + f^{2/3}O( \cE_\beta).
   \eeq
   where
   \beq
      \cE_\beta
          & =  & \nonumber 
               \frac{f^2}{z^{5/2}  } \left( 1+ |z|^{-3/2} \right)
             \left(     e^{2\kappa |\Im \sqrt z|  }     
             \left| e^{i \frac{4 z^{3/2}}{3f} +  i\frac{f\kappa^2}{2\sqrt z}    } \right|
               + 
         1  +  e^{ \kappa 4 \Im \sqrt z}\right)  .
   \eeq
   So the $D^{(\beta)} = 
       \frac{   \beta_+ (\beta_- - \beta_+) }{    \eta^2 \cW^2  }$ term is
   \be\label{Dbetasouth}
     D^{(\beta)} =     \frac{ 1 }{  4  z \eta^2 } 
            \left[ -\left( 1- e^{-i 4\kappa \sqrt z} \right) 
             + 2   \sin(2\kappa \sqrt z) e^{i \frac{4 z^{3/2}}{3f}  +  i\frac{f\kappa^2}{2\sqrt z}   } 
            \left(         1 - ic_1\frac{3f}{z^{3/2}}\right) 
            \right] +  O\left(  \cE_\beta \right).
   \ee

   We now consider the $\alpha_\pm$ terms.
   First we use the (\ref{Ai7}) expansion to obtain,
   \[   - \bar\omega A^2\left(  - \frac{z}{f^{2/3}} \theta_\pm \right)  
     \sim
                     e^{-i\pi /6} \frac{f^{1/3} }{4\pi z^{1/2} }
                         \frac{S^2(e^{i\pi} \theta_\pm z) }{ \theta_\pm^{1/2}} 
                         e^{i \frac{ 4[z\theta_\pm]^{3/2} }{3f}} .
   \]
   The ratio of the $S^2$ term and the $\theta$ terms is approximated as
   \[   
                         \frac{S^2(e^{i\pi} \theta_\pm z) }{ \theta_\pm^{1/2}} 
                             =  
                         1 \pm \frac{\kappa f}{2z} -  i c_1\frac{3f}{  z^{3/2}  }  
                             +  O \left( \frac{f^2}{z^{2}} (1 + |z|^{-1}) \right).   \]
  On the other hand, the exponential term is 
  \[      e^{i \frac{ 4[z\theta_\pm]^{3/2} }{3f}}  =
               e^{i \frac{ 4z^{3/2}  }{3f}    \mp i  2\kappa \sqrt z +  i\frac{f\kappa^2}{2\sqrt z}    }  
               \left(  1 + O \left( \frac{f^2}{z^{3/2}} \right)  \right).  \]
  Thus,
  \be \label{alphasouth1} 
    - \bar\omega A^2\left(  - \frac{z}{f^{2/3}} \theta_\pm \right)   = 
            e^{-i\pi /6} \frac{f^{1/3} }{4\pi z^{1/2} }e^{i \frac{ 4z^{3/2}  }{3f}    \mp i  2\kappa \sqrt z +  i\frac{f\kappa^2}{2\sqrt z}    }  
          \left(  1  \pm \frac{\kappa f}{2z}   -  i c_1\frac{3f}{  z^{3/2}  }  + \frac{f^2}{z^{3/2}} O(1 + |z|^{-3/2})    \right) .
         \ee
   In  the second term, we use (\ref{Ai6}) and (\ref{Ai7}) noting that the exponential terms cancel 
   \beq
   -\omega A\left( -\frac{z}{f^{2/3}}\theta_\pm \right)
      A\left( -\bar \omega\frac{z}{f^{2/3}}\theta_\pm \right)
      & \sim &
      -\omega \frac{ e^{-i\pi /3} f^{1/3}}{4\pi \sqrt z \theta_\pm^{1/2}} 
            S(\omega^{1/2} \theta_\pm z)  S(  e^{i\pi }\theta_\pm z) . 
   \eeq
   The ratio of the $S$ terms to $\theta$ term is,
   \[   \frac{  S(\omega^{1/2} \theta_\pm z)  S(  e^{i\pi }\theta_\pm z) }{\theta_\pm^{1/2}  }=
                 \left( 1\pm \frac{f \kappa}{2z} + O(f^2/z^2) \right)     \left( 1 + \frac{f^2}{z^2} O(1+ |z|^{-1}) \right).\]
  Thus
  \[   -\omega A\left( -\frac{z}{f^{2/3}}\theta_\pm \right)
      A\left( -\bar \omega\frac{z}{f^{2/3}}\theta_\pm \right) 
       =   \frac{\bar \omega f^{1/3}}{4\pi \sqrt z } \left(1 \pm \frac{f  \kappa}{2z}    
                         +\frac{f^2}{z^2} O(1 + |z|^{-1})\right)  . \]
   Therefore, combining the above and (\ref{alphasouth1}), we have, 
   \begin{align}
       \alpha_+ + \alpha_-
       & =  \nonumber
           e^{-i\pi /6}  \frac{ f^{1/3}}{2\pi \sqrt z } 
            \left[ -i +  e^{i\frac{4}{3f}z^{3/2}  +  i\frac{f\kappa^2}{2\sqrt z}   }
            \left( \cos(2\kappa \sqrt z) 
               \left(  1   -  i c_1\frac{3f}{  z^{3/2}  }     \right)
               - i \frac{\kappa f}{2z} \sin(2\kappa \sqrt z)  \right)
                 \right] + f^{1/3} \cE_\alpha   \\
    \intertext{ where the error term is }
      \cE_\alpha  & =  \nonumber
           \frac{f^2}{z^{2}}   e^{ i \frac{4}{3f}z^{3/2} +  i\frac{f\kappa^2}{2\sqrt z}    }
                             O\left( e^{-i2\kappa \sqrt z} +  e^{ i2\kappa \sqrt z}  \right) 
                                   O(1 + |z|^{-3/2})   
                        +    \frac{f^2}{z^{5/2}} O(1+|z|^{-1}).
   \end{align}
   Finally, we have,
   \begin{align}
        \frac{  \alpha_+ + \alpha_-}{ \eta \cW   }
       & = \label{Dalphasouth}
             \frac{  1 }{ \eta \sqrt z } 
            \left[ -i +  e^{i\frac{4}{3f}z^{3/2}  v +  i\frac{f\kappa^2}{2\sqrt z}     }
                 \left( \cos(2\kappa \sqrt z) 
               \left(  1   -  i c_1\frac{3f}{  z^{3/2}  }     \right)
               - i \frac{\kappa f}{2z} \sin(2\kappa \sqrt z)  \right) \right]  +  \cE_\alpha . 
   \end{align}
   Now we combine (\ref{Dbetasouth}) and (\ref{Dalphasouth}) to find,
   \[
     D_f(z)    =
         \frac{1}{4z\eta^2} \left\{\tilde h_0(z)
           +2 e^{i\frac{4}{3f}z^{3/2}  +  i\frac{f\kappa^2}{2\sqrt z}      }
          \left[ \left(  1   -  i c_1\frac{3f}{  z^{3/2}  }     \right)    g_{\eta/\kappa}(2 \kappa \sqrt z)
            - i \frac{  \kappa\eta f}{ z^{1/2}}  \sin(2\kappa \sqrt z)    \right]  \right\} 
           +\cE_\alpha + \cE_\beta
   \]
   where 
   \[ \tilde h_0(z) = \left( \eta 2 \sqrt z   - i\right)^2 + e^{-i 4\kappa \sqrt z} .  \]
   \qed
   \epf
   \paragraph{The Northern patch}
   The third patch  covers the sector $\epsilon <  \arg z  <  \frac{4}{3}\pi - \epsilon$.
   Let $h_0$ be defined as in (\ref{gb}).

    \bl 
        For fixed $\epsilon > 0$, $a  < 1/2$, and sufficiently small $f>0$ we have the approximation 
                \[  D_f =  \frac{h_0(z) }{4z\eta^2} +   
           \frac{f^2}{z^{5/2}}   O\left( 1 + |z|^{-3/2}\right)  \]
        in the set 
       \[     \{ z: |z| > f^{a};    \epsilon  <\arg z <  4\pi/3 - \epsilon      \}.    \]
    \el
   \bpf
   In this patch we can use the (\ref{Ai5}) and (\ref{Ai8})
    expressions  in $\alpha_\pm$ and $\beta_\pm$.
   Observe:
   \[   
                 \frac{S(\omega^{ - 1/2}\theta_\pm z)   S(\omega^{ 1/2}\theta_\mp z)   }{[ \theta_+\theta_-]^{1/4}}
                     =
                 1 +  \frac{f^2}{|z|^2} O\left( 1+ |z|^{-1} \right);\]
 \[                \frac{S(\omega^{1/2}\theta_\pm z)   S(\omega^{- 1/2}\theta_\pm z)   }{[ \theta_\pm]^{1/2}}
               =  
              (    1   +f^2 z^{-5/2}  O\left( 1 + |z|^{-1/2}\right))      ( 1 \pm \frac{\kappa f}{2z} + O(f^2/|z|^2)  )
             =   1 \pm \frac{\kappa f}{2z} + (f^2/|z|^2) O ( 1 + |z|^{-1})  
               \]
    Thus we have the approximations
    \beq\label{betanw1}
      \beta_\pm 
      &=& 
      e^{i\pi/3}\frac{f^{1/3}}{4\pi z^{1/2}} 
                 \frac{S(\omega^{ - 1/2}\theta_\pm z)   S(\omega^{ 1/2}\theta_\mp z)   }{[ \theta_+\theta_-]^{1/4}}
                 e^{\pm i \frac{2 z^{3/2} }{3f}  (\theta_-^{3/2} -\theta_+^{3/2} ) }
                 \\
          & = &\nonumber
      e^{i\pi/3}\frac{f^{1/3}}{4\pi z^{1/2}} 
                 e^{ \pm i 2 \kappa \sqrt z  } 
        \left[
                 1 +  \frac{f^2}{|z|^2} O\left( 1+ |z|^{-1} \right)  \right] 
    \eeq
    and 
    \beq \label{alphanw1}
      \alpha_\pm & =& 
      e^{i\pi/3}\frac{f^{1/3}}{4\pi z^{1/2}} 
                 \frac{S(\omega^{1/2}\theta_\pm z)   S(\omega^{- 1/2}\theta_\pm z)   }{[ \theta_\pm]^{1/2}} \\
                 & = &\nonumber
      e^{i\pi/3}\frac{f^{1/3}}{4\pi z^{1/2}} 
                 \left[ 1 \pm \frac{\kappa f}{2z} + \frac{f^2}{|z|^2}O\left(1 +|z|^{-1}\right) \right].
    \eeq
    
    Thus we have 
    \[  \beta_- - \beta_+
          =  - e^{ i\pi /3 } \frac{f^{1/3} }{4\pi z^{1/2} }   
          \left[ 
          \left(     e^{ + i 2 \kappa \sqrt z  } -   e^{ - i 2 \kappa \sqrt z  }      \right)+  
           \frac{f^2}{|z|^{2}}
          e^{ -i 2 \kappa \sqrt z } O\left(1 +|z|^{-1}\right)    \right]
    \]
    and multiplying by $\beta_+$ and dividing by $(\eta \cW)^2$ we have,
    \beq  \frac{ \beta_+(\beta_- -\beta_+) }{   \eta^2\cW^2}  
       &=&
       \frac{ 1 }{ 4 z \eta^2}  
          \left[ 
          \left(     e^{ + i 4 \kappa \sqrt z  } - 1    \right)+   
              \frac{f^2}{|z|^{2}} O\left(1 +|z|^{-1}\right)   \right] 
    \eeq
    and we have also,
    \[ \frac{ \alpha_+  + \alpha_-}{  \cW \eta } = \frac{   i  }{ \eta   \sqrt z} 
                 \left[ 1   +  \frac{f^2}{|z|^2}O\left(1 + |z|^{-1}\right) \right]. \]
                 
      Thus in this region we have, 
      \beq
         D_f (z) & = &
         D_0(z) 
           + \frac{f^2}{|z|^{5/2}}   O\left( 1 + |z|^{-3/2}\right) \eeq
          where $ D_0(z) =  \frac{h_0(z)}{4z\eta^2} $ as calculated in Section \ref{resonancesection}
            \qed
      
      \epf

\end{document}